%% file: Splitters.tex
\documentclass[preprint,12pt,nonatbib]{elsarticle}

\usepackage{lineno,hyperref}
\modulolinenumbers[5]

\usepackage{amsmath, amssymb}
\usepackage{graphicx}
\usepackage{float}
\usepackage{booktabs}
\usepackage{comment}
\usepackage{rotating}
\usepackage{listings}
\usepackage{subcaption}
\usepackage{placeins}
\usepackage{longtable}
\usepackage{appendix}
\usepackage{multirow}
\usepackage{caption}
\usepackage{makecell}      
\usepackage{siunitx}       
\usepackage{lipsum}        
\usepackage{pdflscape}

\usepackage[style=numeric-comp, backend=biber, sorting=none]{biblatex}

\usepackage{amssymb}
\usepackage{amsmath}

\usepackage{lineno}
\usepackage{booktabs} 
\usepackage{graphicx} 
\usepackage{hyperref} 

\journal{Nuclear Instruments and Methods in Physics Research Section A}

\begin{document}

\begin{frontmatter}


\title{Conceptual Design of Highly-Constrained Splitters\\ for the FFA@CEBAF Energy Upgrade Study}


\author[JLAB]{R.M. Bodenstein\corref{CA}}
\cortext[CA]{Corresponding Author}
\ead{ryanmb@jlab.org}
\author[JLAB]{J.F.~Benesch}
\author[vcu]{A.M.~Coxe}
\author[JLAB]{K.E.~Deitrick}
\author[JLAB]{B.~Freeman}
\author[JLAB]{B.R.~Gamage}
\author[JLAB]{R.~Kazimi}
\author[JLAB]{D.Z.~Khan}
\author[JLAB]{K.E.~Price}
\author[JLAB]{B.~Schaumloffel}

\address[JLAB]{Thomas Jefferson National Accelerator Facility, Newport News, VA, USA}
\address[vcu]{Virginia Commonwealth University, Richmond, VA, USA}

\begin{abstract}
The Continuous Electron Beam Accelerator Facility (CEBAF) at Jefferson Lab is investigating a significant energy upgrade utilizing Fixed-Field Alternating-gradient (FFA) recirculating arcs. This upgrade requires the design of complex horizontal beam splitters to manage up to six concurrent beam passes. This paper presents the conceptual design of these splitters, which are subject to severe physical constraints imposed by the existing accelerator tunnel and multifaceted beam dynamics requirements for matching into the permanent-magnet FFA arcs. The design methodology, centered on multi-pass simulations in the Bmad toolkit, is detailed from the initial geometric layout through the advanced optics matching. Key results include a robust geometric arrangement that fits within the spatial boundaries and the development of multiple, flexible optics matching solutions. Furthermore, the design integrates a viable scheme for extracting high-energy beams for the experimental halls, a critical operational requirement. This work establishes a comprehensive and viable conceptual design, forming a baseline for future engineering and performance optimization studies.
\end{abstract}


\end{frontmatter}


\section{Introduction}
\label{sec:intro}
This section will briefly describe the Continuous Electron Beam Accelerator Facility (CEBAF) to remind the readers of the current machine, as well as its possible future. Relevant background information will be provided, as well as a description of the FFA@CEBAF energy upgrade study.

\subsection{\label{sec:12GeVCEBAF}CEBAF Background}

The Continuous Electron Beam Accelerator Facility (CEBAF) is a recirculating linear accelerator (RLA) capable of accelerating polarized electrons up to a nominal maximum energy of \SI{12}{GeV}. CEBAF consists of two superconducting LINACs, each capable of reaching \SI{1090}{MeV}. Each LINAC is connected by a stack of recirculating arcs, which magnetically spread the beam into energy-separated beamlines, bends the separated beams 180\textdegree~, and recombines them to be launched into the next LINAC.

In order to separate the different energy passes, each recirculating arc starts with a beam separation system, referred to locally (and throughout this paper) as the spreader. The spreader bends the beams in an upward trajectory using common dipoles, sending the lowest-energy beams furthest up due to the beam's lowest rigidity. Each successive pass is bent upward to a lesser degree. Each pass is then further separated through a two-step elevation change. These two steps are used in conjunction with three quadrupoles to control the dispersion and the optics, so that upon entering the arc proper, the optics are achromatic. After matching and transporting, the different passes are recombined in a section called the recombiner, which is designed to be a mirror image of the spreader. The beams are recombined and matched into the LINAC. There are five recirculating arc beamlines on each of the east and west sides of CEBAF, while the LINACs occupy the north and south areas.

CEBAF has four experimental halls: three on the southwest corner named Halls A, B, and C, and one on the northeast corner, named Hall D. To extract the beam to Halls A, B, and C, an RF separator system is used, which provides transverse kicks to the beam at the desired energy for the experiment. This RF kicker system takes advantage of CEBAF's fundamental frequency of \SI{1497}{MHz}, and the timing of the bunch structure for the beams assigned to each experimental hall. As beams pass through the RF kicker system, they are either kicked out toward the extraction system toward Halls A, B, and C, or allowed further recirculations to gain more energy.

Hall D is only capable of taking the top operating energy. Located in the northeast corner of the lab, it was added during the \SI{12}{GeV} Upgrade. Instead of an RF kicker system in the northeast corner, a 750 MHz Separator is used in the southwest corner to allow Hall D beams to pass through the West Arc and North LINAC one additional time, sending $\sim$\SI{12}{GeV} beam to the Hall.

\subsection{\label{sec:UpgradeIntro}The FFA@CEBAF Energy Upgrade}

CEBAF has roughly a decade of planned physics experiments that have been approved and scheduled to occur. After these experiments are completed, the future of CEBAF is uncertain. Furthermore, with the current plans for the Electron Ion Collider (EIC) to be constructed at Brookhaven National Lab (BNL), any future plans for studies at CEBAF should not overlap with the studies that can be performed at EIC. However, the parameter space is broad enough that interesting physics still exists, both by ways of complimentary studies and new, refined studies.

Jefferson Lab is currently looking at two possible upgrades to CEBAF that will allow such studies. It has been proposed that, at the end of the planned physics schedule, CEBAF undergoes an upgrade to allow for positron studies. For more information on this upgrade, please see \cite{gramesPositronBeamsCeBAF2023, voutierJeffersonLabPositron2024}.

After the positron experiments are completed, another upgrade proposal is to increase the energy reach of CEBAF. To accomplish this, rather than major upgrades to the LINACs, a smaller-scale proposal has been put forward: using Fixed-Field Alternating gradient (FFA) recirculating arcs (similar in concept to those of the Cornell BNL ERL Test Accelerator (CBETA)) to increase the total number of passes through the pair of LINACs, thus increasing the total achievable energy. In short, the current highest-energy recirculating arcs in both the east and west sides of CEBAF would be replaced with multipass FFA arcs, each allowing five or six passes, thus adding four or five additional passes through the LINACs, increasing the total achievable energy. This FFA@CEBAF study aims for an energy of over \SI{20}{GeV}, with numbers such as \SI{22}{GeV} being discussed as a possible future goal \cite{nissenDesignProgress222025}. The feasibility of this upgrade is being studied, and the details are discussed \cite{nissenDesignProgress222025, deitrickCEBAF22GeV23}. The total final energy, as well as the science case, is yet to be fully determined.

The FFA arcs will utilize Halbach-style permanent magnet arrays, which offer a compact solution for multipass acceleration \cite{brooksPermanentMagnetsCEBAF2022, brooksOpenmidplaneGradientPermanent23}. While these magnets will include small Panofsky-style multipole correctors for fine-tuning \cite{beneschCorrectorConceptFFA, coxeFFACEBAFAlignmentCorrector, coxeStatusErrorCorrection23, coxeBeamCorrectionMultipass2024, coxeErrorCorrectionAnalysis2024}, they are otherwise non-adjustable. This fixed-field nature necessitates precise control of the beam parameters at the entrance to the FFA arcs to ensure stable transport for all six passes.

\subsection{The Need for a Horizontal Splitter}
To achieve the required precision, dedicated beamlines known as horizontal splitters are essential. Drawing on experience from the CBETA facility \cite{bartnikCBETAFirstMultipass2020}, these splitters are designed to receive the co-linear bunch of up to six FFA-bound passes from the main linac and separate them horizontally into six independent beamlines. Within these lines, the optics for each pass must be individually manipulated to meet the strict matching conditions of the FFA arc.

This independent control must address multiple parameters simultaneously: the Twiss parameters ($\alpha, \beta$), dispersion and its derivative ($\eta, \eta'$), time-of-flight (ToF), and the momentum compaction term ($R_{56}$). Unlike the existing CEBAF arcs, the FFA arcs will not contain dogleg chicanes for path-length correction, shifting this crucial function entirely to the upstream splitters.

\subsection{Design Philosophy and Approach}
The design process was guided by a conservative philosophy best summarized as, ``measure twice, cut once.'' A pessimistic view was maintained, incorporating as many realistic constraints and design restrictions as possible from the outset, as it is far simpler to relax constraints later than to add them to a mature design. The primary focus was on achieving operational robustness, simplicity, and flexibility.

To this end, the conceptual design relies exclusively on conventional linear electromagnets that are either already in use at CEBAF, or are based on detailed designs produced for this project \cite{beneschFirstAttemptConductivelycooled, beneschSecondAttemptConductivelycooled, beneschThirdAttemptConductivelycooleda, beneschExtendedLambertsonFFA, beneschZAModificationConcept, beneschWatercooledCopperSeptum, beneschSimpleModificationDC2023a, beneschRectangularCommonDipoles, beneschConventionalDipoleFFA, beneschMagnetAllow11}. The dipoles were modeled with deliberately large transverse dimensions (\SI{0.5}{m} width in X and Y planes) to ensure ample flexibility for future magnet engineering and to guarantee that the layout could accommodate realistic components. The quadrupoles selected for use are already in use in the current CEBAF machine. The limits placed on the optimization routine were conservative, both to allow for them to be scaled in strength for the higher-energy southwest splitters, and to maintain magnet strength overhead in the worst-case scenarios.

While multifunction magnets present a potential alternative, they were excluded due to both the nonlinearities that real multifunction magnets introduce, and because they add a level of complexity which should only be used when needed. Since CEBAF does not currently use any nonlinear magnets, introducing nonlinear terms will complicate the overall operation of the accelerator.

\subsection{Scope of this Paper}
This paper presents the comprehensive conceptual design of the horizontal splitter for the FFA@CEBAF upgrade. Section \ref{sec:constraints} details the severe physical, geometrical, and beam dynamics constraints that define the design challenge. Section \ref{sec:geometry} describes the development of the geometric layout, including the magnet choices and the multi-pass separation and recombination scheme. Section \ref{sec:optics} presents the results of the optics design and matching simulations, demonstrating multiple viable solutions. Section \ref{sec:extraction} outlines the proposed scheme for high-energy beam extraction. Finally, Section \ref{sec:summary} summarizes the current status and discusses the direction of future work.

\section{Design Constraints and Requirements}
\label{sec:constraints}

\subsection{Physical and Geometrical Constraints}
As an upgrade, the splitter must be installed within the existing CEBAF tunnel, leading to highly restrictive spatial constraints. The design must preserve a mandated personnel access aisle (1.118 m) and accommodate existing tunnel infrastructure. The available space was determined from detailed measurements of the tunnel in the relevant regions. The primary transverse and longitudinal constraints, assuming the splitter occupies the spreader and extraction regions, are summarized in Table \ref{tab:constraints}.

\begin{table}[ht]
\centering
\caption{Summary of the primary physical constraints placed upon the splitter design geometry, derived from engineering measurements of the CEBAF tunnel \cite{bodensteinHorizontalSplitterDesign}.}
\label{tab:constraints}
\resizebox{\textwidth}{!}{%
\begin{tabular}{@{}llcc@{}}
\toprule
Name & Plane & Value & Units \\
\midrule
Wall to Beamline Center & Horizontal & 1.3716 & m \\
Beamline Center to Personnel Clearance Limit & Horizontal & 1.5665 & m \\
\textbf{Total Available Transverse Space} & \textbf{Horizontal} & \textbf{2.939} & \textbf{m} \\
LINAC Beam Height in CEBAF Coordinates & Vertical & y=100 & m \\
\textbf{Total Length in Z (to End of Final Magnet)} & \textbf{Longitudinal} & \textbf{z=92} & \textbf{m} \\
\bottomrule
\end{tabular}
}
\end{table}

An important note is that only the northeast and southwest corners of the CEBAF machine have adequate longitudinal space to accommodate these splitters, as the LINACs start near the beginning of each straight section, but end well before the end of the straight section. Furthermore, there are only two access points into the CEBAF tunnel with crane access for heavy equipment. They are located in the northwest and southeast corners. This limits the ability to place two splitters on each recirculating arc like CBETA, and instead requires one splitter at the beginning of each arc. This also increases the burden on each single splitter.

\begin{figure}[ht]
    \centering
    \includegraphics[width=1\textwidth]{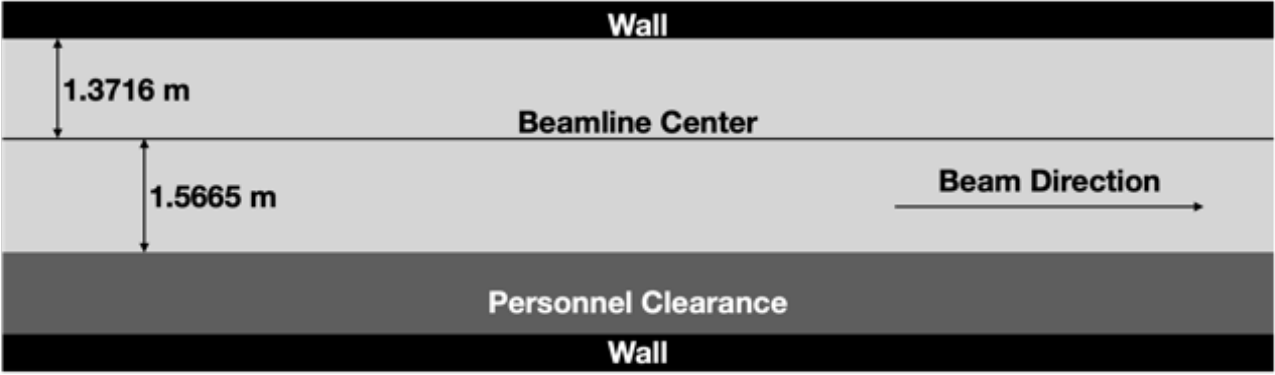} 
    \caption{Diagram of the transverse constraints in the CEBAF tunnel.}
    \label{fig:tunnel_constraints}
\end{figure}

A graphical representation of the transverse space is shown in Figure \ref{fig:tunnel_constraints}, illustrating the asymmetric transverse space available relative to the nominal beamline center.

\subsection{Beam Dynamics Requirements}
Each of the six passes enters the FFA arc at a unique energy and must be delivered to a specific, non-collinear horizontal position with precisely matched optics. The required input parameters for each pass at the entrance of the east FFA arc is listed in Table \ref{tab:match_points_beginning_only}. Please note, these parameters are match points at the beginning of an FFA cell, but the design of the FFA arcs are likely to change, necessitating changes in the matching parameters. Furthermore, different match points within the FFA cell can be used, as each cell is made up of segments, and the starting point can be adjusted as needed. This will be described further later in the paper.

It is also important to note the spread of X values listed in Table \ref{tab:match_points_beginning_only}. There is a $\sim \SI{3}{cm}$ difference between the highest and lowest energy passes. This value changes depending upon the match point within an FFA cell, and version of the FFA arc optics as well. They are also all negative values due to coordinate system definitions.

\begin{table*}[ht!]
\centering
\caption{Example entrance match parameters for the east FFA Arc. These values are at the entrance to the FFA Arc. Other match points in the FFA Cell can be used: this will be described later.}
\label{tab:match_points_beginning_only}
\resizebox{\textwidth}{!}{%
\begin{tabular}{@{}lcccccccc@{}}
\toprule
\textbf{Pass (E)} & \textbf{X (m)} & \textbf{Px} & \textbf{$\beta_{x}$ (m)} & \textbf{$\alpha_{x}$} & \textbf{$\beta_{y}$ (m)} & \textbf{$\alpha_{y}$} & \textbf{$\eta$ (m)} & \textbf{$\eta'$} \\
\midrule
1 (10.55 GeV) & -4.3609E-02 & 1.5555E-02 & 4.1508 & 3.0414 & 6.5160 & -3.1904 & 1.3845E-02 & -2.7713E-03 \\
2 (12.95 GeV) & -4.0953E-02 & 1.4756E-02 & 2.9513 & 1.8208 & 6.4812 & -3.0391 & 3.4001E-02 & -1.1549E-02 \\
3 (14.95 GeV) & -3.6297E-02 & 1.3082E-02 & 2.7191 & 1.5387 & 7.0035 & -3.2108 & 5.0554E-02 & -1.8809E-02 \\
4 (17.15 GeV) & -2.9986E-02 & 1.0682E-02 & 2.6034 & 1.3996 & 8.0521 & -3.6443 & 6.4348E-02 & -2.4885E-02 \\
5 (19.35 GeV) & -2.2289E-02 & 7.6691E-03 & 2.5224 & 1.3113 & 10.173 & -4.5685 & 7.5975E-02 & -3.0021E-02 \\
6 (21.55 GeV) & -1.3418E-02 & 4.1373E-03 & 2.4564 & 1.2473 & 17.058 & -7.6233 & 8.5873E-02 & -3.4401E-02 \\
\bottomrule
\end{tabular}
}
\end{table*}

On top of optics matching requirements, the splitters must manage the ToF and $R_{56}$ for each pass. The ToF must be controlled to within a fraction of an RF degree to ensure proper acceleration in the subsequent linac pass. This requires compensating for the path length differences between the passes within the FFA arcs and also adjusting for thermal expansion and contraction of the accelerator throughout the year, a variation that currently requires daily adjustment of CEBAF's dogleg magnets.

\subsection{Operational Requirements}
Beyond the primary beam dynamics functions, the splitter design must incorporate several key operational features. First, it must provide a mechanism for extracting the high-energy FFA passes to the experimental halls, as no other system is planned for this purpose. Second, the layout must reserve adequate space for essential beamline hardware, including beam position monitors (BPMs), diagnostic viewers, corrector magnets, vacuum pumps, and valves, without which the system cannot be reliably commissioned or operated.

\section{Geometric Layout}
\label{sec:geometry}

The geometric design process prioritized fitting all necessary components into the constrained space before optimizing the optics. The Bmad simulation toolkit \cite{saganBmadRelativisticCharged2006} was used for this task due to its ability to model multiple beamlines with realistic, solid magnet geometries simultaneously.

\subsection{Magnet and Component Design}
The design utilizes two main types of dipoles: \SI{3}{m} long main dipoles and a few shorter \SI{1.5}{m} dipoles for smaller chicanes. All are modeled as \SI{0.5}{m} wide electromagnets, based on conservative engineering analysis which established a field limit of approximately \SI{1.8}{T} \cite{beneschConventionalDipoleFFA}. These designs are specifically made thick as a further design constraint: if the geometry works with magnets of this size but the tolerances are too tight, it is possible to make these dipoles thinner in the relevant plane. Using oversized models for the dipoles ensures that any future, more optimized magnet designs will fit within the established layout. The quadrupoles used in this design are already existing CEBAF QR-type magnets \cite{hiattFabricationMeasurement12}, which are well-understood and capable of providing the required gradients. Care was taken to limit the quadrupole strengths in the simulations so that they can be scaled for the higher-energy southwest splitters and still maintain adequate overhead for the quadrupoles.

\subsection{Multi-Pass Separation Scheme}
The six beam passes enter the splitter co-linearly and are first separated by a common dipole that bends all passes toward the side of the tunnel with more available space (the aisle-side, toward the right from the beam's perspective). Subsequent shared and individual dipoles continue the separation process. Due to the tight packing, avoiding physical collisions between magnets on adjacent beamlines is a major challenge. A key difficulty of the layout is the handling of the two highest-energy passes. To create sufficient clearance from the lower-energy lines, they are bent in the opposite direction, requiring their orbits to cross over and then cross back to ensure they enter the FFA arc in the correct order (lowest energy on the inside, highest on the outside). This would not be necessary if the total number of FFA passes was reduced from six to five, as the beam trajectories would not cross and need to be uncrossed.

\subsection{Time-of-Flight and $R_{56}$ Management}
To provide the necessary range of path-length adjustment, each of the six independent beamlines incorporates at least one large chicane. Simple vertical stacking of the chicanes was not possible due to the 2.939 m transverse space limit. Instead, the chicanes are nested within each other, like Matryoshka dolls. To further enhance operational flexibility, especially for the lowest-energy pass, smaller ``mini-chicanes'' are included in series with the main chicane.

\begin{figure}[ht]
    \centering
    \includegraphics[width=1\textwidth]{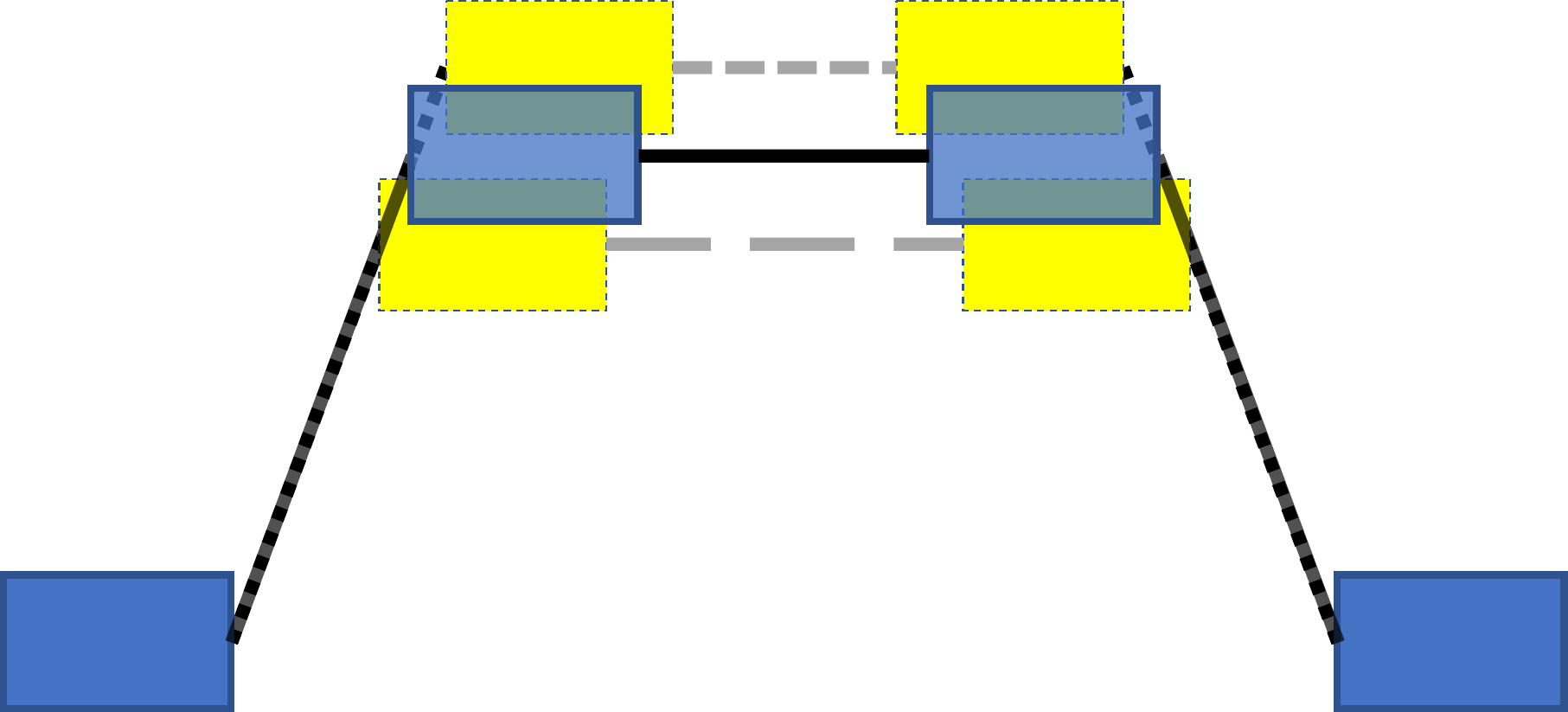} 
    \caption{Diagram of the proposed mechanical movers that may or may not be required for path length correction.}
    \label{fig:mover_graphic}
\end{figure}

To accommodate regular and seasonal path-length changes (which are normally controlled in the doglegs in the current CEBAF machine, but in extreme cases the RF frequency can be adjusted), the design allows for the potential inclusion of mechanical movers on the central dipoles of each chicane. As demonstrated at CBETA, such a system allows for continuous adjustment of the chicane amplitude, providing fine control over the ToF. Figure \ref{fig:mover_graphic} shows graphically how these mechanical movers may work, if needed.

$R_{56}$ adjustments must also occur in the splitters. However, the final $R_{56}$ values are still unknown, as parts of the machine are still under design. The largest unknown contributor to the $R_{56}$ that will require compensation in the splitters is the transition section, which is meant to recombine and match the beams from the FFA Arc into the recombiner and LINAC \cite{gamageResonantMatchingSection2024, gamage:napac2025-tup029}. Currently, $\sim \SI{15}{m}$ are reserved for the transition section. Recent design progress \cite{gamage:napac2025-tup029} indicates that the required space is more likely to be around \SI{50}{m}. Since the length and contribution of this section (and others) are not currently known, only the contributions from the known sections are included in the $R_{56}$ compensation features of the splitters. These values are considered a preliminary ``soft target'' $R_{56}$ value to be used until the design is finished, and then re-matching will be required.

\subsection{Recombination and Matching Section}
After discussions with CEBAF Operations crew, and investigations into beam loss and optics complications in the current CEBAF machine, a symmetric recombination scheme using common dipoles was rejected in favor of an independent recombination and matching section for each pass. One of the main rationales for this is the beam loss that CEBAF currently experiences in the recombiner regions, which uses common dipoles to recombine separate energy lines.

This separate-beamline approach, while requiring more magnets, offers superior operational flexibility. It decouples the matching of each pass from the others and provides a dedicated section at the end of most lines. This final section is critical for placing the quadrupoles needed for the final, precise optics match into the FFA arc. It also allows for finer adjustment and matching into the FFA Arcs, including the requirement of non-collinear recombination of the beams. While a symmetric design may be possible, it will lack the operational robustness and flexibility of the design presented here.

\subsection{Overall Floor Plan}

\begin{figure}[ht]
    \centering
    \includegraphics[width=\textwidth]{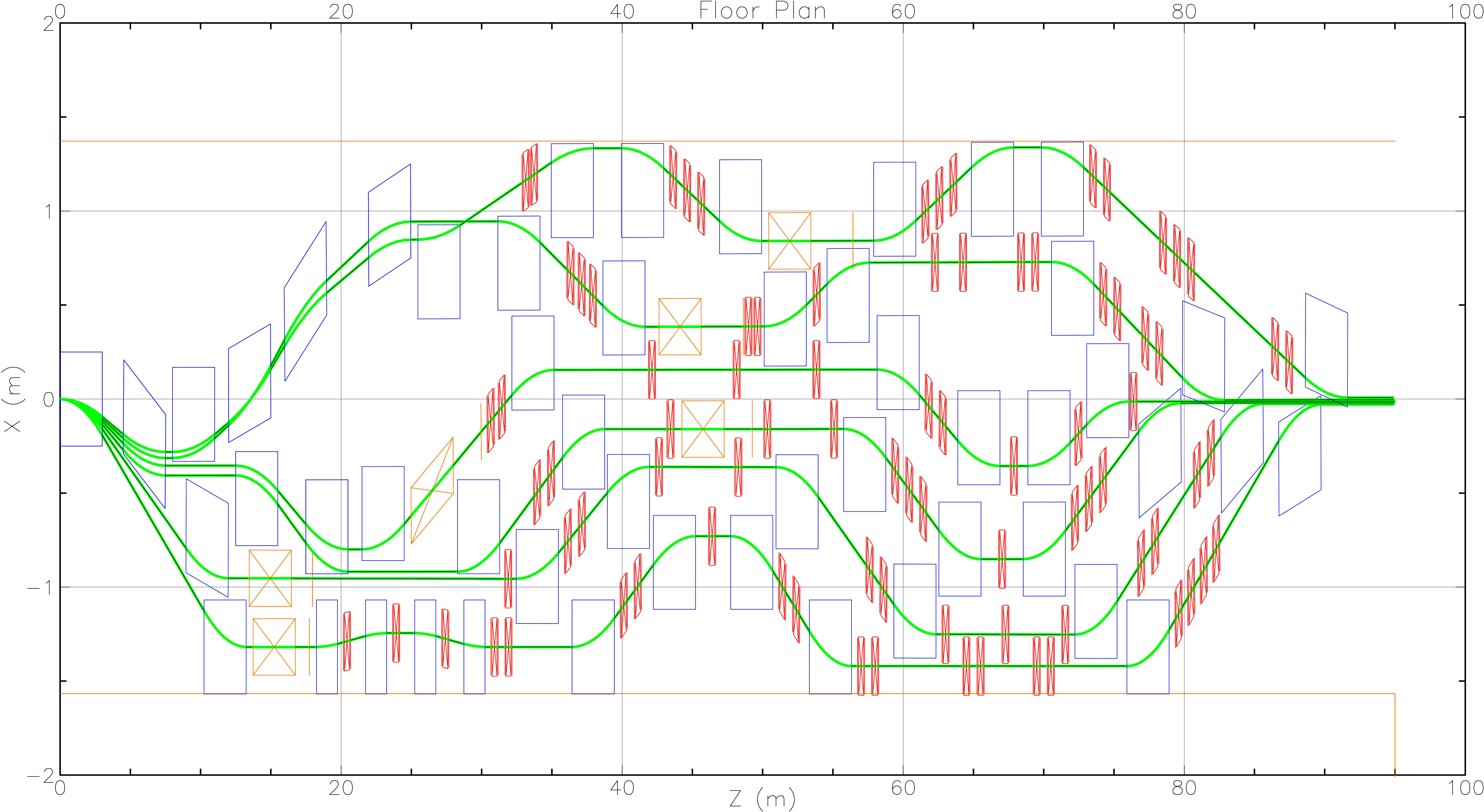}
    \caption{Overhead view of splitter floor plan. Note unequal scales. Dipoles (blue), extraction dipoles (orange), and quadrupoles (red) shown with real dimensions. Horizontal orange lines show transverse spatial limitations with the physical wall at the top and the walkway at the bottom. The aisle-side wall is not shown in this image. Beam travels left to right.}
    \label{fig:floorplan}
\end{figure}

\begin{figure}[ht]
    \centering
    \includegraphics[width=\textwidth]{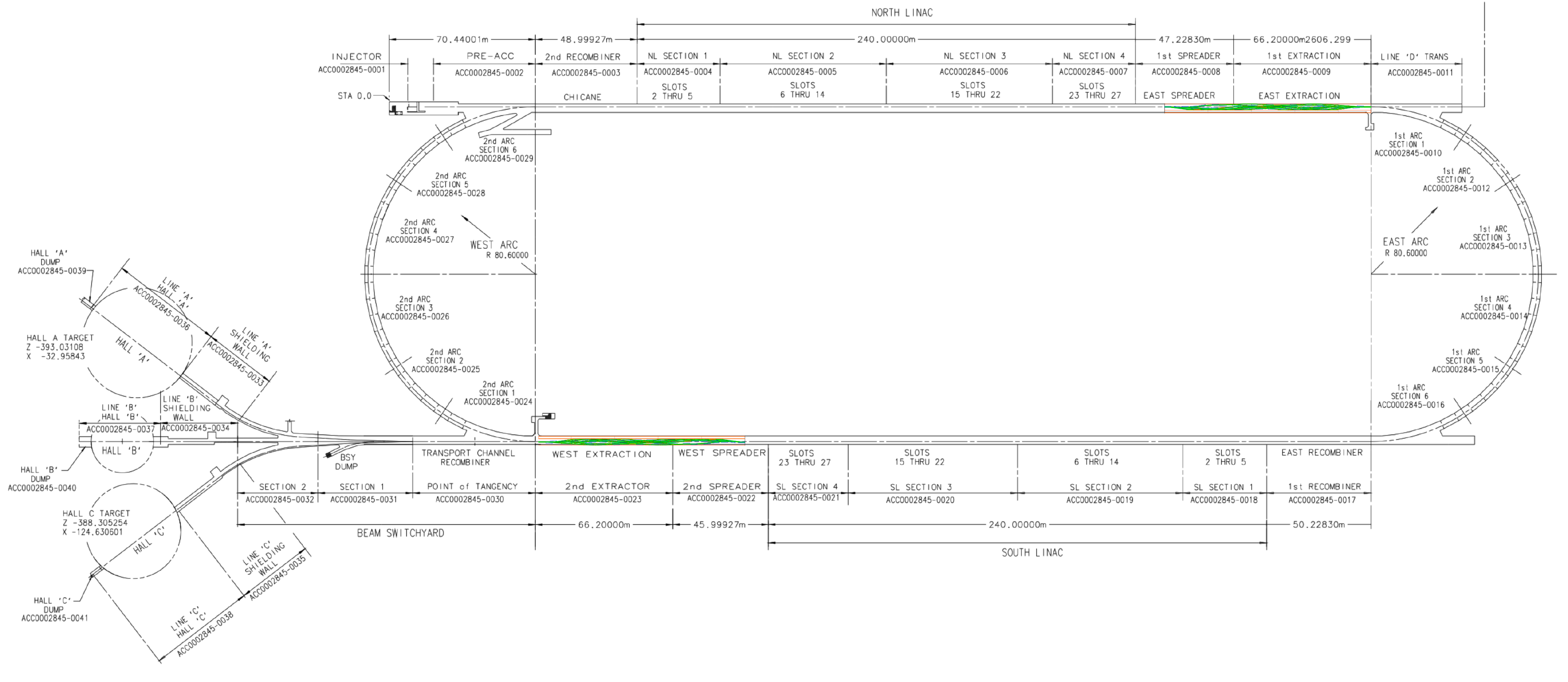}
    \caption{Splitters shown in place in the CEBAF engineering diagram, scaled appropriately. They are located in the northeast and southwest corners. North is up in this image.}
    \label{fig:SpltterInCEBAF}
\end{figure}

The complete geometric layout, accommodating all six passes from the common entrance to their individual entry points into the FFA arc, is shown in Figure \ref{fig:floorplan}. The reader should note that the axes are not equal, but expanded to show more details of the design. The layout includes all transport dipoles (blue), quadrupoles (red), and extraction dipoles/septa (orange), while respecting all physical boundaries and reserving space for correctors, diagnostics, auxiliaries, and other non-optics beamline components. Stay-clear markers are used with the quadrupoles and extraction dipoles/septa to ensure no encroachment by other elements in that beamline. No elements are in contact with any other elements, though one must zoom in to see the separation in some cases. Also, the final dipoles will likely require chamfering to reduce the corners overlapping with the beamline. Figure \ref{fig:SpltterInCEBAF} shows the location of the splitters in the CEBAF enclosure, scaled appropriately.

\section{Optics Design and Matching}
\label{sec:optics}

\subsection{Simulation Framework}
All optics simulations were performed using Bmad \cite{saganBmadRelativisticCharged2006}. The code's ability to model multiple, interacting beamlines was essential. To ensure geometric accuracy for the rectangular dipoles used in the design, the ``patch method'' was employed. This technique uses coordinate system transformations at the entrance and exit of each bend to correctly preserve the magnet geometry, as well as ensure magnets from neighboring beamlines do not collide.

\subsection{Matching Strategy}
The primary challenge of the optics design is to simultaneously satisfy the multiple matching constraints at the FFA arc entrance. The large dispersion generated by the initial separation chicane, which is often of the wrong sign for matching, proved particularly difficult to correct. 

After setting the geometry with the dipoles and adjusting for ToF, the matching process is essentially decoupled from the transport dipoles. Their values are held constant, and only the strengths of the quadrupoles are adjusted. It is important to note that all of the matched solutions that follow only vary the quadrupole strengths: the longitudinal positions are not varied unless otherwise stated. This means that future matches can also start adjusting the placement of the quadrupoles to simplify the process, if needed. But for the work presented here, only the K1 values are adjusted for matching. 

All of the limits placed on the quadrupoles are also conservative. K1 values are limited to $\pm \SI{0.6}{\per\square\meter}$ in the northeast corner to allow for adequate overhead of the magnets. It also allows for scaling of the quadrupole strengths for the southwest corner, where the beams will transit one additional LINAC and have $\sim \SI{1}{GeV}$ higher-energy beams at each pass.

\begin{figure}[!ht]
    \centering
    \includegraphics*[trim=0.cm 0.0cm 0cm 0cm,clip,width=0.8\textwidth]{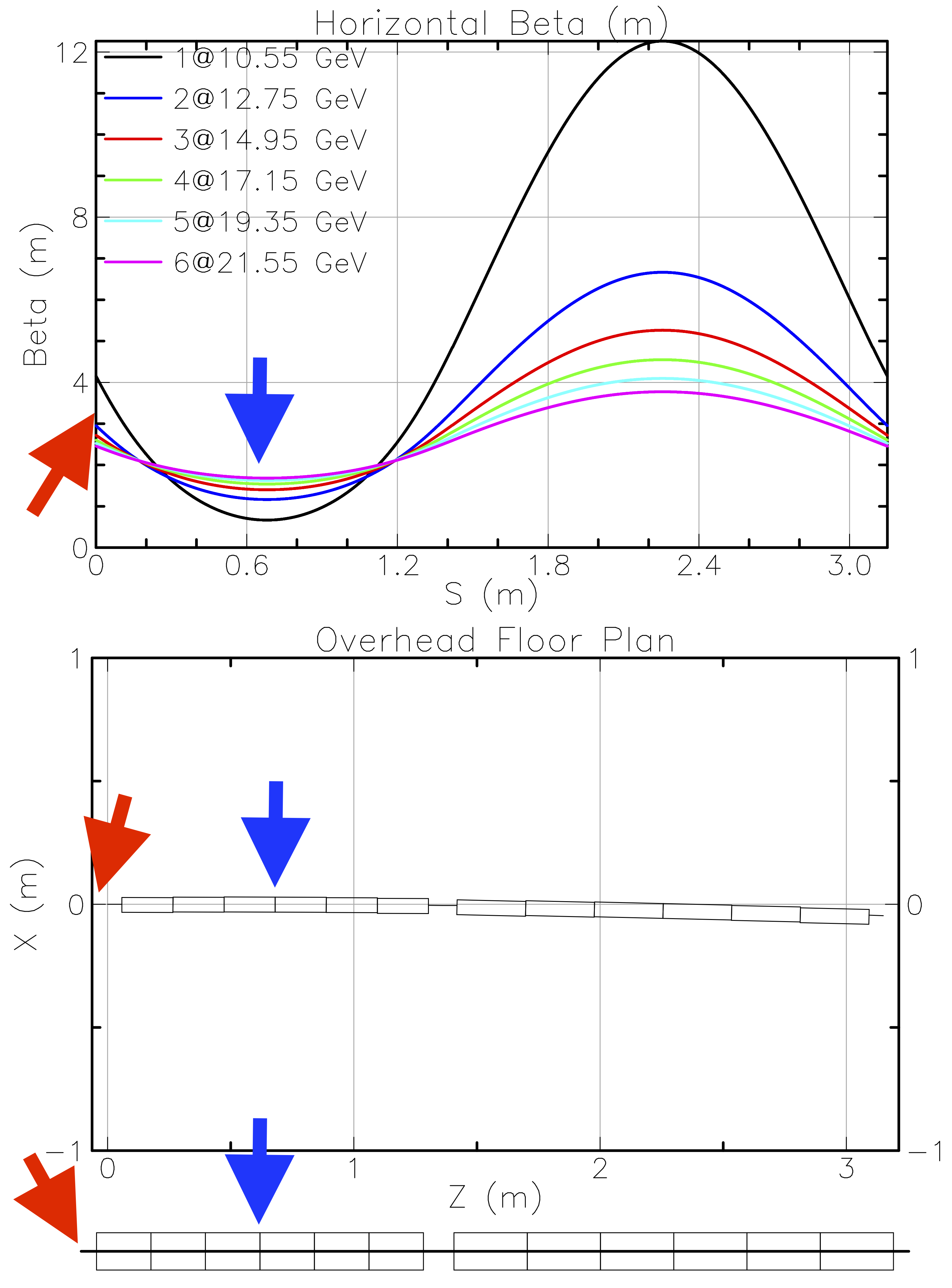}
    \caption{An FFA Cell, including the segmented Bending-Defocusing (left) and Bending-Focusing (right) magnets. $\beta$-functions for each energy pass shown. The match points used for this work are at the beginning of this cell (far left, at the drift indicated by the red arrows), and in the center of the first magnet (after three segments in the first magnet, indicated by the blue arrows).}
    \label{fig:FFA_Cell_Plots}
\end{figure}

To demonstrate the robustness of the geometric layout and explore a wider solution space, a strategy was developed to find multiple distinct matching solutions. This was achieved by varying two key sets of assumptions:
\begin{enumerate}
    \item \textbf{LINAC Input Conditions:} Solutions were generated using input matching beam parameters from both LINAC designs under consideration for the upgrade: the weakly focusing triplet (WFT) and strongly focusing triplet (SFT) optics. The WFT optics use realistic quadrupole strengths in the triplets between the cryomodules in the LINAC, but the $\beta$-functions end up much larger before the splitters (in the hundreds of meters). The SFT optics uses quadrupole settings that are unrealistically strong, and while attempts at addressing this are ongoing, no solutions have yet proven to be feasible. However, the $\beta$-functions exiting the SFT are smaller than those of the WFT, and so provide provide a range of input Twiss parameters from the LINAC into the splitters for matching. This helps to demonstrate the robustness and flexibility of the splitter design.
    \item \textbf{FFA Cell Match Point:} To find more tractable solutions, several distinct matching points within the first FFA cell were chosen as targets. Each magnet in the FFA cell is broken into longitudinal sections, and any of those separations can be used as the match point from the splitters into the FFA arc. The primary points investigated were the very beginning of the cell (labeled ``Beginning'') and a point of symmetry roughly halfway through the first magnet in the cell, a bending-defocusing magnet (labeled BDC for Bending-Defocusing Center). Figure \ref{fig:FFA_Cell_Plots} shows an FFA cell, with the segments and drift spaces indicated in the floor plan and the lattice layout elements of the plot. As shown in Tables \ref{tab:match_points_position_dispersion} and \ref{tab:match_points_twiss}, the ``BDC'' location is advantageous as the $\alpha$ values are nearly zero and the $\beta$-functions are at local extrema, simplifying the matching problem. This also allows demonstration that the splitter design is capable of matching into a variety of different FFA designs, which is important, as the FFA arc design is likely to be updated and changed.
\end{enumerate}

\begin{table*}[ht!]
\centering
\caption{Comparison of orbit and dispersion target matching parameters for all six east FFA Arc passes.}
\label{tab:match_points_position_dispersion}
\begingroup 
\scriptsize
\begin{tabular}{
  l
  l
  S[table-format=-1.4e-1]
  S[table-format=-1.4e-1]
  S[table-format=-1.4e-1]
  S[table-format=-1.4e-1]
}
\toprule
\textbf{Pass (E)} & \textbf{Match Point} & {\textbf{X (m)}} & {\textbf{Px}} & {\textbf{$\eta$ (m)}} & {\textbf{$\eta'$}} \\
\midrule
\multirow{2}{*}{1 (10.55 GeV)} 
& Beginning & -4.3609e-2 & 1.5555e-2 & 1.3845e-2 & -2.7713e-3 \\
& BDC       & -3.1036e-2 & -1.4007e-7 & -7.8352e-3 & 4.3951e-8 \\
\addlinespace
\multirow{2}{*}{2 (12.95 GeV)} 
& Beginning & -4.0953e-2 & 1.4756e-2 & 3.4001e-2 & -1.1549e-2 \\
& BDC       & -3.0635e-2 & -1.7647e-7 & 1.4456e-2 & -1.8690e-7 \\
\addlinespace
\multirow{2}{*}{3 (14.95 GeV)} 
& Beginning & -3.6297e-2 & 1.3082e-2 & 5.0554e-2 & -1.8809e-2 \\
& BDC       & -2.8009e-2 & -1.8064e-7 & 3.2967e-2 & -3.7558e-7 \\
\addlinespace
\multirow{2}{*}{4 (17.15 GeV)} 
& Beginning & -2.9986e-2 & 1.0682e-2 & 6.4348e-2 & -2.4885e-2 \\
& BDC       & -2.3526e-2 & -1.6090e-7 & 4.8495e-2 & -5.3096e-7 \\
\addlinespace
\multirow{2}{*}{5 (19.35 GeV)} 
& Beginning & -2.2289e-2 & 7.6691e-3 & 7.5975e-2 & -3.0021e-2 \\
& BDC       & -1.7479e-2 & -1.4009e-7 & 6.1641e-2 & -6.6020e-7 \\
\addlinespace
\multirow{2}{*}{6 (21.55 GeV)} 
& Beginning & -1.3418e-2 & 4.1373e-3 & 8.5873e-2 & -3.4401e-2 \\
& BDC       & -1.0103e-2 & -1.4475e-7 & 7.2865e-2 & -7.6926e-7 \\
\bottomrule
\end{tabular}
\endgroup
\end{table*}

\begin{table*}[ht!]
\centering
\caption{Comparison of Twiss target matching parameters ($\beta$ and $\alpha$) for all six east FFA Arc passes.}
\label{tab:match_points_twiss}
\begingroup 
\scriptsize
\begin{tabular}{
  l
  l
  S[table-format=1.4e1]
  S[table-format=1.4e1]
  S[table-format=1.4e1]
  S[table-format=-1.4e1]
}
\toprule
\textbf{Pass (E)} & \textbf{Match Point} & {\textbf{$\beta_{x}$ (m)}} & {\textbf{$\alpha_{x}$}} & {\textbf{$\beta_{y}$ (m)}} & {\textbf{$\alpha_{y}$}} \\
\midrule
\multirow{2}{*}{1 (10.55 GeV)} 
& Beginning & 4.1508e0 & 3.0414e0 & 6.5160e0 & -3.1904e0 \\
& BDC       & 6.6586e-1 & 4.3124e-5 & 1.2039e1 & -1.1741e-4 \\
\addlinespace
\multirow{2}{*}{2 (12.95 GeV)} 
& Beginning & 2.9513e0 & 1.8208e0 & 6.4812e0 & -3.0391e0 \\
& BDC       & 1.1641e0 & 2.6829e-5 & 1.0662e1 & -1.0407e-4 \\
\addlinespace
\multirow{2}{*}{3 (14.95 GeV)} 
& Beginning & 2.7191e0 & 1.5387e0 & 7.0035e0 & -3.2108e0 \\
& BDC       & 1.3998e0 & 2.4658e-5 & 1.0689e1 & -1.0462e-4 \\
\addlinespace
\multirow{2}{*}{4 (17.15 GeV)} 
& Beginning & 2.6034e0 & 1.3996e0 & 8.0521e0 & -3.6443e0 \\
& BDC       & 1.5386e0 & 2.4713e-5 & 1.1607e1 & -1.1455e-4 \\
\addlinespace
\multirow{2}{*}{5 (19.35 GeV)} 
& Beginning & 2.5224e0 & 1.3113e0 & 1.0173e1 & -4.5685e0 \\
& BDC       & 1.6261e0 & 2.5401e-5 & 1.4063e1 & -1.3987e-4 \\
\addlinespace
\multirow{2}{*}{6 (21.55 GeV)} 
& Beginning & 2.4564e0 & 1.2473e0 & 1.7058e1 & -7.6233e0 \\
& BDC       & 1.6824e0 & 2.6113e-5 & 2.2819e1 & -2.2893e-4 \\
\bottomrule
\end{tabular}
\endgroup
\end{table*}

For each of these conditions, matching was performed first without considering $R_{56}$, and then again with $R_{56}$ included as a ``soft target.'' As a reminder, the $R_{56}$ term is considered a soft target because significant sections are missing which contribute to the values which must be compensated by the splitters. The largest of these is the transition section \cite{gamageResonantMatchingSection2024, gamage:napac2025-tup029}, which will likely take up far more longitudinal distance than is currently assumed by the available space. But matching without the $R_{56}$ will show one set of solutions, and then using a soft target $R_{56}$ to see what Twiss parameters must be sacrificed helps to bound what the final solutions will likely need to be. It also shows if any changes should be made to the FFA arc design to alter their $R_{56}$ contributions.

\subsection{Presentation of Results}
Thus far, six sets of matching solutions have been proven successful, demonstrating the viability and flexibility of this splitter design. These six solutions can be broken into three pairs of solutions. Each pair consists of one solution which matches all of the Twiss parameters but ignores $R_{56}$ compensation, and one solution which compensates for a ``soft target'' $R_{56}$ and most of the Twiss parameters. This section will present some of these solutions, as well as compare them.

Only one example of each match will be shown in the main body of this text. However, for completeness the reader is encouraged to see \ref{sec:AppA} for plots, match parameter tables, and quadrupole strength tables for all six of the match solutions.

\subsubsection{Match Without $R_{56}$ Compensation}

\begin{figure}[ht]
    \centering
    \includegraphics[width=\textwidth]{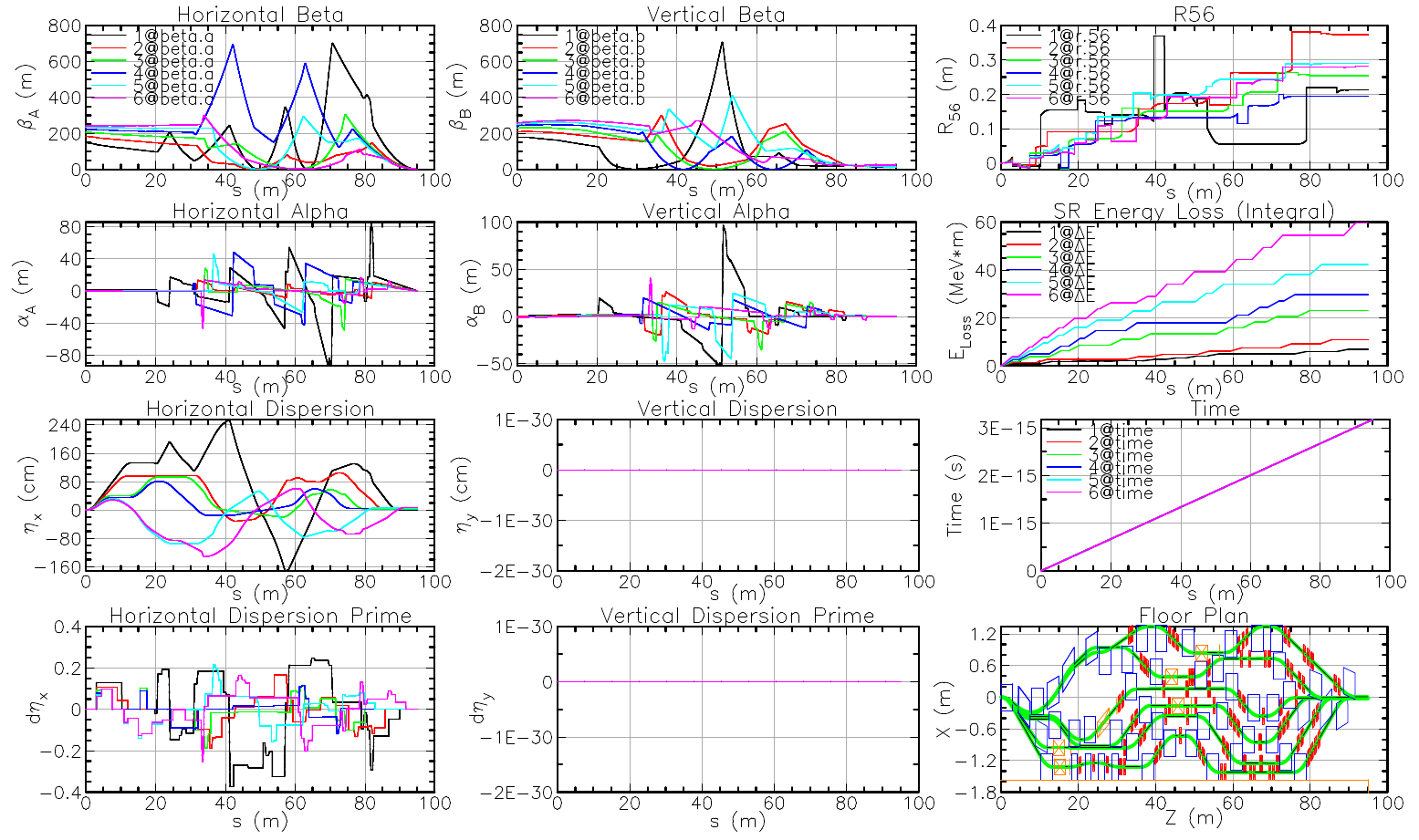}
    \caption{Matching solution from the WFT LINAC input into the BDC FFA match point. In this case, all of the Twiss parameters are matched to within machine precision, and $R_{56}$ compensation is ignored.}
    \label{fig:WFT_match_TwissOnly}
\end{figure}

When ignoring the $R_{56}$ constraint, all other optics parameters can be matched successfully for all six passes. The resulting $\beta$-functions and dispersion along the splitter are well-behaved. Though $\beta$-functions can become large, they are within the expected requirements of the beamline. All quadrupoles were limited to K1 values of \SI{0.6}{m^{-2}} to leave adequate overhead, and allow for scaling of the solution to the southwest corner splitter, which will have higher energies. Plots for the various match parameters are shown in Figure \ref{fig:WFT_match_TwissOnly} for the case of the WFT LINAC input optics being matched into the BDC match point in the FFA Cell.

The merit function for all of the match parameters are effectively zero for all six passes (\ref{sec:AppA_1-2}), indicating a near-perfect match (Table \ref{tab:WFT_Twiss_full_summary}). The quadrupole strengths are all at or below the enforced limits (Table \ref{tab:WFT_Twiss_quad_strengths}). The maximum $\beta$-functions are within the \SI{700}{m} limit placed upon the optimizer.

\FloatBarrier

\subsubsection{Match With $R_{56}$ as a ``Soft Target''}

\begin{figure}[ht]
    \centering
    \includegraphics[width=\textwidth]{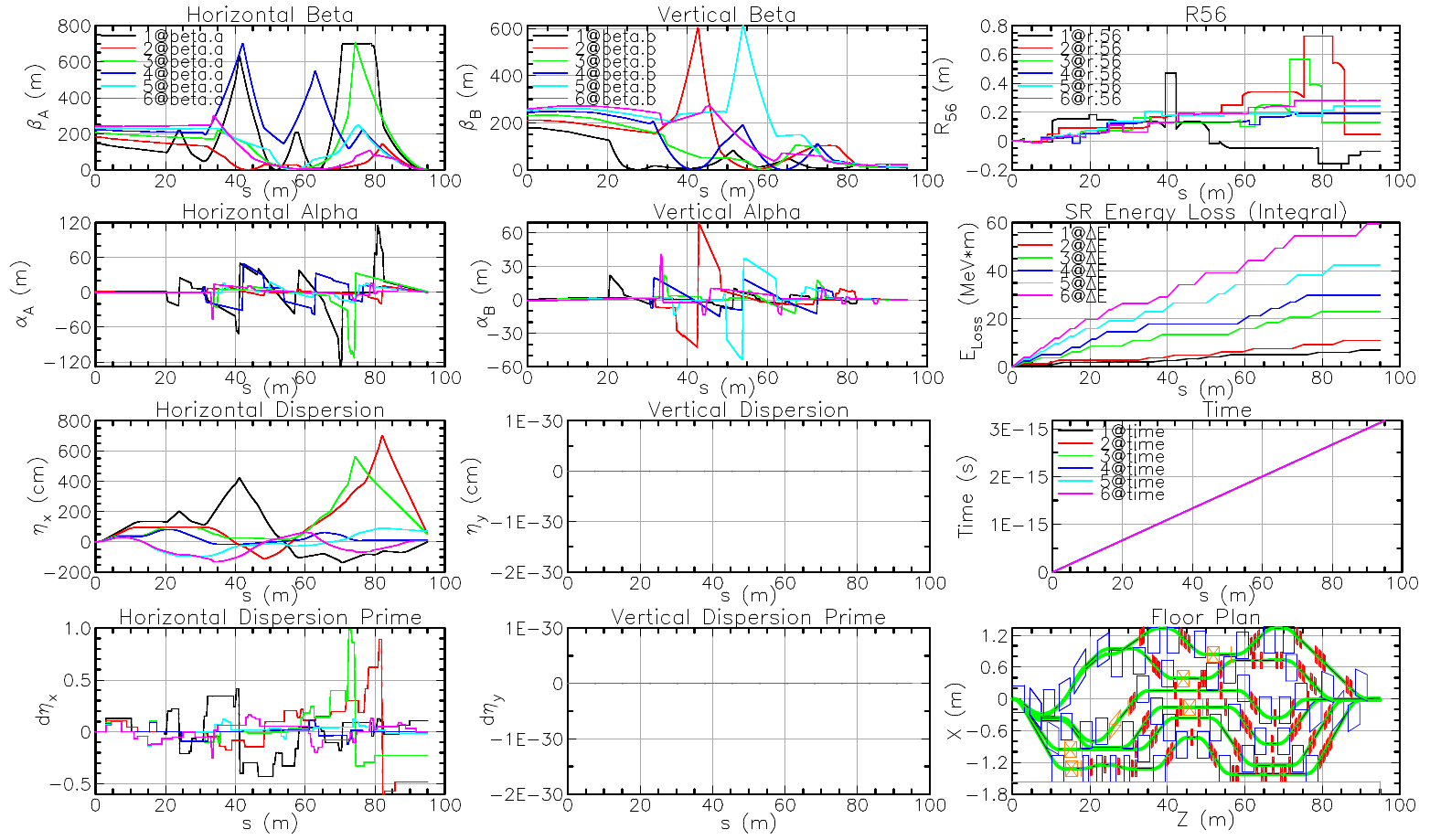}
    \caption{Matching solution from the WFT LINAC input into the BDC FFA match point. In this case, $R_{56}$ and $\beta_{x,y}$ is always matched perfectly, $\alpha_{x,y}$ usually matched very well or perfectly, and $\eta_{x}$ and $\eta'_{x}$ are matched adequately.}
    \label{fig:WFT_match_with_r56}
\end{figure}

Including the $R_{56}$ constraint significantly complicates the problem. The strong quadrupole fields needed to alter the dispersion and $R_{56}$ create large perturbations in the $\beta$-functions. While solutions have been found that bring the $R_{56}$ close to the target value, this often comes at the cost of compromising the match of other parameters.

 Figure \ref{fig:WFT_match_with_r56} shows the match solutions for all six passes from the WFT LINAC input into the BDC match point in the FFA cell. The tables in \ref{sec:AppA_2-2}, specifically Tables \ref{tab:WFT_R56_twiss_full_summary} and \ref{tab:WFT_R56_Twiss_quad_strengths}, show the merit functions for all of the match parameters, and the quad strengths required to achieve said matches, respectively. In order to achieve these matches and compensate $R_{56}$, some small sacrifices were made in other match parameters. Specifically, the horizontal $\eta$ and $\eta'$ matches were de-prioritized, and for the third pass, the $\alpha_{x,y}$ match was not as strong. The merit functions for nearly all other parameters are effectively zero.


\subsubsection{Comparison of Solutions}
The existence of multiple solutions under different input conditions (WFT vs. SFT) and for different match points (Beginning vs. BDC) confirms that the geometric design is not rigidly tied to a single, unique solution. This inherent flexibility is crucial for the success of the overall machine upgrade, as other components of the accelerator are still being designed. Furthermore, this demonstrated robustness will translate to greater operational flexibility and ability to match for non-nominal beams which tend to be the normal operating state of CEBAF.

For the two inputs, there is not a large difference for the results of the matching: the splitters are able to match from both LINAC input options. It could be argued that the Weakly Focusing Triplet option allows for matches which more easily control the maximum $\beta$-functions during transport, but this is only a minor point. In all cases, the maximum $\beta$-functions during transport correspond to beam sizes which fall within the $10\sigma$ beam-stay-clear (BSC) limits of CEBAF. Please see \ref{sec:AppB_Aperture} for further details on this.

More weight depends upon the match point in the FFA Cell. Here, it is very clear that the BDC match point is far simpler for matching, with more solutions being achieved here, and with lower overall merit functions. Further investigations into other FFA Cell match points may find even better options. Additionally, the FFA Arc design is under flux, and it may be that improved solutions can be discovered as the overall design evolves.

Nonetheless, with two very different input and output options, this splitter design is capable of matching all of the optics parameters, or some of the optics parameters and $R_{56}$ compensation. The two best options are those shown in the text above. However, these are both based upon the WFT input option, which is not currently the assumed baseline for the energy upgrade. 

A brief comparison of the merit functions of the six solutions is shown in Table \ref{tab:match_merit_summary}. For more details of the matches, including tables showing all of the match parameters, target values, achieved values, and merit functions, please see \ref{sec:AppA}. This also includes a brief description of how Bmad and Tao calculate the merit functions.

\begin{table*}[ht!]
\centering
\caption{Quantitative comparison of the six matching solutions, demonstrating the flexibility of the geometric design. The 'Merit Function Range' quantifies the final match quality, where a lower value indicates a better match. For the 'No $R_{56}$' cases, the $R_{56}$ merit is excluded from this range to show the intended match quality. The data illustrates that matching $R_{56}$ (bottom half) consistently requires sacrificing the dispersion ($\eta_a$) match. The solutions shown in \textbf{bold} are those presented as primary examples in Section \ref{sec:optics}.}
\label{tab:match_merit_summary}
\small 
\begin{tabular}{@{}lccc@{}}
\toprule
\textbf{\makecell{Solution (Input \\ $\rightarrow$ Match Point)}} & \textbf{\makecell{$R_{56}$ \\ Matched}} & \textbf{\makecell{Merit Function \\ Range}} & \textbf{\makecell{Primary Sacrificed \\ Parameters}} \\
\midrule
SFT $\rightarrow$ Beginning & No & $0 \rightarrow \num{5.76e5}$ & $\max(\beta_a)$, $\max(\beta_b)$ \\
SFT $\rightarrow$ BDC & No & $\approx 0$ & None (All merits $\approx 0$) \\
\textbf{WFT $\rightarrow$ BDC} & \textbf{No} & $\mathbf{\approx 0}$ & \textbf{None (All merits $\approx 0$)} \\
\midrule
SFT $\rightarrow$ Beginning & Yes & $0 \rightarrow \num{1.40e9}$ & $\eta_a$, $\alpha_b$, $\eta'_a$ \\
SFT $\rightarrow$ BDC & Yes & $0 \rightarrow \num{4.06e8}$ & $\eta_a$, $\alpha_b$, $\eta'_a$ \\
\textbf{WFT $\rightarrow$ BDC} & \textbf{Yes} & $\mathbf{0 \rightarrow \num{4.06e7}}$ & \textbf{$\eta_a$, $\eta'_a$} \\
\bottomrule
\end{tabular}
\end{table*}

\subsection{Discussion of Optics Challenges}
While the design of this splitter is robust and flexible, there are some remaining challenges. This section will describe some of the challenges.

\subsubsection{Magnet Crosstalk and Fringe Fields}
When looking at the floor plan of this design, it is reasonable to have concerns with the tight-packed nature of the layout. While no elements are in contact with other elements, they are very close in proximity, and the fields of the magnets will influence the beams in adjacent beamlines. This concern is valid and needs to be fully investigated, along with other error analysis.

\begin{figure}[!ht]
    \centering
    
    \begin{subfigure}[t]{0.31\textwidth}
        \centering
        \includegraphics[width=\textwidth]{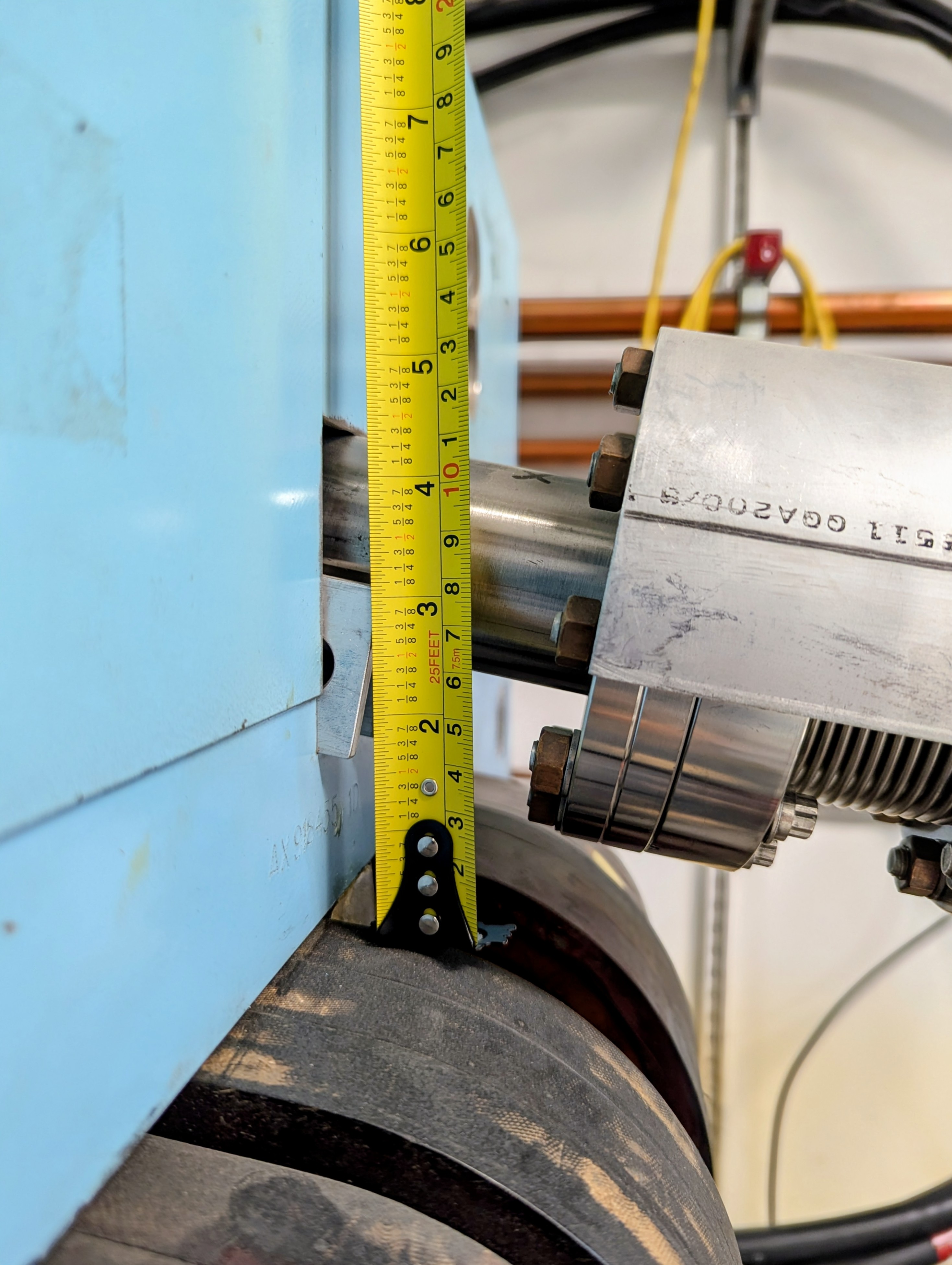}
        \caption{Beam pipe through return steel of common dipole.}
        \label{fig:pipe_through_steel}
    \end{subfigure}%
    \hfill
    \begin{subfigure}[t]{0.55\textwidth}
        \centering
        \includegraphics[width=\textwidth]{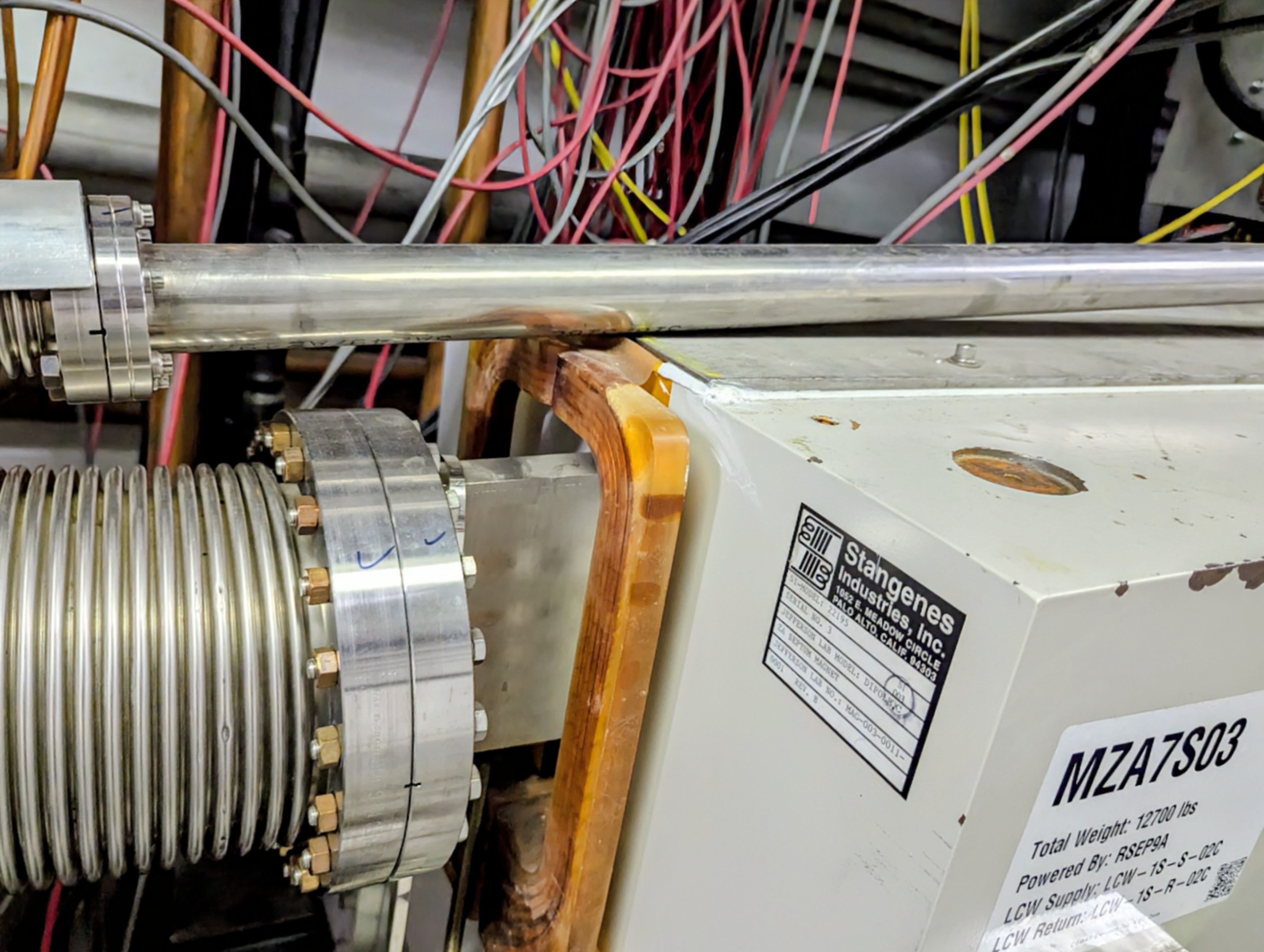}
        \caption{Dipole MZA7S03 with beamline along top.}
        \label{fig:dipole_view1}
    \end{subfigure}%

    \vspace{0.5cm} 

    \begin{subfigure}[t]{0.31\textwidth}
        \centering
        \includegraphics[width=\textwidth]{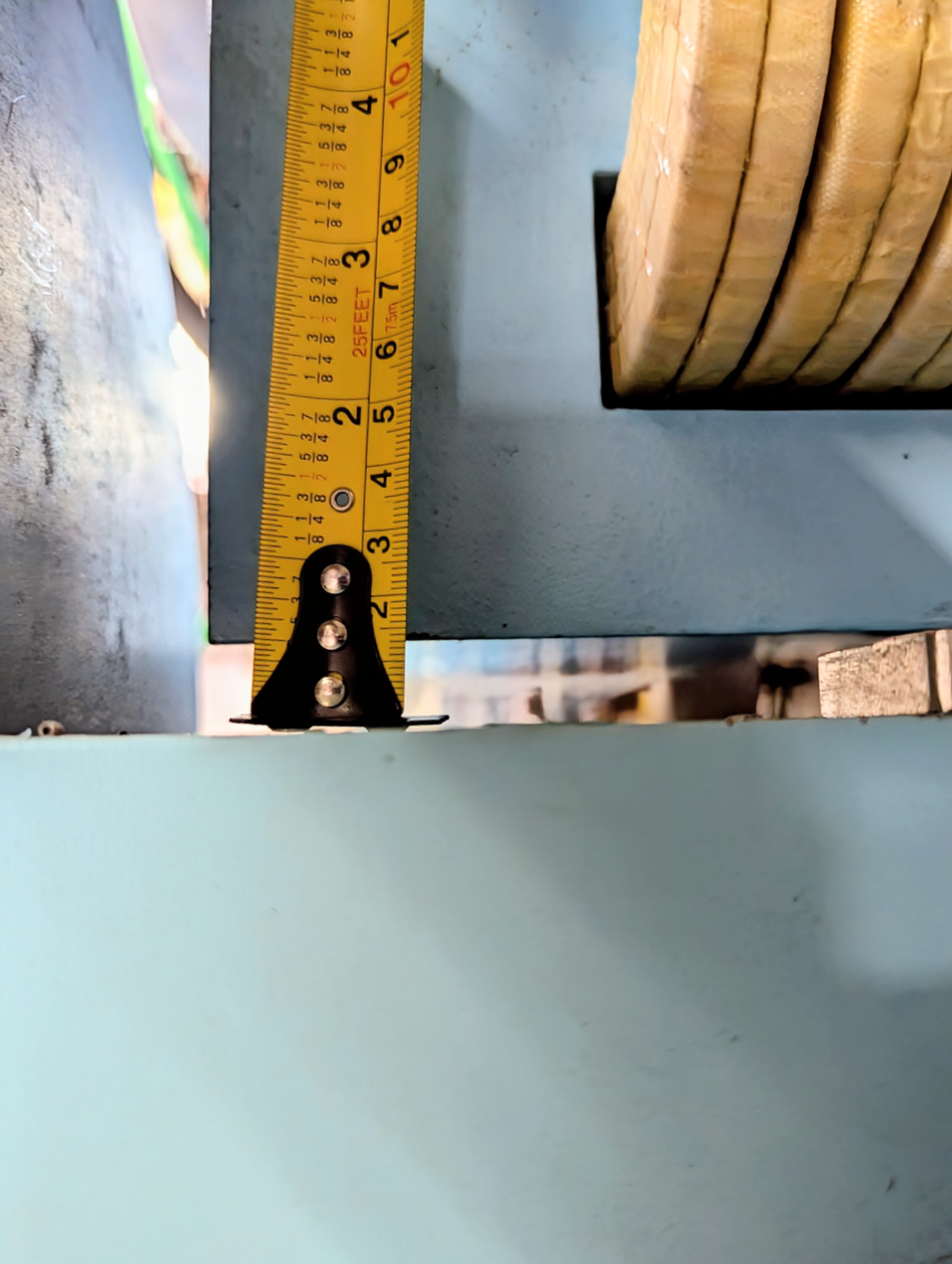}
        \caption{$\SI{1.5}{cm}$ gap between dipoles.}
        \label{fig:steel_gap}
    \end{subfigure}%
    \hfill
    \hfill
    \begin{subfigure}[t]{0.55\textwidth}
        \centering
        \includegraphics[width=\textwidth]{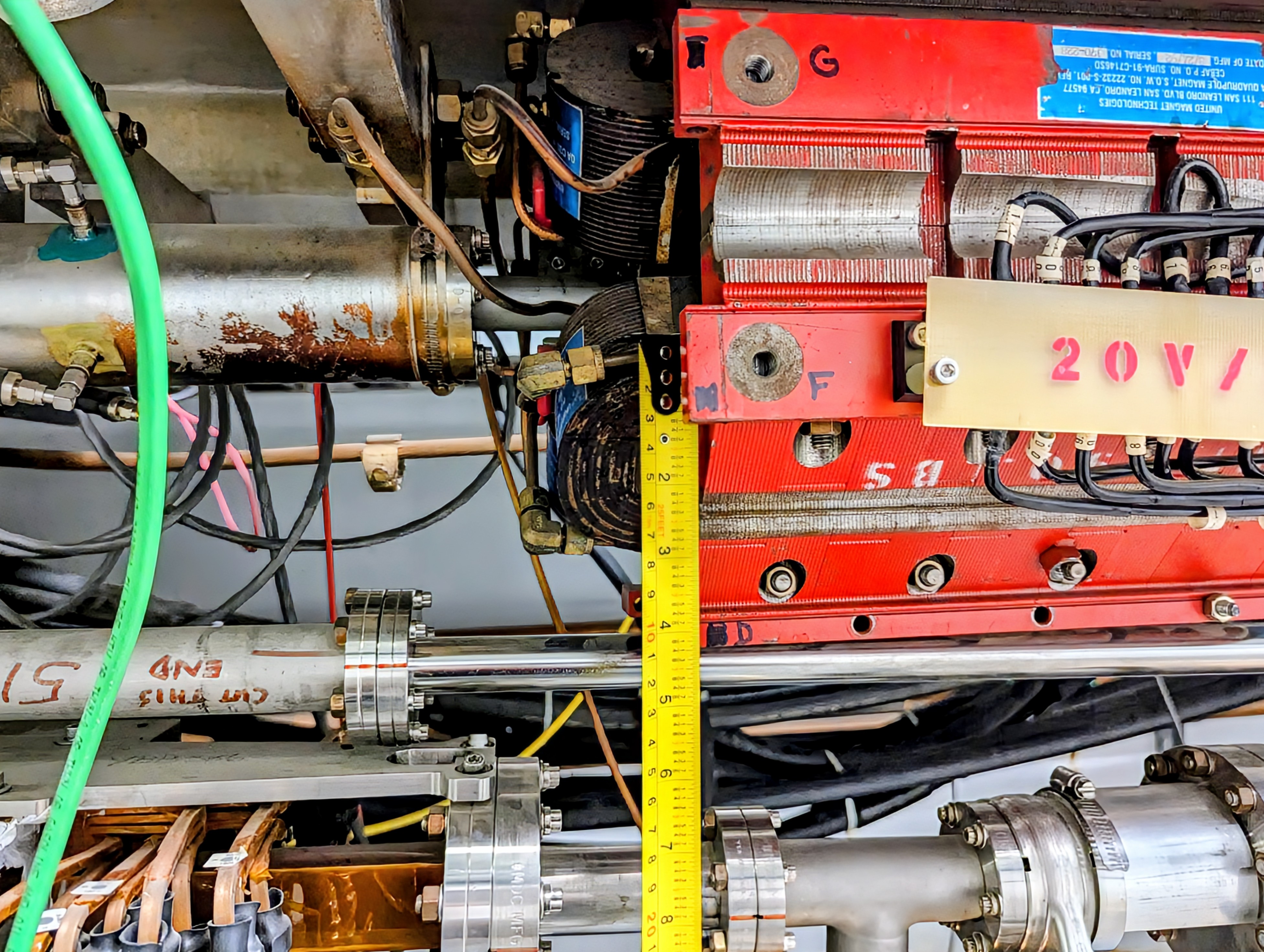}
        \caption{Three beamlines within $\sim \SI{20}{cm}$ .}
        \label{fig:three_lines}
    \end{subfigure}
    
    \caption{Several photographs of CEBAF beamlines with tight clearances.}
    \label{fig:ClearanceIssues}
\end{figure}

However, this is also a problem that is very familiar to CEBAF Operations. In fact, our spreaders, recombiners, and extraction regions all manage multipass beamlines with very tight tolerances. Figure \ref{fig:ClearanceIssues} shows several examples of such tight clearances in the current CEBAF machine. Figure \ref{fig:pipe_through_steel} shows a lower-energy beam pipe passing through the return steel of a common dipole, while remaining outside of the coils. This is also a good example of what this splitter design aims to avoid: recombining beamlines through a common dipole. Figure \ref{fig:dipole_view1} shows a spreader dipole with a lower-energy pass touching and running along the top of it, with some magnetic shielding placed between the beam pipe and the magnet steel. Figure \ref{fig:steel_gap} shows a very small gap of $\sim \SI{1.5}{cm}$ between the return steel of dipoles from adjacent beamlines. Figure \ref{fig:three_lines} shows three beamlines of different energies installed within approximately \SI{20}{cm} from each other, with the middle of these running along the quadrupole return steel from an adjacent line.

The point of displaying these photographs is to demonstrate that, while close proximity of adjacent beamlines to each other is not ideal, it is a common occurrence at CEBAF: one that the Operations crew has the knowledge and experience to accommodate during beam operations. Judicious use of corrector magnets, as well as startup and adjustment procedures (such as magnet hysteresis cycling) allows for strong and accurate control of each energy beam. By ensuring that the crosstalk is reproduced and repeatable, operations can ensure that the steering contribution is consistent.

Furthermore, different choices in magnetic shielding and beam pipe material can reduce the impact of stray fields on adjacent beamlines. For example, carbon steel beam pipes can adequately shield against stray fields. If more shielding is required, a layered/jacketed beamline can also be used. If there is adequate space, magnetic shielding can also be installed between beamlines to further reduce the impact.

Also, should the decision be made to reduce the number of energies transiting the FFA be reduced from six to five, the physical constraints would be reduced for the splitters, and the elements from each adjacent line would be allowed more space.

\subsubsection{Dispersion Control}
Controlling the dispersion of all six beamlines is also challenging. Separating the beamlines requires large, strong bending dipoles. Given the rigidity of the beams, there is not adequate space for placement of independent quadrupoles for several tens of meters into the beamline. In this space, the dispersion due to the bends is uncompensated, and the $\beta$-functions are uncontrolled. Furthermore, the space constraints necessitate bending the beams in a direction which creates dispersion that is counter to what is needed upon entry into the FFA arcs. This combination of factors necessitates strong quadrupoles to be used for dispersion control and Twiss parameter matching.

To include extraction of the beams (see Section \ref{sec:extraction}), dipoles and/or septa need to be installed early in each line as well. They must happen early to allow adequate space for the transfer lines from the splitters to the extraction lines themselves. This further delays placement of quadrupoles.

The final target values for ToF and $R_{56}$ depend on the design of the downstream transition section, which is not yet complete \cite{gamageResonantMatchingSection2024, gamage:napac2025-tup029}. Compensating for the $R_{56}$ and matching the Twiss parameters is further complicated by including dispersion matching. In the matches above, sacrificing the quality of the match for $\eta$ and $\eta'$ tends to allow for matching all of the other parameters. Once the transition section design is finalized, the splitter optics will need to be re-matched in an iterative process with the FFA arc design to reduce the matching burden on the splitter.

\subsubsection{Energy Loss and Emittance}
\label{sec:Emit}
For any high energy transport line, beam degradation due to synchrotron radiation needs to be examined. For the designs presented, this was done by using radiation integrals to calculate beam loss, increase in energy spread, and increase in emittance. The relevant radiation integrals for a transfer line are given below:

\begin{gather*}
    L_2 = \int g^2\gamma_0^4ds \\
    L_3 = \int g^3\gamma_0^7ds \\
    L_{5x} = \int g^3\mathcal{H}_x\gamma_0^6ds,
\end{gather*}%
where $g = 1/\rho$, with $\rho$ being the radius of curvature, $\gamma_0$ being the relativistic factor, and $\mathcal{H}_x = \gamma_x\eta_x^2 + 2\alpha_x\eta_x\eta^{\prime}_x + \beta_x\eta^{\prime 2}_x$, the curly H function. The energy loss, increase in energy spread, and increase in normalized emittance is then given by

\begin{gather*}
    U_0 = \frac{2}{3}r_cmc^2L^2 \\
    \sigma_E^2 = \frac{4}{3}C_qr_c(mc^2)^2L_3 \\
    \epsilon_{nx} = \frac{2}{3}C_qr_cL_{5x}
\end{gather*}%
where $r_c$ is the classical electron radius, $m$ is the mass of the electron, $c$ is the speed of light, and $C_q = \frac{55}{32\sqrt{3}}\frac{\hbar}{mc}$~\cite{helmEvaluationSynchrotronRadiation1973, jowettIntroductoryStatisticalMechanics1987}. 

The energy loss and increase in energy spread for a given splitter line is constant for the solutions presented, depending only on the dipole strengths. While larger than desired, they remain within a range that is typical for high-energy transport lines in constrained geometries. While the induced energy spread will require careful management in the experimental halls, it does not present a fundamental barrier to beam transport. The only way to minimize either of these considerations is by reducing the dipole bend angle, which would require either fewer splitter lines or a larger footprint. 

The emittance growth is substantial, with the final emittance after transport through all six lines being one or two orders of magnitude larger than desired, depending on the solution; this is a natural consequence of the large $\mathcal{H}$-function values through the strong bends required by the constrained footprint. The emittance growth is not spread equally over a given splitter line: while two equal bends will contribute equally to the energy loss, the optics through the dipoles, i.e., the $\mathcal{H}$-function through the dipoles, can mean that one of the two dipoles is the primary driver in emittance growth, while the other is almost negligible. Thus, when considering how to minimize emittance growth for a future design iteration, both bend strengths and optics must be taken into consideration. It may be more fruitful to reduce the bend strength of select dipoles, instead of reducing the bends of all dipoles equally. This becomes even more critical for the first and last dipole in each line; given the fixed optics at both ends of the beamlines, the only real lever at these locations to reduce $L_{5x}$ is the dipole bend strength. However, with sufficient relaxation in requirements and flexibility in optics match parameters, this may be a problem that can be solved in future iterations.

Furthermore, given the number of variables and constraints that are considered in this overall design, it was deemed a higher priority to focus on those previously described parameters. Once more of the Energy Upgrade design is complete, including the FFA arcs, the LINACs, and the transition sections, rematching of the splitters will be required. At this time, the lines can be properly optimized to include reduction of emittance growth.

\section{Beam Extraction Scheme}
\label{sec:extraction}

\subsection{Rationale for Extraction from the Splitter}
The splitter is the only location in the FFA@CEBAF design where the high-energy passes are physically separated by a significant distance. This makes it the only feasible location for installing a system to extract these beams and deliver them to the experimental halls. Therefore, incorporating an extraction scheme was a core requirement of the design. This splitter design is the only one which includes extraction capabilities.

\subsection{Conceptual Design}
The proposed extraction system reserves space in each of the six beamlines for a 3 m-long C-dipole or septum magnet. This allows for two potential modes of operation:
\begin{enumerate}
    \item \textbf{Magnetic Extraction:} By powering the extraction magnet in a given pass, the beam would be bent vertically downwards towards a transfer line leading to the Beam Switchyard (BSY). This method would likely send the same energy beam to Halls A, B, and C simultaneously, while Hall D operates separately.
    \item \textbf{RF-Kicker Extraction:} A timed vertical kick from an upstream RF kicker could be used in conjunction with the septa to extract individual beam bunches from a specific pass, offering more flexibility. This concept is under further investigation \cite{kazimi:ipac2025-tupm114}.
\end{enumerate}

\subsection{Integration into the Layout}
The extraction dipoles are placed within the main chicanes of each beamline, as shown by orange rectangles in the floor plan (Figure \ref{fig:floorplan}). This placement represents a compromise, as it occupies space that could otherwise be used for quadrupoles. However, by placing them on the upstream side of the main chicane, the downstream matching section remains clear for final optics control.

\section{Summary and Future Work}
\label{sec:summary}

\subsection{Summary of Accomplishments}
This work has produced a complete, detailed, conceptual design for a horizontal splitter for the FFA@CEBAF energy upgrade. The design successfully fits all necessary magnetic elements and beam trajectories within the severe physical constraints of the CEBAF tunnel. The robustness of the design has been demonstrated by finding multiple viable optics matching solutions under a variety of input conditions. Finally, the design incorporates a feasible plan for high-energy beam extraction, a critical requirement for the facility's physics program. It remains the only current solution that provides both multiple matching options and beam extraction capability.

\subsection{Future Work and Refinements}
While this conceptual design is a major step forward, significant work remains.
\begin{itemize}
    \item \textbf{Optics:} The immediate next step is to rematch the optics once the downstream transition section design is finalized and the definitive target values for ToF and $R_{56}$ are known. Future optimization studies will also explore varying the longitudinal placement of the quadrupoles, in addition to their strengths, to access a larger solution space and potentially find more robust and efficient matches. Changes to the FFA arc match conditions, as the design evolves, will also be required.

    Given the complexity of this system, error analysis will be required. Specifically, the sensitivity to off-momentum and misalignments must be studied. Chromaticity should be investigated as well, given the size of the $\beta$-functions in some of the matches. Magnet error sensitivities will also need to be investigated. Eventually, particle tracking will also need to be completed to study beam loss and aperture limitations.

    Depending on the complexities of matching solutions once the transition section and the FFA Arc updates and changes are completed, it may necessitate the use of multifunction magnets. The authors do not think they will be necessary, but it is worth investigating their use to simplify the overall design and quantify their impact on the overall design.
    
    \item \textbf{Engineering:} Detailed engineering studies are required to refine the design. While first-order considerations have been addressed, further investigation is needed regarding magnet good-field regions, cross-talk between closely interleaved magnets, and proximity to beamline infrastructure such as pumps, valves, and cabling. Fringe and stray-field mitigation will also be necessary.
    \item \textbf{Performance:} Further beam dynamics studies are needed to evaluate the splitter's impact on beam quality, particularly concerning emittance growth from synchrotron radiation, and to establish error tolerances for magnet misalignments and field errors.
\end{itemize}

\subsection{Potential Simplifications}
The complexity of the six-pass design is significant. Most of the complexity is due to the physical space constraints. One way that this could be alleviated would be to extend the space to include the whole width of the tunnel, and using ramps to allow for equipment to pass over the beamlines. The ramps would need to be capable of holding the weight of very heavy equipment, such as cryomodules, as well as have a low enough angle to allow for heavy equipment to be rolled over the ramps. One would also need to use caution, as the ceiling may present oxygen deficiency hazards (ODH), and the heads of personnel may be in ODH hazards of higher risk.

Many of the challenges related to component spacing and interleaved magnets would be substantially mitigated by reducing the number of FFA passes from six to five. This option would create valuable additional space, simplify the overall design, and likely reduce the synchrotron energy loss and emittance growth, and should be considered as the overall design of the FFA@CEBAF upgrade matures. The authors believe that this reduction of the total number of passes is more appropriate.

\section{Acknowledgments}

Authored by Jefferson Science Associates, LLC under U.S. DOE Contract No. DE-AC05-06OR23177.

Thanks to all who contributed, especially those in CEBAF Operations, who provided insights into the operational complications that come with such a design and suggestions to improve. Special thanks to Scott Berg, who provided the initial recipe for this design based upon his work at CBETA, as well as significant amounts of advice and guidance. Thanks to Todd Satogata and Yves Roblin, who provided meaningful advice and recommendations of the work.


\begin{appendices} 

\input{Appendices/AppendixA.tex}
\input{Appendices/AppendixB.tex}
\input{Appendices/Appendix_AI_Statement.tex}

\end{appendices} 

\printbibliography

\end{document}

%% file: Appendices/AppendixA.tex
\section{Matching Parameters}
\label{sec:AppA}

For completeness, all matching parameters, merit functions, and quadrupole settings for each of the six solutions will be included in this section, as well as relevant plots. There are two main sections, one describing match solutions without the "soft target" $R_{56}$ compensation, and another which sacrifices some Twiss matching to compensate for the $R_{56}$. Within each of those section, it is further broken down into which input conditions are being used: the Strong Focusing Triplet (SFT) LINAC or the Weakly Focusing Triplet (WFT) LINAC. Finally, there may be separate solutions within a section of the above categories, which will be indicated.

With regards to the merit function numbers, the goal is to minimize these values. They are calculated by the Tao simulation program part of Bmad \cite{saganTaoSimulationProgram2025}, and related to the values of the data (targets) and variables, and the weights placed upon the variables. These weights are sometimes adjusted manually by the user in efforts to guide the simulations toward specific goals. For this reason, the authors are including the targets for each match and the model values as well, so that the combination of these and the overall merit function can describe the quality of the matches. Most simply, a smaller the merit function indicates a better match. However, the combination of the variable and target values, as well as the weights of the final matches will influence how low the merit functions may be able to achieve.

\subsection{Matches Without $R_{56}$ Compensation}
\label{sec:AppA_1}
The following will show match solutions without the inclusion of $R_{56}$ compensation. They are broken apart by input LINAC optics; either from the SFT or WFT.

\subsubsection{Strong Focusing Triplet LINAC Input}
\label{sec:AppA_1-1}

There are two matches which use the SFT as inputs: one from SFT into the Beginning match point in the FFA Cell, and another from SFT to the BDC match point in the FFA Cell. Both are shown below, and the reader is requested to take care when reading the following tables.

\textbf{Showing first the SFT matching into the Beginning FFA Cell match point:}

\begin{figure}[!ht]
    \centering
    \includegraphics[width=\textwidth]{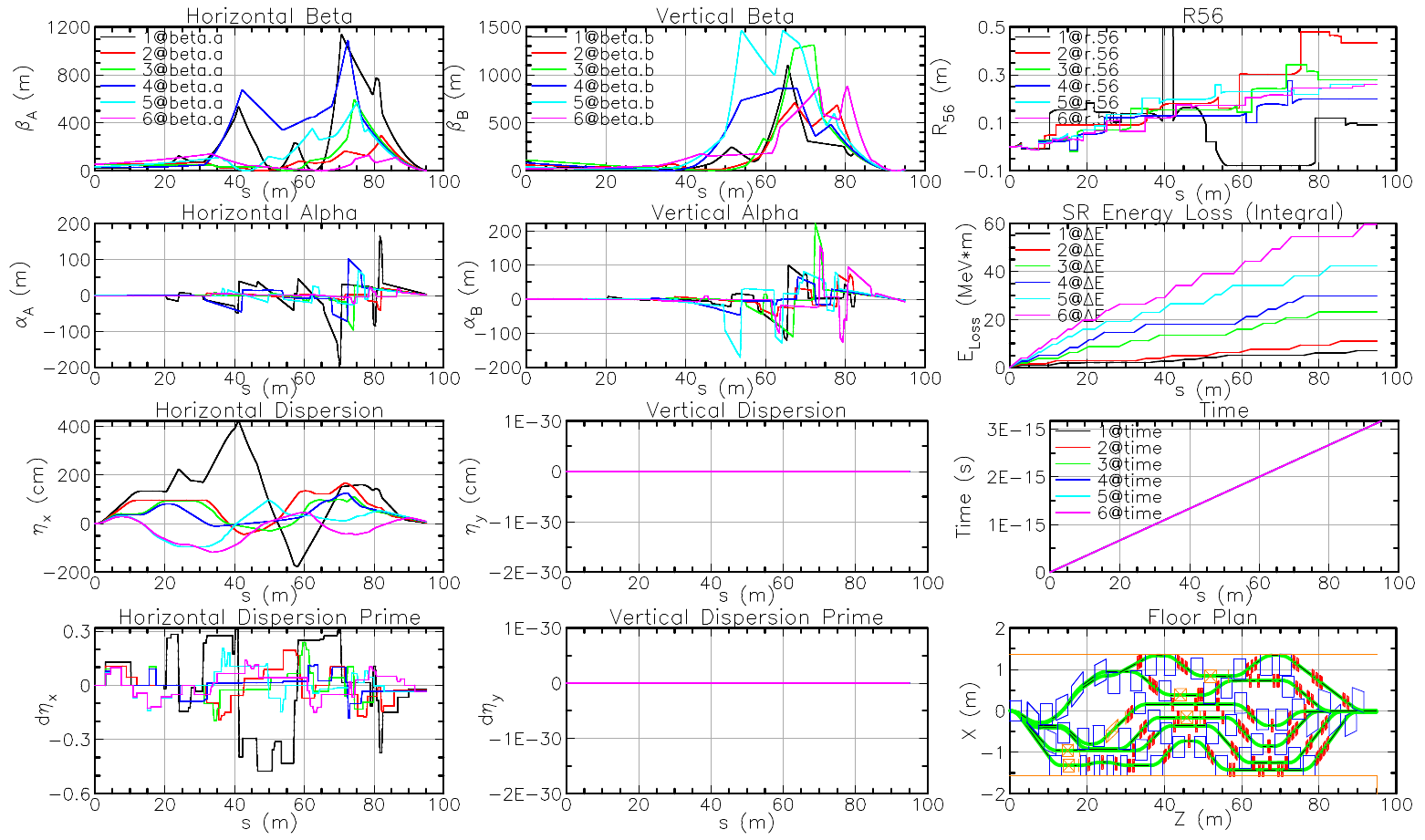}
    \caption{Matching solution from the SFT LINAC input into the Beginning FFA match point. In this case, all or nearly all of the Twiss parameters are matched well, and $R_{56}$ compensation is ignored.}
    \label{fig:SFT_Beginning_Twiss}
\end{figure}

\begin{longtable}{lcccc}
\caption{Complete summary of Twiss parameter optimization constraints for the matching case of the SFT LINAC input and the Beginning match point in the FFA Arc, excluding $R_{56}$ matching. This includes target values for optical functions at the beamline end-point and maximum value limits throughout the line for each of the six energy passes.}
\label{tab:SFT_Beginning_Twiss_full_summary} \\

\toprule
\textbf{Parameter} & \textbf{Target/Limit} & \textbf{Model Value} & \textbf{Merit} & \textbf{Location} \\
\midrule
\endfirsthead

\caption{(Continued)} \\
\toprule
\textbf{Parameter} & \textbf{Target/Limit} & \textbf{Model Value} & \textbf{Merit} & \textbf{Location} \\
\midrule
\endhead

\bottomrule
\multicolumn{5}{r}{\textit{Continued on next page}} \\
\endfoot

\bottomrule
\endlastfoot

\multicolumn{5}{l}{\textbf{First Pass}} \\
$\beta_{a}$[END] & $4.1573 \times 10^{0}$ & $4.1573 \times 10^{0}$ & $9.373 \times 10^{-7}$ & - \\
$\beta_{b}$[END] & $6.5151 \times 10^{0}$ & $6.5151 \times 10^{0}$ & $2.916 \times 10^{-7}$ & - \\
$\alpha_{a}$[END] & $3.0489 \times 10^{0}$ & $3.0489 \times 10^{0}$ & $7.836 \times 10^{-8}$ & - \\
$\alpha_{b}$[END] & $-3.1896 \times 10^{0}$ & $-3.1896 \times 10^{0}$ & $5.412 \times 10^{-8}$ & - \\
$\eta_{a}$[END] & $2.7200 \times 10^{-2}$ & $2.7200 \times 10^{-2}$ & $1.883 \times 10^{-11}$ & - \\
$\eta'_{a}$[END] & $-1.9800 \times 10^{-2}$ & $-1.9800 \times 10^{-2}$ & $9.289 \times 10^{-13}$ & - \\
$R_{56}$[END] & $-7.0886 \times 10^{-2}$ & $9.1707 \times 10^{-2}$ & $2.644 \times 10^{13}$ & - \\
max($\beta_a$) & $7.0000 \times 10^{2}$ & $1.1342 \times 10^{3}$ & $1.885 \times 10^{5}$ & M9Q16 \\
max($\beta_b$) & $7.0000 \times 10^{2}$ & $1.0804 \times 10^{3}$ & $1.447 \times 10^{5}$ & M9Q14 \\
\midrule
\multicolumn{5}{l}{\textbf{Second Pass}} \\
$\beta_{a}$[END] & $2.9506 \times 10^{0}$ & $2.9506 \times 10^{0}$ & $5.130 \times 10^{-16}$ & - \\
$\beta_{b}$[END] & $6.4768 \times 10^{0}$ & $6.4768 \times 10^{0}$ & $2.840 \times 10^{-19}$ & - \\
$\alpha_{a}$[END] & $1.8215 \times 10^{0}$ & $1.8215 \times 10^{0}$ & $8.780 \times 10^{-21}$ & - \\
$\alpha_{b}$[END] & $-3.0366 \times 10^{0}$ & $-3.0366 \times 10^{0}$ & $5.049 \times 10^{-23}$ & - \\
$\eta_{a}$[END] & $4.5700 \times 10^{-2}$ & $4.5700 \times 10^{-2}$ & $2.053 \times 10^{-21}$ & - \\
$\eta'_{a}$[END] & $-2.6400 \times 10^{-2}$ & $-2.6400 \times 10^{-2}$ & $1.157 \times 10^{-22}$ & - \\
$R_{56}$[END] & $4.4694 \times 10^{-2}$ & $4.3370 \times 10^{-1}$ & $1.513 \times 10^{14}$ & - \\
max($\beta_a$) & $7.0000 \times 10^{2}$ & $2.9197 \times 10^{2}$ & 0 & M11Q14 \\
max($\beta_b$) & $7.0000 \times 10^{2}$ & $7.0000 \times 10^{2}$ & 0 & D1105\#5 \\
\midrule
\multicolumn{5}{l}{\textbf{Third Pass}} \\
$\beta_{a}$[END] & $2.7180 \times 10^{0}$ & $2.7180 \times 10^{0}$ & $2.827 \times 10^{-3}$ & - \\
$\beta_{b}$[END] & $6.9948 \times 10^{0}$ & $6.9948 \times 10^{0}$ & $1.848 \times 10^{0}$ & - \\
$\alpha_{a}$[END] & $1.5388 \times 10^{0}$ & $1.5387 \times 10^{0}$ & $1.164 \times 10^{-2}$ & - \\
$\alpha_{b}$[END] & $-3.2063 \times 10^{0}$ & $-3.2063 \times 10^{0}$ & $2.432 \times 10^{-3}$ & - \\
$\eta_{a}$[END] & $6.0800 \times 10^{-2}$ & $6.0800 \times 10^{-2}$ & $3.006 \times 10^{-13}$ & - \\
$\eta'_{a}$[END] & $-3.0900 \times 10^{-2}$ & $-3.0900 \times 10^{-2}$ & $1.627 \times 10^{-9}$ & - \\
$R_{56}$[END] & $1.2809 \times 10^{-1}$ & $2.7849 \times 10^{-1}$ & $2.262 \times 10^{13}$ & - \\
max($\beta_a$) & $7.0000 \times 10^{2}$ & $5.9036 \times 10^{2}$ & 0 & M13Q12 \\
max($\beta_b$) & $7.0000 \times 10^{2}$ & $1.3097 \times 10^{3}$ & $3.718 \times 10^{5}$ & D1308\#2 \\
\midrule
\multicolumn{5}{l}{\textbf{Fourth Pass}} \\
$\beta_{a}$[END] & $2.6023 \times 10^{0}$ & $2.6023 \times 10^{0}$ & $1.965 \times 10^{0}$ & - \\
$\beta_{b}$[END] & $8.0350 \times 10^{0}$ & $8.0350 \times 10^{0}$ & $3.452 \times 10^{-2}$ & - \\
$\alpha_{a}$[END] & $1.3995 \times 10^{0}$ & $1.3995 \times 10^{0}$ & $1.419 \times 10^{-3}$ & - \\
$\alpha_{b}$[END] & $-3.6359 \times 10^{0}$ & $-3.6359 \times 10^{0}$ & $2.086 \times 10^{-4}$ & - \\
$\eta_{a}$[END] & $7.3500 \times 10^{-2}$ & $7.3500 \times 10^{-2}$ & $1.252 \times 10^{-8}$ & - \\
$\eta'_{a}$[END] & $-3.3600 \times 10^{-2}$ & $-3.3600 \times 10^{-2}$ & $3.325 \times 10^{-6}$ & - \\
$R_{56}$[END] & $1.9190 \times 10^{-1}$ & $1.9951 \times 10^{-1}$ & $5.797 \times 10^{10}$ & - \\
max($\beta_a$) & $7.0000 \times 10^{2}$ & $1.0755 \times 10^{3}$ & $1.410 \times 10^{5}$ & D1508\#2 \\
max($\beta_b$) & $7.0000 \times 10^{2}$ & $8.5944 \times 10^{2}$ & $2.542 \times 10^{4}$ & D1507\#2 \\
\midrule
\multicolumn{5}{l}{\textbf{Fifth Pass}} \\
$\beta_{a}$[END] & $2.5213 \times 10^{0}$ & $2.5213 \times 10^{0}$ & $2.067 \times 10^{0}$ & - \\
$\beta_{b}$[END] & $1.0132 \times 10^{1}$ & $1.0132 \times 10^{1}$ & $6.994 \times 10^{-1}$ & - \\
$\alpha_{a}$[END] & $1.3111 \times 10^{0}$ & $1.3111 \times 10^{0}$ & $1.458 \times 10^{-3}$ & - \\
$\alpha_{b}$[END] & $-4.5492 \times 10^{0}$ & $-4.5492 \times 10^{0}$ & $1.077 \times 10^{-3}$ & - \\
$\eta_{a}$[END] & $8.4100 \times 10^{-2}$ & $8.4101 \times 10^{-2}$ & $3.896 \times 10^{-5}$ & - \\
$\eta'_{a}$[END] & $-3.4900 \times 10^{-2}$ & $-3.4900 \times 10^{-2}$ & $1.323 \times 10^{-5}$ & - \\
$R_{56}$[END] & $2.4233 \times 10^{-1}$ & $2.6042 \times 10^{-1}$ & $3.272 \times 10^{11}$ & - \\
max($\beta_a$) & $7.0000 \times 10^{2}$ & $5.6689 \times 10^{2}$ & 0 & D1710\#5 \\
max($\beta_b$) & $7.0000 \times 10^{2}$ & $1.4590 \times 10^{3}$ & $5.760 \times 10^{5}$ & M17Q6 \\
\midrule
\multicolumn{5}{l}{\textbf{Sixth Pass}} \\
$\beta_{a}$[END] & $2.4552 \times 10^{0}$ & $2.4552 \times 10^{0}$ & $6.660 \times 10^{-3}$ & - \\
$\beta_{b}$[END] & $1.6840 \times 10^{1}$ & $1.6840 \times 10^{1}$ & $1.409 \times 10^{-2}$ & - \\
$\alpha_{a}$[END] & $1.2471 \times 10^{0}$ & $1.2471 \times 10^{0}$ & $1.888 \times 10^{-6}$ & - \\
$\alpha_{b}$[END] & $-7.5245 \times 10^{0}$ & $-7.5245 \times 10^{0}$ & $2.055 \times 10^{-5}$ & - \\
$\eta_{a}$[END] & $9.3200 \times 10^{-2}$ & $9.3200 \times 10^{-2}$ & $5.910 \times 10^{-9}$ & - \\
$\eta'_{a}$[END] & $-3.5000 \times 10^{-2}$ & $-3.5000 \times 10^{-2}$ & $1.126 \times 10^{-8}$ & - \\
$R_{56}$[END] & $2.8081 \times 10^{-1}$ & $2.6071 \times 10^{-1}$ & $4.040 \times 10^{11}$ & - \\
max($\beta_a$) & $7.0000 \times 10^{2}$ & $1.3979 \times 10^{2}$ & 0 & D1906\#2 \\
max($\beta_b$) & $7.0000 \times 10^{2}$ & $8.6919 \times 10^{2}$ & $2.863 \times 10^{4}$ & D1912\#2 \\

\end{longtable}

\begin{longtable}{lS[table-format=2.2]S[table-format=-1.4] @{\hskip 1cm} lS[table-format=2.2]S[table-format=-1.4]}
\caption{Optimized integrated quadrupole strengths ($K_1$) for each of the six passes. The longitudinal position of the center of each magnet is given by the S-coordinate. All quadrupole strengths were limited to a range of $\pm \SI{0.6}{\per\meter\squared}$ during the optimization.}
\label{tab:SFT_Beginning_Twiss_quad_strengths} \\

\toprule
\textbf{Element} & {\textbf{S-Pos. (m)}} & {\textbf{$K_1$ (\si{\per\meter\squared})}} & \textbf{Element} & {\textbf{S-Pos. (m)}} & {\textbf{$K_1$ (\si{\per\meter\squared})}} \\
\midrule
\endfirsthead

\caption{(Continued)} \\
\toprule
\textbf{Element} & {\textbf{S-Pos. (m)}} & {\textbf{$K_1$ (\si{\per\meter\squared})}} & \textbf{Element} & {\textbf{S-Pos. (m)}} & {\textbf{$K_1$ (\si{\per\meter\squared})}} \\
\midrule
\endhead

\bottomrule
\endfoot

\multicolumn{6}{c}{\textbf{First Pass (P9) Quadrupoles}} \\
\midrule
M9Q1 & 20.66 & -0.6000 & M9Q11 & 57.28 & 0.5811 \\
M9Q2 & 24.16 & 0.5405 & M9Q12 & 58.28 & 0.5347 \\
M9Q3 & 27.66 & -0.1765 & M9Q13 & 64.78 & 0.0693 \\
M9Q4 & 31.16 & -0.4388 & M9Q14 & 65.78 & -0.5702 \\
M9Q5 & 32.16 & -0.0596 & M9Q15 & 69.78 & -0.0990 \\
M9Q6 & 40.36 & -0.0980 & M9Q16 & 70.78 & 0.5675 \\
M9Q7 & 41.36 & 0.4725 & M9Q17 & 79.88 & -0.4501 \\
M9Q8 & 46.66 & 0.2116 & M9Q18 & 80.78 & 0.3674 \\
M9Q9 & 51.66 & -0.3877 & M9Q19 & 81.68 & 0.6000 \\
M9Q10 & 52.66 & -0.0463 & M9Q20 & 82.58 & -0.6000 \\
\midrule
\multicolumn{6}{c}{\textbf{Second Pass (P11) Quadrupoles}} \\
\midrule
M11Q1 & 32.11 & 0.2937 & M11Q8 & 63.28 & -0.1620 \\
M11Q2 & 36.36 & -0.2157 & M11Q9 & 67.54 & -0.2364 \\
M11Q3 & 37.36 & -0.2807 & M11Q10 & 71.80 & 0.2626 \\
M11Q4 & 42.86 & 0.6000 & M11Q11 & 77.21 & 0.0153 \\
M11Q5 & 48.50 & 0.6000 & M11Q12 & 78.21 & -0.3844 \\
M11Q6 & 57.86 & 0.0721 & M11Q13 & 81.21 & -0.2265 \\
M11Q7 & 58.86 & 0.2828 & M11Q14 & 82.21 & 0.6000 \\
\midrule
\multicolumn{6}{c}{\textbf{Third Pass (P13) Quadrupoles}} \\
\midrule
M13Q1 & 34.16 & 0.5247 & M13Q7 & 60.66 & 0.1792 \\
M13Q2 & 35.16 & -0.6000 & M13Q8 & 61.66 & 0.4476 \\
M13Q3 & 43.66 & 0.0405 & M13Q9 & 67.28 & -0.2390 \\
M13Q4 & 50.55 & 0.6000 & M13Q10 & 72.46 & -0.4977 \\
M13Q5 & 55.30 & 0.6000 & M13Q11 & 73.46 & 0.0037 \\
M13Q6 & 59.66 & -0.5072 & M13Q12 & 74.46 & 0.6000 \\
\midrule
\multicolumn{6}{c}{\textbf{Fourth Pass (P15) Quadrupoles}} \\
\midrule
M15Q1 & 30.86 & -0.0460 & M15Q6 & 62.04 & 0.0391 \\
M15Q2 & 31.66 & -0.5962 & M15Q7 & 63.04 & -0.0441 \\
M15Q3 & 42.35 & 0.2561 & M15Q8 & 68.11 & -0.2162 \\
M15Q4 & 48.36 & 0.0148 & M15Q9 & 72.70 & 0.4529 \\
M15Q5 & 54.05 & -0.1609 & M15Q10 & 76.62 & -0.2683 \\
\midrule
\multicolumn{6}{c}{\textbf{Fifth Pass (P17) Quadrupoles}} \\
\midrule
M17Q1 & 36.56 & 0.5500 & M17Q8 & 64.50 & -0.2833 \\
M17Q2 & 37.36 & -0.5500 & M17Q9 & 68.63 & -0.0718 \\
M17Q3 & 38.16 & 0.5088 & M17Q10 & 69.63 & -0.0766 \\
M17Q4 & 49.23 & -0.0375 & M17Q11 & 74.48 & 0.0286 \\
M17Q5 & 49.87 & 0.5500 & M17Q12 & 75.48 & 0.4606 \\
M17Q6 & 54.09 & -0.3913 & M17Q13 & 77.48 & -0.5500 \\
M17Q7 & 62.50 & 0.3729 & M17Q14 & 78.48 & 0.2466 \\
\midrule
\multicolumn{6}{c}{\textbf{Sixth Pass (P19) Quadrupoles}} \\
\midrule
M19Q1 & 33.36 & 0.1070 & M19Q9 & 73.75 & -0.0585 \\
M19Q2 & 33.96 & 0.0541 & M19Q10 & 74.75 & 0.0570 \\
M19Q3 & 43.86 & -0.1464 & M19Q11 & 78.75 & -0.0251 \\
M19Q4 & 44.86 & -0.1317 & M19Q12 & 79.75 & -0.1948 \\
M19Q5 & 45.86 & 0.1897 & M19Q13 & 80.75 & -0.1513 \\
M19Q6 & 61.81 & 0.6000 & M19Q14 & 86.75 & 0.5286 \\
M19Q7 & 62.81 & 0.3166 & M19Q15 & 87.75 & -0.0635 \\
M19Q8 & 63.81 & -0.5285 & & & \\

\end{longtable}
\FloatBarrier

\textbf{Showing next the SFT matching into the BDC FFA Cell match point:}

\begin{figure}[!ht]
    \centering
    \includegraphics[width=\textwidth]{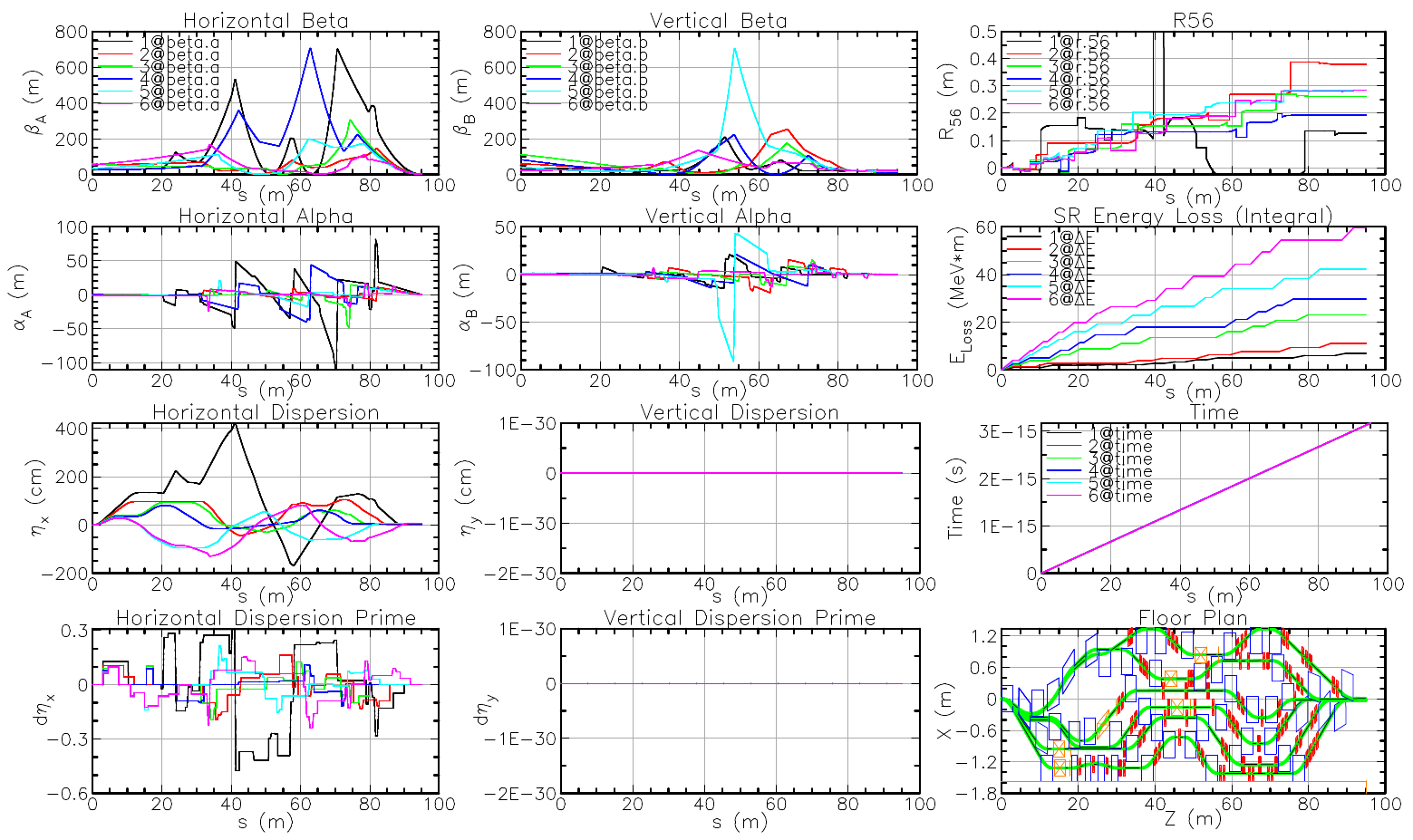}
    \caption{Matching solution from the SFT LINAC input into the BDC FFA match point. In this case, all of the Twiss parameters are always matched perfectly, and $R_{56}$ compensation is ignored.}
    \label{fig:SFT_Twiss}
\end{figure}

\begin{longtable}{lcccc}
\caption{Complete summary of Twiss parameter optimization constraints for the matching case of the SFT LINAC input and the BDC match point in the FFA Arc, excluding $R_{56}$ matching. This includes target values for optical functions at the beamline end-point and maximum value limits throughout the line for each of the six energy passes.}
\label{tab:SFT_Twiss_full_summary} \\

\toprule
\textbf{Parameter} & \textbf{Target/Limit} & \textbf{Model Value} & \textbf{Merit} & \textbf{Location} \\
\midrule
\endfirsthead

\caption{(Continued)} \\
\toprule
\textbf{Parameter} & \textbf{Target/Limit} & \textbf{Model Value} & \textbf{Merit} & \textbf{Location} \\
\midrule
\endhead

\bottomrule
\multicolumn{5}{r}{\textit{Continued on next page}} \\
\endfoot

\bottomrule
\endlastfoot

\multicolumn{5}{l}{\textbf{First Pass}} \\
$\beta_{a}$[END] & $6.6586 \times 10^{-1}$ & $6.6586 \times 10^{-1}$ & $2.840 \times 10^{-19}$ & - \\
$\beta_{b}$[END] & $1.2039 \times 10^{1}$ & $1.2039 \times 10^{1}$ & $6.185 \times 10^{-18}$ & - \\
$\alpha_{a}$[END] & $4.3124 \times 10^{-5}$ & $4.3124 \times 10^{-5}$ & $1.180 \times 10^{-21}$ & - \\
$\alpha_{b}$[END] & $-1.1741 \times 10^{-4}$ & $-1.1741 \times 10^{-4}$ & $4.793 \times 10^{-23}$ & - \\
$\eta_{a}$[END] & $-7.8352 \times 10^{-3}$ & $-7.8352 \times 10^{-3}$ & $1.516 \times 10^{-20}$ & - \\
$\eta'_{a}$[END] & $4.3951 \times 10^{-8}$ & $4.3951 \times 10^{-8}$ & $1.309 \times 10^{-22}$ & - \\
$R_{56}$[END] & $-7.0886 \times 10^{-2}$ & $1.2731 \times 10^{-1}$ & $3.928 \times 10^{13}$ & - \\
max($\beta_a$) & $7.0000 \times 10^{2}$ & $7.0000 \times 10^{2}$ & 0 & M9Q16 \\
max($\beta_b$) & $7.0000 \times 10^{2}$ & $2.0685 \times 10^{2}$ & 0 & D908\#2 \\
\midrule
\multicolumn{5}{l}{\textbf{Second Pass}} \\
$\beta_{a}$[END] & $1.1641 \times 10^{0}$ & $1.1641 \times 10^{0}$ & $6.390 \times 10^{-19}$ & - \\
$\beta_{b}$[END] & $1.0662 \times 10^{1}$ & $1.0662 \times 10^{1}$ & $3.032 \times 10^{-17}$ & - \\
$\alpha_{a}$[END] & $2.6829 \times 10^{-5}$ & $2.6829 \times 10^{-5}$ & $9.488 \times 10^{-22}$ & - \\
$\alpha_{b}$[END] & $-1.0407 \times 10^{-4}$ & $-1.0407 \times 10^{-4}$ & $5.382 \times 10^{-24}$ & - \\
$\eta_{a}$[END] & $1.4456 \times 10^{-2}$ & $1.4456 \times 10^{-2}$ & $2.359 \times 10^{-25}$ & - \\
$\eta'_{a}$[END] & $-1.8690 \times 10^{-7}$ & $-1.8690 \times 10^{-7}$ & $1.007 \times 10^{-22}$ & - \\
$R_{56}$[END] & $4.4694 \times 10^{-2}$ & $3.7945 \times 10^{-1}$ & $1.121 \times 10^{14}$ & - \\
max($\beta_a$) & $7.0000 \times 10^{2}$ & $1.4403 \times 10^{2}$ & 0 & D1106\#11 \\
max($\beta_b$) & $7.0000 \times 10^{2}$ & $2.5212 \times 10^{2}$ & 0 & D1105\#5 \\
\midrule
\multicolumn{5}{l}{\textbf{Third Pass}} \\
$\beta_{a}$[END] & $1.3998 \times 10^{0}$ & $1.3998 \times 10^{0}$ & $7.704 \times 10^{-18}$ & - \\
$\beta_{b}$[END] & $1.0669 \times 10^{1}$ & $1.0669 \times 10^{1}$ & $5.834 \times 10^{-17}$ & - \\
$\alpha_{a}$[END] & $2.4658 \times 10^{-5}$ & $2.4658 \times 10^{-5}$ & $6.668 \times 10^{-22}$ & - \\
$\alpha_{b}$[END] & $-1.0462 \times 10^{-4}$ & $-1.0462 \times 10^{-4}$ & $3.689 \times 10^{-22}$ & - \\
$\eta_{a}$[END] & $3.2967 \times 10^{-2}$ & $3.2967 \times 10^{-2}$ & $8.517 \times 10^{-23}$ & - \\
$\eta'_{a}$[END] & $-3.7558 \times 10^{-7}$ & $-3.7558 \times 10^{-7}$ & $3.522 \times 10^{-23}$ & - \\
$R_{56}$[END] & $1.2809 \times 10^{-1}$ & $2.6117 \times 10^{-1}$ & $1.771 \times 10^{13}$ & - \\
max($\beta_a$) & $7.0000 \times 10^{2}$ & $3.0368 \times 10^{2}$ & 0 & M13Q12 \\
max($\beta_b$) & $7.0000 \times 10^{2}$ & $1.7258 \times 10^{2}$ & 0 & M13Q9 \\
\midrule
\multicolumn{5}{l}{\textbf{Fourth Pass}} \\
$\beta_{a}$[END] & $1.5386 \times 10^{0}$ & $1.5386 \times 10^{0}$ & $1.333 \times 10^{-18}$ & - \\
$\beta_{b}$[END] & $1.1607 \times 10^{1}$ & $1.1607 \times 10^{1}$ & $6.109 \times 10^{-17}$ & - \\
$\alpha_{a}$[END] & $2.4713 \times 10^{-5}$ & $2.4713 \times 10^{-5}$ & $1.503 \times 10^{-22}$ & - \\
$\alpha_{b}$[END] & $-1.1455 \times 10^{-4}$ & $-1.1455 \times 10^{-4}$ & $2.483 \times 10^{-26}$ & - \\
$\eta_{a}$[END] & $4.8495 \times 10^{-2}$ & $4.8495 \times 10^{-2}$ & $2.644 \times 10^{-21}$ & - \\
$\eta'_{a}$[END] & $-5.3096 \times 10^{-7}$ & $-5.3096 \times 10^{-7}$ & $1.567 \times 10^{-23}$ & - \\
$R_{56}$[END] & $1.9190 \times 10^{-1}$ & $1.9246 \times 10^{-1}$ & $3.144 \times 10^{8}$ & - \\
max($\beta_a$) & $7.0000 \times 10^{2}$ & $7.0000 \times 10^{2}$ & 0 & D1506\#5 \\
max($\beta_b$) & $7.0000 \times 10^{2}$ & $2.2086 \times 10^{2}$ & 0 & D1505\#8 \\
\midrule
\multicolumn{5}{l}{\textbf{Fifth Pass}} \\
$\beta_{a}$[END] & $1.6261 \times 10^{0}$ & $1.6261 \times 10^{0}$ & $4.450 \times 10^{-18}$ & - \\
$\beta_{b}$[END] & $1.4063 \times 10^{1}$ & $1.4063 \times 10^{1}$ & $1.392 \times 10^{-17}$ & - \\
$\alpha_{a}$[END] & $2.5401 \times 10^{-5}$ & $2.5401 \times 10^{-5}$ & $2.412 \times 10^{-21}$ & - \\
$\alpha_{b}$[END] & $-1.3987 \times 10^{-4}$ & $-1.3987 \times 10^{-4}$ & $1.722 \times 10^{-23}$ & - \\
$\eta_{a}$[END] & $6.1641 \times 10^{-2}$ & $6.1641 \times 10^{-2}$ & $1.282 \times 10^{-20}$ & - \\
$\eta'_{a}$[END] & $-6.6020 \times 10^{-7}$ & $-6.6020 \times 10^{-7}$ & $4.329 \times 10^{-23}$ & - \\
$R_{56}$[END] & $2.4233 \times 10^{-1}$ & $2.8420 \times 10^{-1}$ & $1.753 \times 10^{12}$ & - \\
max($\beta_a$) & $7.0000 \times 10^{2}$ & $1.9860 \times 10^{2}$ & 0 & M17Q7 \\
max($\beta_b$) & $7.0000 \times 10^{2}$ & $7.0000 \times 10^{2}$ & 0 & M17Q6 \\
\midrule
\multicolumn{5}{l}{\textbf{Sixth Pass}} \\
$\beta_{a}$[END] & $1.6824 \times 10^{0}$ & $1.6824 \times 10^{0}$ & $1.546 \times 10^{-18}$ & - \\
$\beta_{b}$[END] & $2.2819 \times 10^{1}$ & $2.2819 \times 10^{1}$ & $4.089 \times 10^{-17}$ & - \\
$\alpha_{a}$[END] & $2.6113 \times 10^{-5}$ & $2.6113 \times 10^{-5}$ & $4.884 \times 10^{-23}$ & - \\
$\alpha_{b}$[END] & $-2.2893 \times 10^{-4}$ & $-2.2893 \times 10^{-4}$ & $1.541 \times 10^{-23}$ & - \\
$\eta_{a}$[END] & $7.2865 \times 10^{-2}$ & $7.2865 \times 10^{-2}$ & $1.454 \times 10^{-20}$ & - \\
$\eta'_{a}$[END] & $-7.6926 \times 10^{-7}$ & $-7.6926 \times 10^{-7}$ & $1.388 \times 10^{-22}$ & - \\
$R_{56}$[END] & $2.8081 \times 10^{-1}$ & $2.8375 \times 10^{-1}$ & $8.646 \times 10^{9}$ & - \\
max($\beta_a$) & $7.0000 \times 10^{2}$ & $1.6566 \times 10^{2}$ & 0 & M19Q2 \\
max($\beta_b$) & $7.0000 \times 10^{2}$ & $1.3377 \times 10^{2}$ & 0 & M19Q4 \\

\end{longtable}

\begin{longtable}{lS[table-format=2.2]S[table-format=-1.4] @{\hskip 1cm} lS[table-format=2.2]S[table-format=-1.4]}
\caption{Optimized integrated quadrupole strengths ($K_1$) for each of the six passes. The longitudinal position of the center of each magnet is given by the S-coordinate. All quadrupole strengths were limited to a range of $\pm \SI{0.6}{\per\meter\squared}$ during the optimization.}
\label{tab:SFT_Twiss_quad_strengths} \\

\toprule
\textbf{Element} & {\textbf{S-Pos. (m)}} & {\textbf{$K_1$ (\si{\per\meter\squared})}} & \textbf{Element} & {\textbf{S-Pos. (m)}} & {\textbf{$K_1$ (\si{\per\meter\squared})}} \\
\midrule
\endfirsthead

\caption{(Continued)} \\
\toprule
\textbf{Element} & {\textbf{S-Pos. (m)}} & {\textbf{$K_1$ (\si{\per\meter\squared})}} & \textbf{Element} & {\textbf{S-Pos. (m)}} & {\textbf{$K_1$ (\si{\per\meter\squared})}} \\
\midrule
\endhead

\bottomrule
\endfoot

\multicolumn{6}{c}{\textbf{First Pass (P9) Quadrupoles}} \\
\midrule
M9Q1 & 20.66 & -0.6000 & M9Q11 & 57.28 & 0.5311 \\
M9Q2 & 24.16 & 0.5405 & M9Q12 & 58.28 & 0.5315 \\
M9Q3 & 27.66 & -0.1765 & M9Q13 & 64.78 & 0.1258 \\
M9Q4 & 31.16 & -0.4388 & M9Q14 & 65.78 & -0.5395 \\
M9Q5 & 32.16 & -0.0596 & M9Q15 & 69.78 & -0.0459 \\
M9Q6 & 40.36 & -0.0969 & M9Q16 & 70.78 & 0.5483 \\
M9Q7 & 41.36 & 0.5243 & M9Q17 & 79.88 & -0.2963 \\
M9Q8 & 46.66 & 0.0667 & M9Q18 & 80.78 & 0.2021 \\
M9Q9 & 51.66 & -0.4195 & M9Q19 & 81.68 & 0.6000 \\
M9Q10 & 52.66 & -0.0911 & M9Q20 & 82.58 & -0.5658 \\
\midrule
\multicolumn{6}{c}{\textbf{Second Pass (P11) Quadrupoles}} \\
\midrule
M11Q1 & 32.11 & 0.2937 & M11Q8 & 63.28 & -0.2033 \\
M11Q2 & 36.36 & -0.2157 & M11Q9 & 67.54 & -0.2161 \\
M11Q3 & 37.36 & -0.2807 & M11Q10 & 71.80 & 0.1064 \\
M11Q4 & 42.86 & 0.6000 & M11Q11 & 77.21 & 0.1190 \\
M11Q5 & 48.50 & 0.2439 & M11Q12 & 78.21 & -0.2094 \\
M11Q6 & 57.86 & 0.5904 & M11Q13 & 81.21 & -0.0447 \\
M11Q7 & 58.86 & -0.1868 & M11Q14 & 82.21 & 0.3856 \\
\midrule
\multicolumn{6}{c}{\textbf{Third Pass (P13) Quadrupoles}} \\
\midrule
M13Q1 & 34.16 & 0.5247 & M13Q7 & 60.66 & 0.6000 \\
M13Q2 & 35.16 & -0.6000 & M13Q8 & 61.66 & -0.1637 \\
M13Q3 & 43.66 & 0.0405 & M13Q9 & 67.28 & -0.3358 \\
M13Q4 & 50.55 & 0.6000 & M13Q10 & 72.46 & -0.0471 \\
M13Q5 & 55.30 & -0.4086 & M13Q11 & 73.46 & -0.2897 \\
M13Q6 & 59.66 & 0.0059 & M13Q12 & 74.46 & 0.6000 \\
\midrule
\multicolumn{6}{c}{\textbf{Fourth Pass (P15) Quadrupoles}} \\
\midrule
M15Q1 & 30.86 & 0.1734 & M15Q6 & 62.04 & 0.0240 \\
M15Q2 & 31.66 & -0.6000 & M15Q7 & 63.04 & 0.3249 \\
M15Q3 & 42.35 & 0.3072 & M15Q8 & 68.11 & -0.0233 \\
M15Q4 & 48.36 & -0.1554 & M15Q9 & 72.70 & -0.6000 \\
M15Q5 & 54.05 & -0.3893 & M15Q10 & 76.62 & 0.3266 \\
\midrule
\multicolumn{6}{c}{\textbf{Fifth Pass (P17) Quadrupoles}} \\
\midrule
M17Q1 & 36.56 & 0.6000 & M17Q8 & 64.50 & 0.0142 \\
M17Q2 & 37.36 & -0.4601 & M17Q9 & 68.63 & -0.0002 \\
M17Q3 & 38.16 & 0.0734 & M17Q10 & 69.63 & -0.0652 \\
M17Q4 & 49.23 & 0.1396 & M17Q11 & 74.48 & -0.0613 \\
M17Q5 & 49.87 & 0.6000 & M17Q12 & 75.48 & 0.0164 \\
M17Q6 & 54.09 & -0.5447 & M17Q13 & 77.48 & 0.0232 \\
M17Q7 & 62.50 & 0.2968 & M17Q14 & 78.48 & 0.1903 \\
\midrule
\multicolumn{6}{c}{\textbf{Sixth Pass (P19) Quadrupoles}} \\
\midrule
M19Q1 & 33.36 & -0.4061 & M19Q9 & 73.75 & -0.5500 \\
M19Q2 & 33.96 & 0.5500 & M19Q10 & 74.75 & 0.5479 \\
M19Q3 & 43.86 & 0.0054 & M19Q11 & 78.75 & 0.5500 \\
M19Q4 & 44.86 & -0.1742 & M19Q12 & 79.75 & -0.5500 \\
M19Q5 & 45.86 & -0.0168 & M19Q13 & 80.75 & 0.1361 \\
M19Q6 & 61.81 & 0.5877 & M19Q14 & 86.75 & 0.5500 \\
M19Q7 & 62.81 & 0.2850 & M19Q15 & 87.75 & -0.4598 \\
M19Q8 & 63.81 & -0.5652 & & & \\

\end{longtable}
\FloatBarrier

\subsubsection{Weakly Focusing Triplet LINAC Input}
\label{sec:AppA_1-2}

\begin{figure}[!ht]
    \centering
    \includegraphics[width=\textwidth]{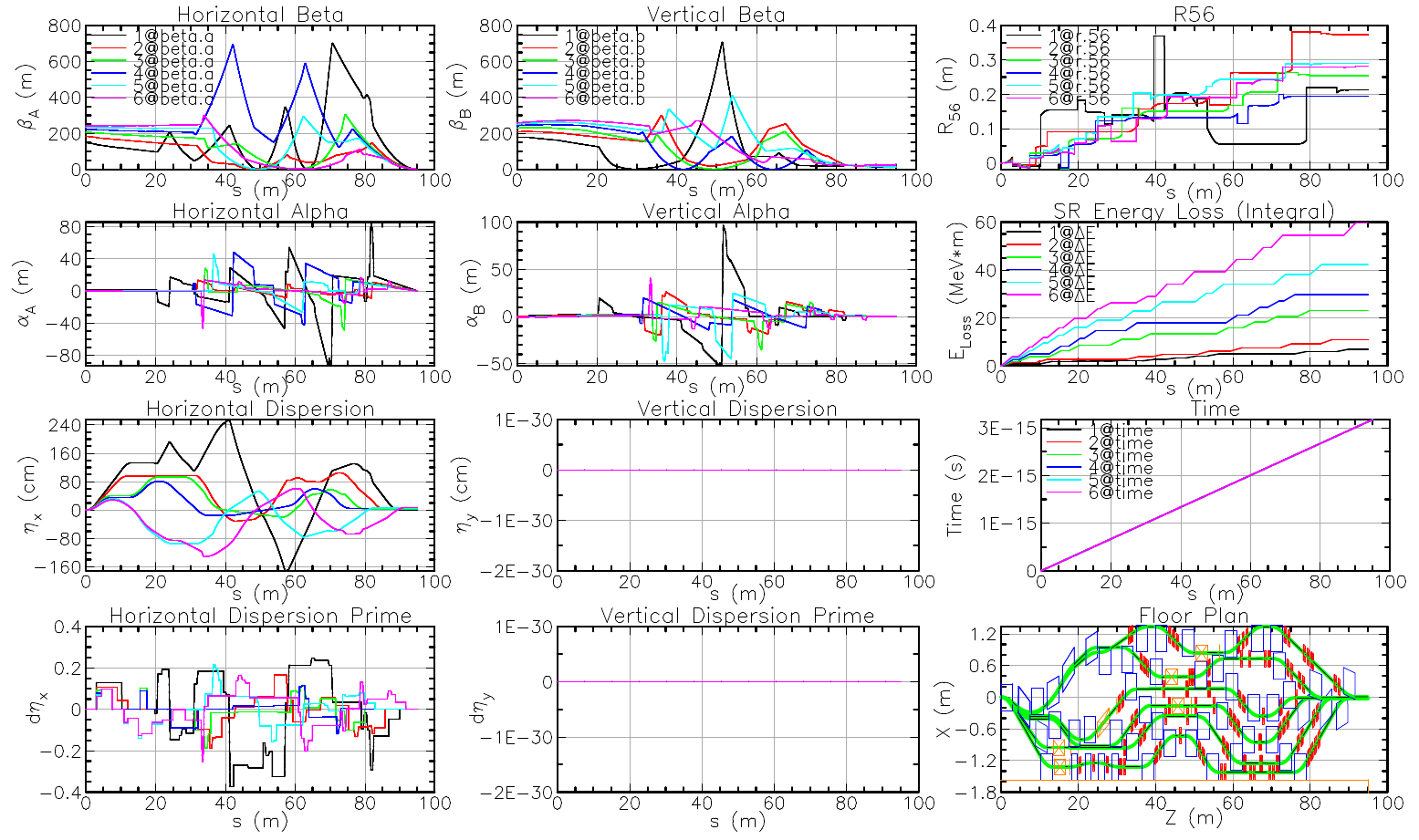}
    \caption{Matching solution from the WFT LINAC input into the BDC FFA match point. In this case, all of the Twiss parameters are always matched perfectly, and $R_{56}$ compensation is ignored.}
    \label{fig:WFT_Twiss}
\end{figure}

\begin{longtable}{lcccc}
\caption{Complete summary of Twiss parameter optimization constraints for the matching case of the WFT LINAC input and the BDC match point in the FFA Arc, excluding $R_{56}$ matching. This includes target values for optical functions at the beamline end-point and maximum value limits throughout the line for each of the six energy passes.}
\label{tab:WFT_Twiss_full_summary} \\

\toprule
\textbf{Parameter} & \textbf{Target/Limit} & \textbf{Model Value} & \textbf{Merit} & \textbf{Location} \\
\midrule
\endfirsthead

\caption{(Continued)} \\
\toprule
\textbf{Parameter} & \textbf{Target/Limit} & \textbf{Model Value} & \textbf{Merit} & \textbf{Location} \\
\midrule
\endhead

\bottomrule
\multicolumn{5}{r}{\textit{Continued on next page}} \\
\endfoot

\bottomrule
\endlastfoot

\multicolumn{5}{l}{\textbf{First Pass}} \\
$\beta_{a}$[END] & $6.6586 \times 10^{-1}$ & $6.6586 \times 10^{-1}$ & $3.462 \times 10^{-19}$ & - \\
$\beta_{b}$[END] & $1.2039 \times 10^{1}$ & $1.2039 \times 10^{1}$ & $1.262 \times 10^{-19}$ & - \\
$\alpha_{a}$[END] & $4.3124 \times 10^{-5}$ & $4.3124 \times 10^{-5}$ & $1.149 \times 10^{-20}$ & - \\
$\alpha_{b}$[END] & $-1.1741 \times 10^{-4}$ & $-1.1741 \times 10^{-4}$ & $5.772 \times 10^{-24}$ & - \\
$\eta_{a}$[END] & $-7.8352 \times 10^{-3}$ & $-7.8352 \times 10^{-3}$ & $6.415 \times 10^{-24}$ & - \\
$\eta'_{a}$[END] & $4.3951 \times 10^{-8}$ & $4.3951 \times 10^{-8}$ & $2.939 \times 10^{-23}$ & - \\
$R_{56}$[END] & $-7.0886 \times 10^{-2}$ & $2.1209 \times 10^{-1}$ & $8.008 \times 10^{13}$ & - \\
max($\beta_a$) & $7.0000 \times 10^{2}$ & $7.0000 \times 10^{2}$ & 0 & M9Q16 \\
max($\beta_b$) & $7.0000 \times 10^{2}$ & $7.0000 \times 10^{2}$ & 0 & D908\#2 \\
\midrule
\multicolumn{5}{l}{\textbf{Second Pass}} \\
$\beta_{a}$[END] & $1.1641 \times 10^{0}$ & $1.1641 \times 10^{0}$ & $2.174 \times 10^{-19}$ & - \\
$\beta_{b}$[END] & $1.0662 \times 10^{1}$ & $1.0662 \times 10^{1}$ & 0 & - \\
$\alpha_{a}$[END] & $2.6829 \times 10^{-5}$ & $2.6829 \times 10^{-5}$ & $8.040 \times 10^{-26}$ & - \\
$\alpha_{b}$[END] & $-1.0407 \times 10^{-4}$ & $-1.0407 \times 10^{-4}$ & $1.358 \times 10^{-28}$ & - \\
$\eta_{a}$[END] & $1.4456 \times 10^{-2}$ & $1.4456 \times 10^{-2}$ & $2.113 \times 10^{-23}$ & - \\
$\eta'_{a}$[END] & $-1.8690 \times 10^{-7}$ & $-1.8690 \times 10^{-7}$ & $4.234 \times 10^{-25}$ & - \\
$R_{56}$[END] & $4.4694 \times 10^{-2}$ & $3.7272 \times 10^{-1}$ & $1.076 \times 10^{14}$ & - \\
max($\beta_a$) & $7.0000 \times 10^{2}$ & $1.8362 \times 10^{2}$ & 0 & BEGINNING \\
max($\beta_b$) & $7.0000 \times 10^{2}$ & $2.9570 \times 10^{2}$ & 0 & D1102\#2 \\
\midrule
\multicolumn{5}{l}{\textbf{Third Pass}} \\
$\beta_{a}$[END] & $1.3998 \times 10^{0}$ & $1.3998 \times 10^{0}$ & $5.966 \times 10^{-20}$ & - \\
$\beta_{b}$[END] & $1.0669 \times 10^{1}$ & $1.0669 \times 10^{1}$ & $1.546 \times 10^{-18}$ & - \\
$\alpha_{a}$[END] & $2.4658 \times 10^{-5}$ & $2.4658 \times 10^{-5}$ & $6.756 \times 10^{-25}$ & - \\
$\alpha_{b}$[END] & $-1.0462 \times 10^{-4}$ & $-1.0462 \times 10^{-4}$ & $1.343 \times 10^{-23}$ & - \\
$\eta_{a}$[END] & $3.2967 \times 10^{-2}$ & $3.2967 \times 10^{-2}$ & $4.333 \times 10^{-26}$ & - \\
$\eta'_{a}$[END] & $-3.7558 \times 10^{-7}$ & $-3.7558 \times 10^{-7}$ & $2.938 \times 10^{-26}$ & - \\
$R_{56}$[END] & $1.2809 \times 10^{-1}$ & $2.5442 \times 10^{-1}$ & $1.596 \times 10^{13}$ & - \\
max($\beta_a$) & $7.0000 \times 10^{2}$ & $3.0368 \times 10^{2}$ & 0 & M13Q12 \\
max($\beta_b$) & $7.0000 \times 10^{2}$ & $2.3230 \times 10^{2}$ & 0 & MCB1 \\
\midrule
\multicolumn{5}{l}{\textbf{Fourth Pass}} \\
$\beta_{a}$[END] & $1.5386 \times 10^{0}$ & $1.5386 \times 10^{0}$ & $4.437 \times 10^{-19}$ & - \\
$\beta_{b}$[END] & $1.1607 \times 10^{1}$ & $1.1607 \times 10^{1}$ & $1.262 \times 10^{-17}$ & - \\
$\alpha_{a}$[END] & $2.4713 \times 10^{-5}$ & $2.4713 \times 10^{-5}$ & $1.011 \times 10^{-21}$ & - \\
$\alpha_{b}$[END] & $-1.1455 \times 10^{-4}$ & $-1.1455 \times 10^{-4}$ & $2.577 \times 10^{-22}$ & - \\
$\eta_{a}$[END] & $4.8495 \times 10^{-2}$ & $4.8495 \times 10^{-2}$ & $2.123 \times 10^{-24}$ & - \\
$\eta'_{a}$[END] & $-5.3096 \times 10^{-7}$ & $-5.3096 \times 10^{-7}$ & $1.637 \times 10^{-28}$ & - \\
$R_{56}$[END] & $1.9190 \times 10^{-1}$ & $1.9366 \times 10^{-1}$ & $3.085 \times 10^{9}$ & - \\
max($\beta_a$) & $7.0000 \times 10^{2}$ & $6.8792 \times 10^{2}$ & 0 & D1505\#2 \\
max($\beta_b$) & $7.0000 \times 10^{2}$ & $2.4895 \times 10^{2}$ & 0 & MCB2 \\
\midrule
\multicolumn{5}{l}{\textbf{Fifth Pass}} \\
$\beta_{a}$[END] & $1.6261 \times 10^{0}$ & $1.6261 \times 10^{0}$ & $7.499 \times 10^{-19}$ & - \\
$\beta_{b}$[END] & $1.4063 \times 10^{1}$ & $1.4063 \times 10^{1}$ & $3.818 \times 10^{-18}$ & - \\
$\alpha_{a}$[END] & $2.5401 \times 10^{-5}$ & $2.5401 \times 10^{-5}$ & $2.474 \times 10^{-22}$ & - \\
$\alpha_{b}$[END] & $-1.3987 \times 10^{-4}$ & $-1.3987 \times 10^{-4}$ & $3.005 \times 10^{-23}$ & - \\
$\eta_{a}$[END] & $6.1641 \times 10^{-2}$ & $6.1641 \times 10^{-2}$ & $2.123 \times 10^{-24}$ & - \\
$\eta'_{a}$[END] & $-6.6020 \times 10^{-7}$ & $-6.6020 \times 10^{-7}$ & $1.595 \times 10^{-26}$ & - \\
$R_{56}$[END] & $2.4233 \times 10^{-1}$ & $2.8971 \times 10^{-1}$ & $2.245 \times 10^{12}$ & - \\
max($\beta_a$) & $7.0000 \times 10^{2}$ & $2.9238 \times 10^{2}$ & 0 & M17Q7 \\
max($\beta_b$) & $7.0000 \times 10^{2}$ & $4.0781 \times 10^{2}$ & 0 & M17Q6 \\
\midrule
\multicolumn{5}{l}{\textbf{Sixth Pass}} \\
$\beta_{a}$[END] & $1.6824 \times 10^{0}$ & $1.6824 \times 10^{0}$ & $2.848 \times 10^{-18}$ & - \\
$\beta_{b}$[END] & $2.2819 \times 10^{1}$ & $2.2819 \times 10^{1}$ & $4.544 \times 10^{-18}$ & - \\
$\alpha_{a}$[END] & $2.6113 \times 10^{-5}$ & $2.6113 \times 10^{-5}$ & $5.525 \times 10^{-23}$ & - \\
$\alpha_{b}$[END] & $-2.2893 \times 10^{-4}$ & $-2.2893 \times 10^{-4}$ & $1.274 \times 10^{-23}$ & - \\
$\eta_{a}$[END] & $7.2865 \times 10^{-2}$ & $7.2865 \times 10^{-2}$ & $4.391 \times 10^{-22}$ & - \\
$\eta'_{a}$[END] & $-7.6926 \times 10^{-7}$ & $-7.6926 \times 10^{-7}$ & $7.236 \times 10^{-24}$ & - \\
$R_{56}$[END] & $2.8081 \times 10^{-1}$ & $2.8153 \times 10^{-1}$ & $5.159 \times 10^{8}$ & - \\
max($\beta_a$) & $7.0000 \times 10^{2}$ & $2.9888 \times 10^{2}$ & 0 & M19Q2 \\
max($\beta_b$) & $7.0000 \times 10^{2}$ & $2.7170 \times 10^{2}$ & 0 & MCB4B \\

\end{longtable}

\begin{longtable}{lS[table-format=2.2]S[table-format=-1.4] @{\hskip 1cm} lS[table-format=2.2]S[table-format=-1.4]}
\caption{Optimized integrated quadrupole strengths ($K_1$) for each of the six passes. The longitudinal position of the center of each magnet is given by the S-coordinate. All quadrupole strengths were limited to a range of $\pm \SI{0.6}{\per\meter\squared}$ during the optimization.}
\label{tab:WFT_Twiss_quad_strengths} \\

\toprule
\textbf{Element} & {\textbf{S-Pos. (m)}} & {\textbf{$K_1$ (\si{\per\meter\squared})}} & \textbf{Element} & {\textbf{S-Pos. (m)}} & {\textbf{$K_1$ (\si{\per\meter\squared})}} \\
\midrule
\endfirsthead

\caption{(Continued)} \\
\toprule
\textbf{Element} & {\textbf{S-Pos. (m)}} & {\textbf{$K_1$ (\si{\per\meter\squared})}} & \textbf{Element} & {\textbf{S-Pos. (m)}} & {\textbf{$K_1$ (\si{\per\meter\squared})}} \\
\midrule
\endhead

\bottomrule
\endfoot

\multicolumn{6}{c}{\textbf{First Pass (P9) Quadrupoles}} \\
\midrule
M9Q1 & 20.66 & -0.4107 & M9Q11 & 57.28 & 0.5916 \\
M9Q2 & 24.16 & 0.4979 & M9Q12 & 58.28 & 0.3052 \\
M9Q3 & 27.66 & -0.0627 & M9Q13 & 64.78 & 0.3239 \\
M9Q4 & 31.16 & -0.6000 & M9Q14 & 65.78 & -0.6000 \\
M9Q5 & 32.16 & -0.1398 & M9Q15 & 69.78 & 0.0797 \\
M9Q6 & 40.36 & 0.0112 & M9Q16 & 70.78 & 0.4543 \\
M9Q7 & 41.36 & 0.4937 & M9Q17 & 79.88 & -0.1930 \\
M9Q8 & 46.66 & 0.0444 & M9Q18 & 80.78 & 0.2105 \\
M9Q9 & 51.66 & -0.5985 & M9Q19 & 81.68 & 0.5356 \\
M9Q10 & 52.66 & 0.1025 & M9Q20 & 82.58 & -0.6000 \\
\midrule
\multicolumn{6}{c}{\textbf{Second Pass (P11) Quadrupoles}} \\
\midrule
M11Q1 & 32.11 & 0.2819 & M11Q8 & 63.28 & -0.1974 \\
M11Q2 & 36.36 & -0.3954 & M11Q9 & 67.54 & -0.2161 \\
M11Q3 & 37.36 & -0.0597 & M11Q10 & 71.80 & 0.1064 \\
M11Q4 & 42.86 & 0.4588 & M11Q11 & 77.21 & 0.1190 \\
M11Q5 & 48.50 & 0.5436 & M11Q12 & 78.21 & -0.2094 \\
M11Q6 & 57.86 & 0.6000 & M11Q13 & 81.21 & -0.0447 \\
M11Q7 & 58.86 & -0.1919 & M11Q14 & 82.21 & 0.3856 \\
\midrule
\multicolumn{6}{c}{\textbf{Third Pass (P13) Quadrupoles}} \\
\midrule
M13Q1 & 34.16 & 0.4772 & M13Q7 & 60.66 & 0.4468 \\
M13Q2 & 35.16 & -0.6000 & M13Q8 & 61.66 & -0.6000 \\
M13Q3 & 43.66 & 0.1954 & M13Q9 & 67.28 & -0.2555 \\
M13Q4 & 50.55 & 0.0150 & M13Q10 & 72.46 & 0.0224 \\
M13Q5 & 55.30 & -0.0578 & M13Q11 & 73.46 & -0.2882 \\
M13Q6 & 59.66 & 0.3134 & M13Q12 & 74.46 & 0.6000 \\
\midrule
\multicolumn{6}{c}{\textbf{Fourth Pass (P15) Quadrupoles}} \\
\midrule
M15Q1 & 30.86 & 0.1306 & M15Q6 & 62.04 & -0.0548 \\
M15Q2 & 31.66 & -0.3726 & M15Q7 & 63.04 & 0.3687 \\
M15Q3 & 42.35 & 0.3236 & M15Q8 & 68.11 & -0.0075 \\
M15Q4 & 48.36 & -0.2390 & M15Q9 & 72.70 & -0.6000 \\
M15Q5 & 54.05 & -0.4206 & M15Q10 & 76.62 & 0.3420 \\
\midrule
\multicolumn{6}{c}{\textbf{Fifth Pass (P17) Quadrupoles}} \\
\midrule
M17Q1 & 36.56 & 0.6000 & M17Q8 & 64.50 & 0.0091 \\
M17Q2 & 37.36 & -0.1334 & M17Q9 & 68.63 & -0.0419 \\
M17Q3 & 38.16 & -0.4361 & M17Q10 & 69.63 & -0.1255 \\
M17Q4 & 49.23 & 0.1614 & M17Q11 & 74.48 & -0.0613 \\
M17Q5 & 49.87 & 0.5989 & M17Q12 & 75.48 & 0.0164 \\
M17Q6 & 54.09 & -0.4792 & M17Q13 & 77.48 & 0.0232 \\
M17Q7 & 62.50 & 0.3699 & M17Q14 & 78.48 & 0.1903 \\
\midrule
\multicolumn{6}{c}{\textbf{Sixth Pass (P19) Quadrupoles}} \\
\midrule
M19Q1 & 33.36 & -0.4819 & M19Q9 & 73.75 & -0.4768 \\
M19Q2 & 33.96 & 0.5784 & M19Q10 & 74.75 & 0.4529 \\
M19Q3 & 43.86 & 0.1211 & M19Q11 & 78.75 & 0.5500 \\
M19Q4 & 44.86 & -0.1310 & M19Q12 & 79.75 & -0.5500 \\
M19Q5 & 45.86 & -0.1182 & M19Q13 & 80.75 & 0.1361 \\
M19Q6 & 61.81 & 0.6000 & M19Q14 & 86.75 & 0.5500 \\
M19Q7 & 62.81 & 0.3919 & M19Q15 & 87.75 & -0.4598 \\
M19Q8 & 63.81 & -0.5783 & & & \\

\end{longtable}
\FloatBarrier

\subsection{Matches Including $R_{56}$ Compensation}
\label{sec:AppA_2}
The following will show match solutions, including $R_{56}$ compensation. They are broken apart by input LINAC optics; either from the SFT or WFT.

\subsubsection{Strong Focusing Triplet LINAC Input}
\label{sec:AppA_2-1}

There are two matches which use the SFT as inputs: one from SFT into the Beginning match point in the FFA Cell, and another from SFT to the BDC match point in the FFA Cell. Both are shown below, and the reader is requested to take care when reading the following tables.

\textbf{Showing first the SFT matching into the Beginning FFA Cell match point:}

\begin{figure}[!ht]
    \centering
    \includegraphics[width=\textwidth]{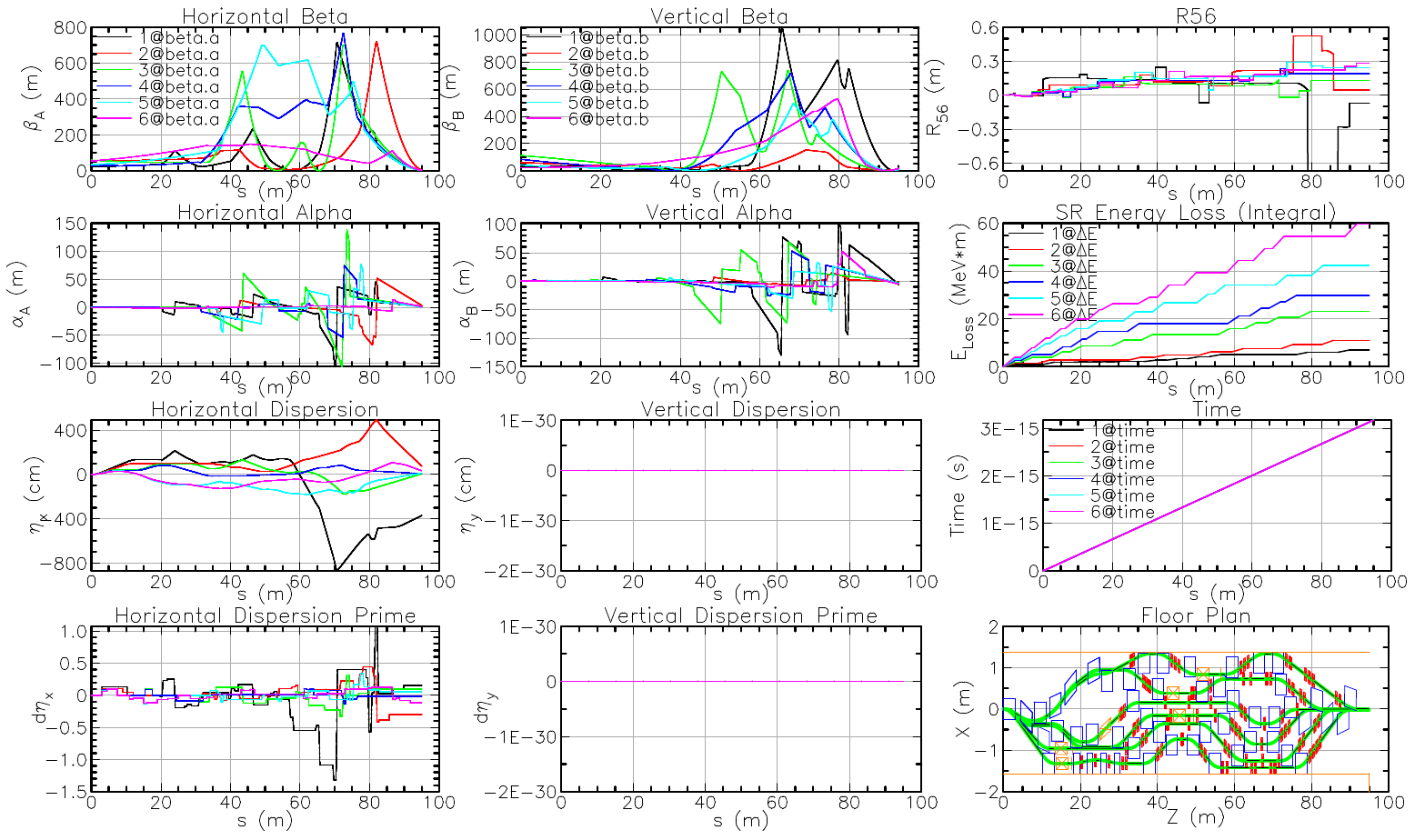}
    \caption{Matching solution from the SFT LINAC input into the Beginning FFA match point. In this case, $R_{56}$ and $\beta_{x,y}$ is always matched perfectly, while $\alpha_{x,y}$ and $\eta_{a}$ and $\eta'_{a}$ have matches of varying strength.}
    \label{fig:SFT_Beginning_R56_Twiss}
\end{figure}

\begin{longtable}{lcccc}
\caption{Complete summary of Twiss parameter optimization constraints for the matching case of the SFT LINAC input and the Beginning match point in the FFA Arc, including $R_{56}$ matching. This includes target values for optical functions at the beamline end-point and maximum value limits throughout the line for each of the six energy passes.}
\label{tab:SFT_Beginning_R56_twiss_full_summary} \\

\toprule
\textbf{Parameter} & \textbf{Target/Limit} & \textbf{Model Value} & \textbf{Merit} & \textbf{Location} \\
\midrule
\endfirsthead

\caption{(Continued)} \\
\toprule
\textbf{Parameter} & \textbf{Target/Limit} & \textbf{Model Value} & \textbf{Merit} & \textbf{Location} \\
\midrule
\endhead

\bottomrule
\multicolumn{5}{r}{\textit{Continued on next page}} \\
\endfoot

\bottomrule
\endlastfoot

\multicolumn{5}{l}{\textbf{First Pass}} \\
$\beta_{a}$[END] & $4.1573 \times 10^{0}$ & $4.1573 \times 10^{0}$ & $1.138 \times 10^{1}$ & - \\
$\beta_{b}$[END] & $6.5151 \times 10^{0}$ & $6.5151 \times 10^{0}$ & $7.290 \times 10^{0}$ & - \\
$\alpha_{a}$[END] & $3.0489 \times 10^{0}$ & $1.2360 \times 10^{0}$ & $3.286 \times 10^{6}$ & - \\
$\alpha_{b}$[END] & $-3.1896 \times 10^{0}$ & $-6.0894 \times 10^{0}$ & $8.409 \times 10^{6}$ & - \\
$\eta_{a}$[END] & $2.7200 \times 10^{-2}$ & $-3.7165 \times 10^{0}$ & $1.402 \times 10^{9}$ & - \\
$\eta'_{a}$[END] & $-1.9800 \times 10^{-2}$ & $1.5193 \times 10^{-1}$ & $2.949 \times 10^{6}$ & - \\
$R_{56}$[END] & $-7.0886 \times 10^{-2}$ & $-7.0886 \times 10^{-2}$ & $1.308 \times 10^{-13}$ & - \\
max($\beta_a$) & $7.0000 \times 10^{2}$ & $7.1127 \times 10^{2}$ & $1.269 \times 10^{2}$ & M9Q16 \\
max($\beta_b$) & $7.0000 \times 10^{2}$ & $1.0447 \times 10^{3}$ & $1.188 \times 10^{5}$ & M9Q14 \\
\midrule
\multicolumn{5}{l}{\textbf{Second Pass}} \\
$\beta_{a}$[END] & $2.9506 \times 10^{0}$ & $2.9506 \times 10^{0}$ & $1.596 \times 10^{1}$ & - \\
$\beta_{b}$[END] & $6.4768 \times 10^{0}$ & $6.4768 \times 10^{0}$ & $5.170 \times 10^{0}$ & - \\
$\alpha_{a}$[END] & $1.8215 \times 10^{0}$ & $3.1861 \times 10^{0}$ & $1.862 \times 10^{6}$ & - \\
$\alpha_{b}$[END] & $-3.0366 \times 10^{0}$ & $-7.1030 \times 10^{-1}$ & $5.412 \times 10^{6}$ & - \\
$\eta_{a}$[END] & $4.5700 \times 10^{-2}$ & $7.4439 \times 10^{-1}$ & $4.882 \times 10^{7}$ & - \\
$\eta'_{a}$[END] & $-2.6400 \times 10^{-2}$ & $-3.0046 \times 10^{-1}$ & $7.511 \times 10^{6}$ & - \\
$R_{56}$[END] & $4.4694 \times 10^{-2}$ & $4.4694 \times 10^{-2}$ & $1.531 \times 10^{-13}$ & - \\
max($\beta_a$) & $7.0000 \times 10^{2}$ & $7.1015 \times 10^{2}$ & $1.031 \times 10^{2}$ & M11Q14 \\
max($\beta_b$) & $7.0000 \times 10^{2}$ & $1.5171 \times 10^{2}$ & 0 & M11Q10 \\
\midrule
\multicolumn{5}{l}{\textbf{Third Pass}} \\
$\beta_{a}$[END] & $2.7180 \times 10^{0}$ & $2.7180 \times 10^{0}$ & $1.330 \times 10^{1}$ & - \\
$\beta_{b}$[END] & $6.9948 \times 10^{0}$ & $6.9948 \times 10^{0}$ & $1.427 \times 10^{1}$ & - \\
$\alpha_{a}$[END] & $1.5388 \times 10^{0}$ & $1.1993 \times 10^{0}$ & $1.152 \times 10^{5}$ & - \\
$\alpha_{b}$[END] & $-3.2063 \times 10^{0}$ & $-2.2045 \times 10^{0}$ & $1.004 \times 10^{6}$ & - \\
$\eta_{a}$[END] & $6.0800 \times 10^{-2}$ & $4.8292 \times 10^{-2}$ & $1.564 \times 10^{4}$ & - \\
$\eta'_{a}$[END] & $-3.0900 \times 10^{-2}$ & $9.3643 \times 10^{-2}$ & $1.551 \times 10^{6}$ & - \\
$R_{56}$[END] & $1.2809 \times 10^{-1}$ & $1.2809 \times 10^{-1}$ & $1.926 \times 10^{-17}$ & - \\
max($\beta_a$) & $7.0000 \times 10^{2}$ & $7.0047 \times 10^{2}$ & $2.218 \times 10^{-1}$ & M13Q10 \\
max($\beta_b$) & $7.0000 \times 10^{2}$ & $7.2508 \times 10^{2}$ & $6.292 \times 10^{2}$ & M13Q4 \\
\midrule
\multicolumn{5}{l}{\textbf{Fourth Pass}} \\
$\beta_{a}$[END] & $2.6023 \times 10^{0}$ & $2.6023 \times 10^{0}$ & $1.485 \times 10^{1}$ & - \\
$\beta_{b}$[END] & $8.0350 \times 10^{0}$ & $8.0350 \times 10^{0}$ & $2.944 \times 10^{0}$ & - \\
$\alpha_{a}$[END] & $1.3995 \times 10^{0}$ & $9.8180 \times 10^{-1}$ & $1.745 \times 10^{5}$ & - \\
$\alpha_{b}$[END] & $-3.6359 \times 10^{0}$ & $-3.5727 \times 10^{0}$ & $4.000 \times 10^{3}$ & - \\
$\eta_{a}$[END] & $7.3500 \times 10^{-2}$ & $7.4246 \times 10^{-2}$ & $5.566 \times 10^{1}$ & - \\
$\eta'_{a}$[END] & $-3.3600 \times 10^{-2}$ & $-1.6806 \times 10^{-2}$ & $2.820 \times 10^{4}$ & - \\
$R_{56}$[END] & $1.9190 \times 10^{-1}$ & $1.9190 \times 10^{-1}$ & $1.926 \times 10^{-17}$ & - \\
max($\beta_a$) & $7.0000 \times 10^{2}$ & $7.5858 \times 10^{2}$ & $3.432 \times 10^{3}$ & D1508\#2 \\
max($\beta_b$) & $7.0000 \times 10^{2}$ & $7.1406 \times 10^{2}$ & $1.978 \times 10^{2}$ & D1507\#2 \\
\midrule
\multicolumn{5}{l}{\textbf{Fifth Pass}} \\
$\beta_{a}$[END] & $2.5213 \times 10^{0}$ & $2.5213 \times 10^{0}$ & $1.257 \times 10^{1}$ & - \\
$\beta_{b}$[END] & $1.0132 \times 10^{1}$ & $1.0132 \times 10^{1}$ & $5.861 \times 10^{0}$ & - \\
$\alpha_{a}$[END] & $1.3111 \times 10^{0}$ & $8.9904 \times 10^{-1}$ & $1.698 \times 10^{5}$ & - \\
$\alpha_{b}$[END] & $-4.5492 \times 10^{0}$ & $-4.2207 \times 10^{0}$ & $1.079 \times 10^{5}$ & - \\
$\eta_{a}$[END] & $8.4100 \times 10^{-2}$ & $1.0620 \times 10^{-1}$ & $4.883 \times 10^{4}$ & - \\
$\eta'_{a}$[END] & $-3.4900 \times 10^{-2}$ & $5.0860 \times 10^{-2}$ & $7.355 \times 10^{5}$ & - \\
$R_{56}$[END] & $2.4233 \times 10^{-1}$ & $2.4233 \times 10^{-1}$ & $2.496 \times 10^{-16}$ & - \\
max($\beta_a$) & $7.0000 \times 10^{2}$ & $7.0030 \times 10^{2}$ & $8.964 \times 10^{-2}$ & M17Q4 \\
max($\beta_b$) & $7.0000 \times 10^{2}$ & $4.8923 \times 10^{2}$ & 0 & M17Q9 \\
\midrule
\multicolumn{5}{l}{\textbf{Sixth Pass}} \\
$\beta_{a}$[END] & $2.4552 \times 10^{0}$ & $2.4552 \times 10^{0}$ & $4.437 \times 10^{-19}$ & - \\
$\beta_{b}$[END] & $1.6840 \times 10^{1}$ & $1.6840 \times 10^{1}$ & $2.840 \times 10^{-17}$ & - \\
$\alpha_{a}$[END] & $1.2471 \times 10^{0}$ & $1.3639 \times 10^{0}$ & $1.365 \times 10^{4}$ & - \\
$\alpha_{b}$[END] & $-7.5245 \times 10^{0}$ & $-5.3041 \times 10^{0}$ & $4.930 \times 10^{6}$ & - \\
$\eta_{a}$[END] & $9.3200 \times 10^{-2}$ & $3.3108 \times 10^{-1}$ & $5.659 \times 10^{6}$ & - \\
$\eta'_{a}$[END] & $-3.5000 \times 10^{-2}$ & $-1.1699 \times 10^{-1}$ & $6.723 \times 10^{5}$ & - \\
$R_{56}$[END] & $2.8081 \times 10^{-1}$ & $2.8081 \times 10^{-1}$ & $5.208 \times 10^{-16}$ & - \\
max($\beta_a$) & $7.0000 \times 10^{2}$ & $1.4796 \times 10^{2}$ & 0 & M19Q5 \\
max($\beta_b$) & $7.0000 \times 10^{2}$ & $5.2637 \times 10^{2}$ & 0 & D1912\#11 \\

\end{longtable}

\begin{longtable}{lS[table-format=2.2]S[table-format=-1.4] @{\hskip 1cm} lS[table-format=2.2]S[table-format=-1.4]}
\caption{Optimized integrated quadrupole strengths ($K_1$) for each of the six passes. The longitudinal position of the center of each magnet is given by the S-coordinate. All quadrupole strengths were limited to a range of $\pm \SI{0.6}{\per\meter\squared}$ during the optimization.}
\label{tab:SFT_Beginning_R56_Twiss_quad_strengths} \\

\toprule
\textbf{Element} & {\textbf{S-Pos. (m)}} & {\textbf{$K_1$ (\si{\per\meter\squared})}} & \textbf{Element} & {\textbf{S-Pos. (m)}} & {\textbf{$K_1$ (\si{\per\meter\squared})}} \\
\midrule
\endfirsthead

\caption{(Continued)} \\
\toprule
\textbf{Element} & {\textbf{S-Pos. (m)}} & {\textbf{$K_1$ (\si{\per\meter\squared})}} & \textbf{Element} & {\textbf{S-Pos. (m)}} & {\textbf{$K_1$ (\si{\per\meter\squared})}} \\
\midrule
\endhead

\bottomrule
\endfoot

\multicolumn{6}{c}{\textbf{First Pass (P9) Quadrupoles}} \\
\midrule
M9Q1 & 20.66 & -0.5323 & M9Q11 & 57.28 & 0.4988 \\
M9Q2 & 24.16 & 0.5935 & M9Q12 & 58.28 & 0.4633 \\
M9Q3 & 27.66 & -0.0619 & M9Q13 & 64.78 & 0.0642 \\
M9Q4 & 31.16 & -0.3722 & M9Q14 & 65.78 & -0.5706 \\
M9Q5 & 32.16 & -0.2316 & M9Q15 & 69.78 & -0.1051 \\
M9Q6 & 40.36 & -0.4595 & M9Q16 & 70.78 & 0.5469 \\
M9Q7 & 41.36 & 0.1046 & M9Q17 & 79.88 & -0.4392 \\
M9Q8 & 46.66 & 0.5083 & M9Q18 & 80.78 & 0.2952 \\
M9Q9 & 51.66 & -0.1289 & M9Q19 & 81.68 & 0.5041 \\
M9Q10 & 52.66 & -0.0512 & M9Q20 & 82.58 & -0.5999 \\
\midrule
\multicolumn{6}{c}{\textbf{Second Pass (P11) Quadrupoles}} \\
\midrule
M11Q1 & 32.11 & -0.0790 & M11Q8 & 63.28 & -0.0599 \\
M11Q2 & 36.36 & -0.2096 & M11Q9 & 67.54 & 0.0295 \\
M11Q3 & 37.36 & 0.2669 & M11Q10 & 71.80 & -0.1801 \\
M11Q4 & 42.86 & 0.3198 & M11Q11 & 77.21 & -0.2314 \\
M11Q5 & 48.50 & -0.6000 & M11Q12 & 78.21 & -0.1088 \\
M11Q6 & 57.86 & -0.0739 & M11Q13 & 81.21 & 0.0944 \\
M11Q7 & 58.86 & -0.0665 & M11Q14 & 82.21 & 0.4155 \\
\midrule
\multicolumn{6}{c}{\textbf{Third Pass (P13) Quadrupoles}} \\
\midrule
M13Q1 & 34.16 & -0.5596 & M13Q7 & 60.66 & 0.3326 \\
M13Q2 & 35.16 & -0.1211 & M13Q8 & 61.66 & 0.6000 \\
M13Q3 & 43.66 & 0.5268 & M13Q9 & 67.28 & -0.5479 \\
M13Q4 & 50.55 & -0.3757 & M13Q10 & 72.46 & 0.4494 \\
M13Q5 & 55.30 & -0.1940 & M13Q11 & 73.46 & 0.5829 \\
M13Q6 & 59.66 & 0.2639 & M13Q12 & 74.46 & -0.6000 \\
\midrule
\multicolumn{6}{c}{\textbf{Fourth Pass (P15) Quadrupoles}} \\
\midrule
M15Q1 & 30.86 & 0.1741 & M15Q6 & 62.04 & 0.0987 \\
M15Q2 & 31.66 & -0.5962 & M15Q7 & 63.04 & -0.0284 \\
M15Q3 & 42.35 & 0.2561 & M15Q8 & 68.11 & -0.3036 \\
M15Q4 & 48.36 & 0.0415 & M15Q9 & 72.70 & 0.4761 \\
M15Q5 & 54.05 & -0.1118 & M15Q10 & 76.62 & -0.2987 \\
\midrule
\multicolumn{6}{c}{\textbf{Fifth Pass (P17) Quadrupoles}} \\
\midrule
M17Q1 & 36.56 & -0.1813 & M17Q8 & 64.50 & -0.0004 \\
M17Q2 & 37.36 & -0.6000 & M17Q9 & 68.63 & -0.2730 \\
M17Q3 & 38.16 & 0.4435 & M17Q10 & 69.63 & -0.0258 \\
M17Q4 & 49.23 & 0.1266 & M17Q11 & 74.48 & -0.1663 \\
M17Q5 & 49.87 & 0.0484 & M17Q12 & 75.48 & 0.4173 \\
M17Q6 & 54.09 & -0.0690 & M17Q13 & 77.48 & 0.3817 \\
M17Q7 & 62.50 & 0.1435 & M17Q14 & 78.48 & -0.6000 \\
\midrule
\multicolumn{6}{c}{\textbf{Sixth Pass (P19) Quadrupoles}} \\
\midrule
M19Q1 & 33.36 & 0.0703 & M19Q9 & 73.75 & -0.0569 \\
M19Q2 & 33.96 & -0.0348 & M19Q10 & 74.75 & 0.0583 \\
M19Q3 & 43.86 & -0.0335 & M19Q11 & 78.75 & -0.0235 \\
M19Q4 & 44.86 & 0.0024 & M19Q12 & 79.75 & -0.1932 \\
M19Q5 & 45.86 & 0.0542 & M19Q13 & 80.75 & -0.1504 \\
M19Q6 & 61.81 & 0.0730 & M19Q14 & 86.75 & 0.5230 \\
M19Q7 & 62.81 & 0.0047 & M19Q15 & 87.75 & -0.0678 \\
M19Q8 & 63.81 & -0.0322 & & & \\

\end{longtable}

\FloatBarrier

\textbf{Showing next the SFT matching into the BDC FFA Cell match point:}

\begin{figure}[!ht]
    \centering
    \includegraphics[width=\textwidth]{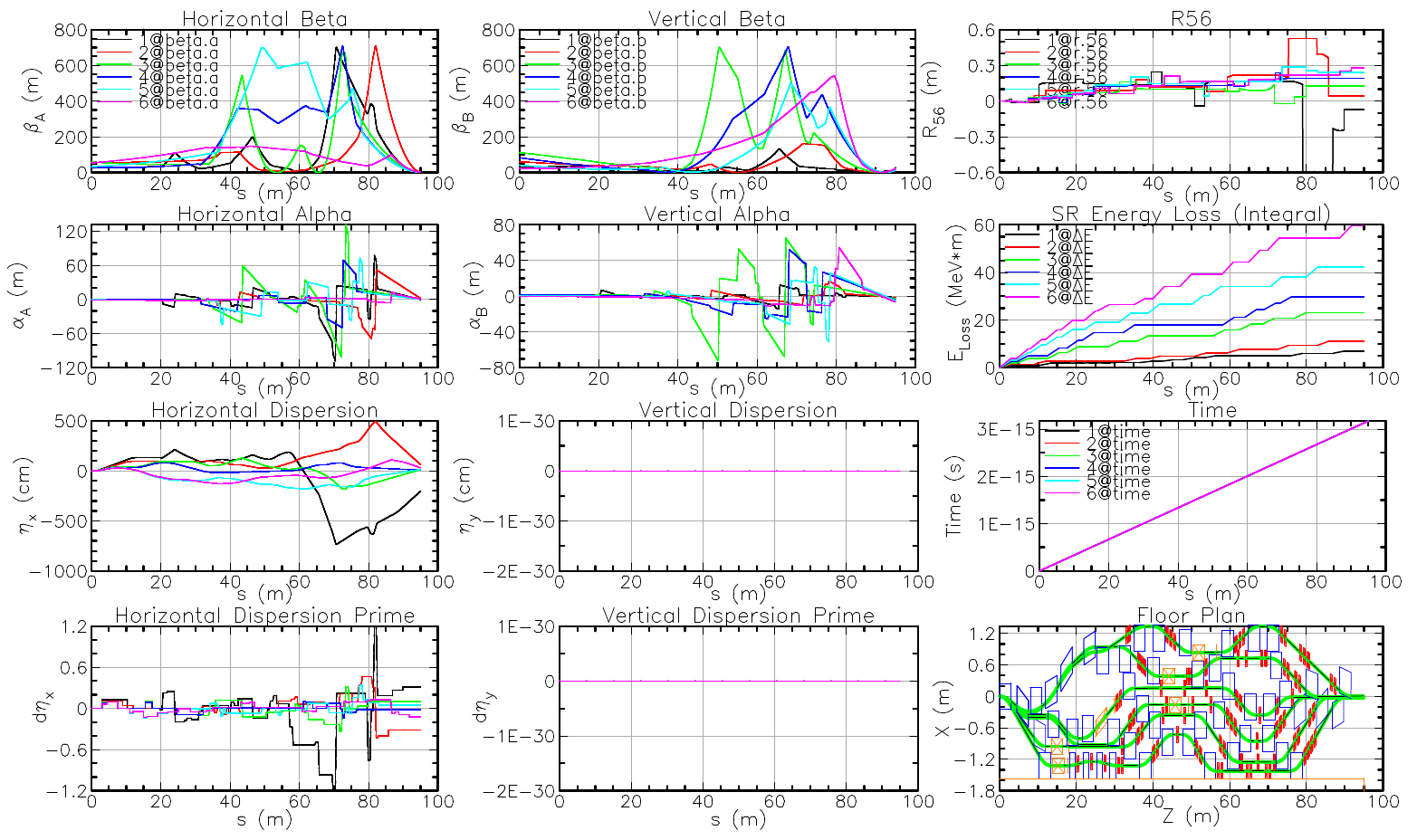}
    \caption{Matching solution from the SFT LINAC input into the BDC FFA match point. In this case, $R_{56}$ and $\beta_{x,y}$ is always matched perfectly, while $\alpha_{x,y}$ and $\eta_{a}$ and $\eta'_{a}$ have matches of varying strength.}
    \label{fig:SFT_R56_Twiss}
\end{figure}

\begin{longtable}{lcccc}
\caption{Complete summary of Twiss parameter optimization constraints for the matching case of the SFT LINAC input and the BDC match point in the FFA Arc, including $R_{56}$ matching. This includes target values for optical functions at the beamline end-point and maximum value limits throughout the line for each of the six energy passes.}
\label{tab:SFT_R56_twiss_full_summary} \\

\toprule
\textbf{Parameter} & \textbf{Target/Limit} & \textbf{Model Value} & \textbf{Merit} & \textbf{Location} \\
\midrule
\endfirsthead

\caption{(Continued)} \\
\toprule
\textbf{Parameter} & \textbf{Target/Limit} & \textbf{Model Value} & \textbf{Merit} & \textbf{Location} \\
\midrule
\endhead

\bottomrule
\multicolumn{5}{r}{\textit{Continued on next page}} \\
\endfoot

\bottomrule
\endlastfoot

\multicolumn{5}{l}{\textbf{First Pass}} \\
$\beta_{a}$[END] & $6.6586 \times 10^{-1}$ & $6.6586 \times 10^{-1}$ & $2.381 \times 10^{-18}$ & - \\
$\beta_{b}$[END] & $1.2039 \times 10^{1}$ & $1.2039 \times 10^{1}$ & $2.788 \times 10^{-16}$ & - \\
$\alpha_{a}$[END] & $4.3124 \times 10^{-5}$ & $-4.2697 \times 10^{-2}$ & $1.827 \times 10^{3}$ & - \\
$\alpha_{b}$[END] & $-1.1741 \times 10^{-4}$ & $-1.1382 \times 10^{0}$ & $1.295 \times 10^{6}$ & - \\
$\eta_{a}$[END] & $-7.8352 \times 10^{-3}$ & $-2.0235 \times 10^{0}$ & $4.063 \times 10^{8}$ & - \\
$\eta'_{a}$[END] & $4.3951 \times 10^{-8}$ & $3.1545 \times 10^{-1}$ & $9.951 \times 10^{6}$ & - \\
$R_{56}$[END] & $-7.0886 \times 10^{-2}$ & $-7.0886 \times 10^{-2}$ & $5.268 \times 10^{-14}$ & - \\
max($\beta_a$) & $7.0000 \times 10^{2}$ & $7.0000 \times 10^{2}$ & 0 & M9Q16 \\
max($\beta_b$) & $7.0000 \times 10^{2}$ & $1.3102 \times 10^{2}$ & 0 & M9Q14 \\
\midrule
\multicolumn{5}{l}{\textbf{Second Pass}} \\
$\beta_{a}$[END] & $1.1641 \times 10^{0}$ & $1.1641 \times 10^{0}$ & $8.332 \times 10^{-20}$ & - \\
$\beta_{b}$[END] & $1.0662 \times 10^{1}$ & $1.0662 \times 10^{1}$ & $1.136 \times 10^{-16}$ & - \\
$\alpha_{a}$[END] & $2.6829 \times 10^{-5}$ & $1.8910 \times 10^{0}$ & $3.576 \times 10^{6}$ & - \\
$\alpha_{b}$[END] & $-1.0407 \times 10^{-4}$ & $-1.6655 \times 10^{0}$ & $2.773 \times 10^{6}$ & - \\
$\eta_{a}$[END] & $1.4456 \times 10^{-2}$ & $6.3057 \times 10^{-1}$ & $3.796 \times 10^{7}$ & - \\
$\eta'_{a}$[END] & $-1.8690 \times 10^{-7}$ & $-3.1314 \times 10^{-1}$ & $9.806 \times 10^{6}$ & - \\
$R_{56}$[END] & $4.4694 \times 10^{-2}$ & $4.4694 \times 10^{-2}$ & $1.628 \times 10^{-13}$ & - \\
max($\beta_a$) & $7.0000 \times 10^{2}$ & $7.0000 \times 10^{2}$ & 0 & M11Q14 \\
max($\beta_b$) & $7.0000 \times 10^{2}$ & $1.6213 \times 10^{2}$ & 0 & M11Q10 \\
\midrule
\multicolumn{5}{l}{\textbf{Third Pass}} \\
$\beta_{a}$[END] & $1.3998 \times 10^{0}$ & $1.3998 \times 10^{0}$ & $1.804 \times 10^{-15}$ & - \\
$\beta_{b}$[END] & $1.0669 \times 10^{1}$ & $1.0669 \times 10^{1}$ & $2.300 \times 10^{-17}$ & - \\
$\alpha_{a}$[END] & $2.4658 \times 10^{-5}$ & $5.5600 \times 10^{-1}$ & $3.091 \times 10^{5}$ & - \\
$\alpha_{b}$[END] & $-1.0462 \times 10^{-4}$ & $-2.6784 \times 10^{0}$ & $7.173 \times 10^{6}$ & - \\
$\eta_{a}$[END] & $3.2967 \times 10^{-2}$ & $1.2496 \times 10^{-1}$ & $8.463 \times 10^{5}$ & - \\
$\eta'_{a}$[END] & $-3.7558 \times 10^{-7}$ & $9.9008 \times 10^{-2}$ & $9.803 \times 10^{5}$ & - \\
$R_{56}$[END] & $1.2809 \times 10^{-1}$ & $1.2809 \times 10^{-1}$ & $6.240 \times 10^{-17}$ & - \\
max($\beta_a$) & $7.0000 \times 10^{2}$ & $6.7114 \times 10^{2}$ & 0 & D1308\#5 \\
max($\beta_b$) & $7.0000 \times 10^{2}$ & $7.0000 \times 10^{2}$ & 0 & M13Q4 \\
\midrule
\multicolumn{5}{l}{\textbf{Fourth Pass}} \\
$\beta_{a}$[END] & $1.5386 \times 10^{0}$ & $1.5386 \times 10^{0}$ & $4.378 \times 10^{-17}$ & - \\
$\beta_{b}$[END] & $1.1607 \times 10^{1}$ & $1.1607 \times 10^{1}$ & $2.908 \times 10^{-16}$ & - \\
$\alpha_{a}$[END] & $2.4713 \times 10^{-5}$ & $3.8831 \times 10^{-1}$ & $1.508 \times 10^{5}$ & - \\
$\alpha_{b}$[END] & $-1.1455 \times 10^{-4}$ & $-4.3379 \times 10^{0}$ & $1.882 \times 10^{7}$ & - \\
$\eta_{a}$[END] & $4.8495 \times 10^{-2}$ & $7.4420 \times 10^{-2}$ & $6.721 \times 10^{4}$ & - \\
$\eta'_{a}$[END] & $-5.3096 \times 10^{-7}$ & $-1.7070 \times 10^{-2}$ & $2.914 \times 10^{4}$ & - \\
$R_{56}$[END] & $1.9190 \times 10^{-1}$ & $1.9190 \times 10^{-1}$ & $1.233 \times 10^{-17}$ & - \\
max($\beta_a$) & $7.0000 \times 10^{2}$ & $7.0000 \times 10^{2}$ & 0 & D1508\#2 \\
max($\beta_b$) & $7.0000 \times 10^{2}$ & $7.0000 \times 10^{2}$ & 0 & D1507\#2 \\
\midrule
\multicolumn{5}{l}{\textbf{Fifth Pass}} \\
$\beta_{a}$[END] & $1.6261 \times 10^{0}$ & $1.6261 \times 10^{0}$ & $2.935 \times 10^{-17}$ & - \\
$\beta_{b}$[END] & $1.4063 \times 10^{1}$ & $1.4063 \times 10^{1}$ & $4.089 \times 10^{-17}$ & - \\
$\alpha_{a}$[END] & $2.5401 \times 10^{-5}$ & $4.2785 \times 10^{-1}$ & $1.830 \times 10^{5}$ & - \\
$\alpha_{b}$[END] & $-1.3987 \times 10^{-4}$ & $-5.0845 \times 10^{0}$ & $2.585 \times 10^{7}$ & - \\
$\eta_{a}$[END] & $6.1641 \times 10^{-2}$ & $1.4104 \times 10^{-1}$ & $6.304 \times 10^{5}$ & - \\
$\eta'_{a}$[END] & $-6.6020 \times 10^{-7}$ & $5.1050 \times 10^{-2}$ & $2.606 \times 10^{5}$ & - \\
$R_{56}$[END] & $2.4233 \times 10^{-1}$ & $2.4233 \times 10^{-1}$ & $3.775 \times 10^{-17}$ & - \\
max($\beta_a$) & $7.0000 \times 10^{2}$ & $7.0000 \times 10^{2}$ & 0 & M17Q4 \\
max($\beta_b$) & $7.0000 \times 10^{2}$ & $4.9647 \times 10^{2}$ & 0 & M17Q9 \\
\midrule
\multicolumn{5}{l}{\textbf{Sixth Pass}} \\
$\beta_{a}$[END] & $1.6824 \times 10^{0}$ & $1.6824 \times 10^{0}$ & $1.972 \times 10^{-21}$ & - \\
$\beta_{b}$[END] & $2.2819 \times 10^{1}$ & $2.2819 \times 10^{1}$ & $2.133 \times 10^{-17}$ & - \\
$\alpha_{a}$[END] & $2.6113 \times 10^{-5}$ & $9.5193 \times 10^{-1}$ & $9.061 \times 10^{5}$ & - \\
$\alpha_{b}$[END] & $-2.2893 \times 10^{-4}$ & $-6.5168 \times 10^{0}$ & $4.247 \times 10^{7}$ & - \\
$\eta_{a}$[END] & $7.2865 \times 10^{-2}$ & $3.1295 \times 10^{-1}$ & $5.764 \times 10^{6}$ & - \\
$\eta'_{a}$[END] & $-7.6926 \times 10^{-7}$ & $-1.2359 \times 10^{-1}$ & $1.528 \times 10^{6}$ & - \\
$R_{56}$[END] & $2.8081 \times 10^{-1}$ & $2.8081 \times 10^{-1}$ & $6.933 \times 10^{-16}$ & - \\
max($\beta_a$) & $7.0000 \times 10^{2}$ & $1.4796 \times 10^{2}$ & 0 & M19Q5 \\
max($\beta_b$) & $7.0000 \times 10^{2}$ & $5.4139 \times 10^{2}$ & 0 & D1912\#11 \\

\end{longtable}

\begin{longtable}{lS[table-format=2.2]S[table-format=-1.4] @{\hskip 1cm} lS[table-format=2.2]S[table-format=-1.4]}
\caption{Optimized integrated quadrupole strengths ($K_1$) for each of the six passes. The longitudinal position of the center of each magnet is given by the S-coordinate. All quadrupole strengths were limited to a range of $\pm \SI{0.6}{\per\meter\squared}$ during the optimization.}
\label{tab:SFT_R56_Twiss_quad_strengths} \\

\toprule
\textbf{Element} & {\textbf{S-Pos. (m)}} & {\textbf{$K_1$ (\si{\per\meter\squared})}} & \textbf{Element} & {\textbf{S-Pos. (m)}} & {\textbf{$K_1$ (\si{\per\meter\squared})}} \\
\midrule
\endfirsthead

\caption{(Continued)} \\
\toprule
\textbf{Element} & {\textbf{S-Pos. (m)}} & {\textbf{$K_1$ (\si{\per\meter\squared})}} & \textbf{Element} & {\textbf{S-Pos. (m)}} & {\textbf{$K_1$ (\si{\per\meter\squared})}} \\
\midrule
\endhead

\bottomrule
\endfoot

\multicolumn{6}{c}{\textbf{First Pass (P9) Quadrupoles}} \\
\midrule
M9Q1 & 20.66 & -0.5323 & M9Q11 & 57.28 & 0.4988 \\
M9Q2 & 24.16 & 0.5935 & M9Q12 & 58.28 & 0.4633 \\
M9Q3 & 27.66 & -0.0619 & M9Q13 & 64.78 & 0.0642 \\
M9Q4 & 31.16 & -0.3722 & M9Q14 & 65.78 & -0.5706 \\
M9Q5 & 32.16 & -0.2316 & M9Q15 & 69.78 & -0.1051 \\
M9Q6 & 40.36 & -0.4237 & M9Q16 & 70.78 & 0.5469 \\
M9Q7 & 41.36 & 0.1392 & M9Q17 & 79.88 & -0.4223 \\
M9Q8 & 46.66 & 0.4987 & M9Q18 & 80.78 & 0.3262 \\
M9Q9 & 51.66 & -0.3068 & M9Q19 & 81.68 & 0.5520 \\
M9Q10 & 52.66 & -0.1841 & M9Q20 & 82.58 & -0.5336 \\
\midrule
\multicolumn{6}{c}{\textbf{Second Pass (P11) Quadrupoles}} \\
\midrule
M11Q1 & 32.11 & -0.0790 & M11Q8 & 63.28 & -0.0599 \\
M11Q2 & 36.36 & -0.2096 & M11Q9 & 67.54 & 0.0295 \\
M11Q3 & 37.36 & 0.2669 & M11Q10 & 71.80 & -0.1801 \\
M11Q4 & 42.86 & 0.3198 & M11Q11 & 77.21 & -0.2314 \\
M11Q5 & 48.50 & -0.6000 & M11Q12 & 78.21 & -0.1088 \\
M11Q6 & 57.86 & -0.0899 & M11Q13 & 81.21 & 0.0944 \\
M11Q7 & 58.86 & -0.0918 & M11Q14 & 82.21 & 0.4208 \\
\midrule
\multicolumn{6}{c}{\textbf{Third Pass (P13) Quadrupoles}} \\
\midrule
M13Q1 & 34.16 & -0.5537 & M13Q7 & 60.66 & 0.3299 \\
M13Q2 & 35.16 & -0.1167 & M13Q8 & 61.66 & 0.6000 \\
M13Q3 & 43.66 & 0.5254 & M13Q9 & 67.28 & -0.5533 \\
M13Q4 & 50.55 & -0.3758 & M13Q10 & 72.46 & 0.4333 \\
M13Q5 & 55.30 & -0.1960 & M13Q11 & 73.46 & 0.5880 \\
M13Q6 & 59.66 & 0.2637 & M13Q12 & 74.46 & -0.5772 \\
\midrule
\multicolumn{6}{c}{\textbf{Fourth Pass (P15) Quadrupoles}} \\
\midrule
M15Q1 & 30.86 & 0.1741 & M15Q6 & 62.04 & 0.1009 \\
M15Q2 & 31.66 & -0.6000 & M15Q7 & 63.04 & -0.0274 \\
M15Q3 & 42.35 & 0.1756 & M15Q8 & 68.11 & -0.3062 \\
M15Q4 & 48.36 & 0.0529 & M15Q9 & 72.70 & 0.4771 \\
M15Q5 & 54.05 & -0.1244 & M15Q10 & 76.62 & -0.2979 \\
\midrule
\multicolumn{6}{c}{\textbf{Fifth Pass (P17) Quadrupoles}} \\
\midrule
M17Q1 & 36.56 & -0.1813 & M17Q8 & 64.50 & 0.0018 \\
M17Q2 & 37.36 & -0.6000 & M17Q9 & 68.63 & -0.2765 \\
M17Q3 & 38.16 & 0.4435 & M17Q10 & 69.63 & -0.0061 \\
M17Q4 & 49.23 & 0.1266 & M17Q11 & 74.48 & -0.1674 \\
M17Q5 & 49.87 & 0.0484 & M17Q12 & 75.48 & 0.4178 \\
M17Q6 & 54.09 & -0.0690 & M17Q13 & 77.48 & 0.3821 \\
M17Q7 & 62.50 & 0.1478 & M17Q14 & 78.48 & -0.5999 \\
\midrule
\multicolumn{6}{c}{\textbf{Sixth Pass (P19) Quadrupoles}} \\
\midrule
M19Q1 & 33.36 & 0.0703 & M19Q9 & 73.75 & -0.0585 \\
M19Q2 & 33.96 & -0.0348 & M19Q10 & 74.75 & 0.0570 \\
M19Q3 & 43.86 & -0.0335 & M19Q11 & 78.75 & -0.0251 \\
M19Q4 & 44.86 & 0.0024 & M19Q12 & 79.75 & -0.1948 \\
M19Q5 & 45.86 & 0.0542 & M19Q13 & 80.75 & -0.1513 \\
M19Q6 & 61.81 & 0.0754 & M19Q14 & 86.75 & 0.5286 \\
M19Q7 & 62.81 & 0.0064 & M19Q15 & 87.75 & -0.0635 \\
M19Q8 & 63.81 & -0.0312 & & & \\

\end{longtable}
\FloatBarrier

\subsubsection{Weakly Focusing Triplet LINAC Input}
\label{sec:AppA_2-2}

\begin{figure}[!ht]
    \centering
    \includegraphics[width=\textwidth]{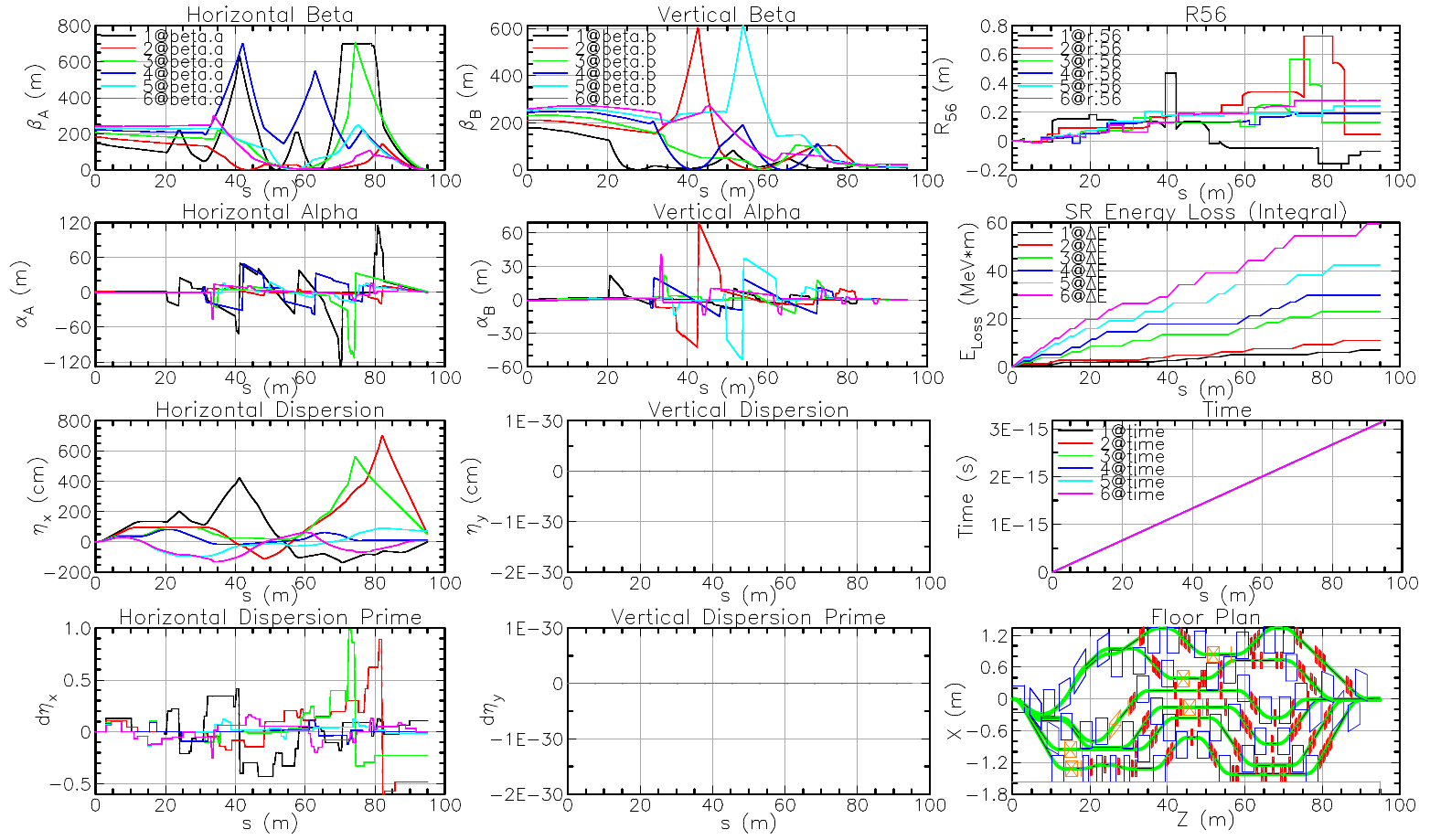}
    \caption{Matching solution from the WFT LINAC input into the BDC FFA match point. In this case, $R_{56}$ and $\beta_{x,y}$ is always matched perfectly, $\alpha_{x,y}$ usually matched very well or perfectly, and $\eta_{a}$ and $\eta'_{a}$ are matched adequately.}
    \label{fig:WFT_match_with_r56_appendix}
\end{figure}

\begin{longtable}{lcccc}
\caption{Complete summary of Twiss parameter optimization constraints for the matching case of the WFT LINAC input and the BDC match point in the FFA Arc, including $R_{56}$ matching. This includes target values for optical functions at the beamline end-point and maximum value limits throughout the line for each of the six energy passes.}
\label{tab:WFT_R56_twiss_full_summary} \\

\toprule
\textbf{Parameter} & \textbf{Target/Limit} & \textbf{Model Value} & \textbf{Merit} & \textbf{Location} \\
\midrule
\endfirsthead

\caption{(Continued)} \\
\toprule
\textbf{Parameter} & \textbf{Target/Limit} & \textbf{Model Value} & \textbf{Merit} & \textbf{Location} \\
\midrule
\endhead

\bottomrule
\multicolumn{5}{r}{\textit{Continued on next page}} \\
\endfoot

\bottomrule
\endlastfoot

\multicolumn{5}{l}{\textbf{First Pass}} \\
$\beta_{a}$[END]  & $6.6586 \times 10^{-1}$ & $6.6586 \times 10^{-1}$ & $4.437 \times 10^{-21}$ & - \\
$\beta_{b}$[END]  & $1.2039 \times 10^{1}$  & $1.2039 \times 10^{1}$  & $1.546 \times 10^{-18}$ & - \\
$\alpha_{a}$[END] & $4.3124 \times 10^{-5}$ & $4.3124 \times 10^{-5}$ & $3.585 \times 10^{-22}$ & - \\
$\alpha_{b}$[END] & $-1.1741 \times 10^{-4}$ & $-1.1741 \times 10^{-4}$ & $4.641 \times 10^{-23}$ & - \\
$\eta_{a}$[END]   & $-7.8352 \times 10^{-3}$ & $-2.7903 \times 10^{-3}$ & $2.545 \times 10^{3}$ & - \\
$\eta'_{a}$[END]  & $4.3951 \times 10^{-8}$  & $1.0875 \times 10^{-1}$  & $1.183 \times 10^{6}$ & - \\
$R_{56}$[END]     & $-7.0886 \times 10^{-2}$ & $-7.0886 \times 10^{-2}$ & $1.733 \times 10^{-18}$ & - \\
max($\beta_a$) & $7.0000 \times 10^{2}$ & $7.0000 \times 10^{2}$ & $1.837 \times 10^{-21}$ & M9B11 \\
max($\beta_b$) & $7.0000 \times 10^{2}$ & $1.7815 \times 10^{2}$ & 0 & BEGINNING \\
\midrule
\multicolumn{5}{l}{\textbf{Second Pass}} \\
$\beta_{a}$[END]  & $1.1641 \times 10^{0}$ & $1.1641 \times 10^{0}$ & $1.008 \times 10^{-17}$ & - \\
$\beta_{b}$[END]  & $1.0662 \times 10^{1}$ & $1.0662 \times 10^{1}$ & $2.300 \times 10^{-17}$ & - \\
$\alpha_{a}$[END] & $2.6829 \times 10^{-5}$ & $2.6829 \times 10^{-5}$ & $2.228 \times 10^{-24}$ & - \\
$\alpha_{b}$[END] & $-1.0407 \times 10^{-4}$ & $-1.0407 \times 10^{-4}$ & $1.351 \times 10^{-23}$ & - \\
$\eta_{a}$[END]   & $1.4456 \times 10^{-2}$ & $5.0798 \times 10^{-1}$ & $2.436 \times 10^{7}$ & - \\
$\eta'_{a}$[END]  & $-1.8690 \times 10^{-7}$ & $-4.8296 \times 10^{-1}$ & $2.333 \times 10^{7}$ & - \\
$R_{56}$[END]     & $4.4694 \times 10^{-2}$ & $4.4694 \times 10^{-2}$ & $1.204 \times 10^{-18}$ & - \\
max($\beta_a$) & $7.0000 \times 10^{2}$ & $1.8362 \times 10^{2}$ & 0 & BEGINNING \\
max($\beta_b$) & $7.0000 \times 10^{2}$ & $5.9605 \times 10^{2}$ & 0 & D1103\#2 \\
\midrule
\multicolumn{5}{l}{\textbf{Third Pass}} \\
$\beta_{a}$[END]  & $1.3998 \times 10^{0}$  & $1.3998 \times 10^{0}$  & $1.098 \times 10^{-16}$ & - \\
$\beta_{b}$[END]  & $1.0669 \times 10^{1}$  & $1.0669 \times 10^{1}$  & $5.099 \times 10^{-15}$ & - \\
$\alpha_{a}$[END] & $2.4658 \times 10^{-5}$ & $1.0789 \times 10^{0}$  & $1.164 \times 10^{6}$ & - \\
$\alpha_{b}$[END] & $-1.0462 \times 10^{-4}$ & $-4.7091 \times 10^{-1}$ & $2.217 \times 10^{5}$ & - \\
$\eta_{a}$[END]   & $3.2967 \times 10^{-2}$  & $5.0604 \times 10^{-1}$  & $2.238 \times 10^{7}$ & - \\
$\eta'_{a}$[END]  & $-3.7558 \times 10^{-7}$ & $-2.2781 \times 10^{-1}$ & $5.190 \times 10^{6}$ & - \\
$R_{56}$[END]     & $1.2809 \times 10^{-1}$  & $1.2809 \times 10^{-1}$  & $1.850 \times 10^{-15}$ & - \\
max($\beta_a$) & $7.0000 \times 10^{2}$ & $7.0000 \times 10^{2}$ & $3.158 \times 10^{-17}$ & M13Q12 \\
max($\beta_b$) & $7.0000 \times 10^{2}$ & $2.3230 \times 10^{2}$ & 0 & MCB1 \\
\midrule
\multicolumn{5}{l}{\textbf{Fourth Pass}} \\
$\beta_{a}$[END]  & $1.5386 \times 10^{0}$  & $1.5386 \times 10^{0}$  & $3.333 \times 10^{-19}$ & - \\
$\beta_{b}$[END]  & $1.1607 \times 10^{1}$  & $1.1607 \times 10^{1}$  & $1.262 \times 10^{-19}$ & - \\
$\alpha_{a}$[END] & $2.4713 \times 10^{-5}$ & $2.4713 \times 10^{-5}$ & $2.500 \times 10^{-22}$ & - \\
$\alpha_{b}$[END] & $-1.1455 \times 10^{-4}$ & $-1.1455 \times 10^{-4}$ & $6.285 \times 10^{-24}$ & - \\
$\eta_{a}$[END]   & $4.8495 \times 10^{-2}$  & $1.1647 \times 10^{-1}$  & $4.620 \times 10^{5}$ & - \\
$\eta'_{a}$[END]  & $-5.3096 \times 10^{-7}$ & $9.7679 \times 10^{-4}$ & $9.552 \times 10^{1}$ & - \\
$R_{56}$[END]     & $1.9190 \times 10^{-1}$  & $1.9190 \times 10^{-1}$  & $7.704 \times 10^{-19}$ & - \\
max($\beta_a$) & $7.0000 \times 10^{2}$ & $6.9511 \times 10^{2}$ & 0 & D1505\#2 \\
max($\beta_b$) & $7.0000 \times 10^{2}$ & $2.4895 \times 10^{2}$ & 0 & MCB2 \\
\midrule
\multicolumn{5}{l}{\textbf{Fifth Pass}} \\
$\beta_{a}$[END]  & $1.6261 \times 10^{0}$  & $1.6261 \times 10^{0}$  & $3.865 \times 10^{-19}$ & - \\
$\beta_{b}$[END]  & $1.4063 \times 10^{1}$  & $1.4063 \times 10^{1}$  & $1.818 \times 10^{-17}$ & - \\
$\alpha_{a}$[END] & $2.5401 \times 10^{-5}$ & $2.5401 \times 10^{-5}$ & $4.689 \times 10^{-23}$ & - \\
$\alpha_{b}$[END] & $-1.3987 \times 10^{-4}$ & $-1.3987 \times 10^{-4}$ & $2.454 \times 10^{-23}$ & - \\
$\eta_{a}$[END]   & $6.1641 \times 10^{-2}$  & $6.9842 \times 10^{-1}$  & $4.055 \times 10^{7}$ & - \\
$\eta'_{a}$[END]  & $-6.6020 \times 10^{-7}$ & $-1.6724 \times 10^{-2}$ & $2.797 \times 10^{4}$ & - \\
$R_{56}$[END]     & $2.4233 \times 10^{-1}$  & $2.4233 \times 10^{-1}$  & $7.704 \times 10^{-19}$ & - \\
max($\beta_a$) & $7.0000 \times 10^{2}$ & $2.4720 \times 10^{2}$ & 0 & D1710\#5 \\
max($\beta_b$) & $7.0000 \times 10^{2}$ & $6.0681 \times 10^{2}$ & 0 & M17Q6 \\
\midrule
\multicolumn{5}{l}{\textbf{Sixth Pass}} \\
$\beta_{a}$[END]  & $1.6824 \times 10^{0}$  & $1.6824 \times 10^{0}$  & $3.155 \times 10^{-20}$ & - \\
$\beta_{b}$[END]  & $2.2819 \times 10^{1}$  & $2.2819 \times 10^{1}$  & $1.262 \times 10^{-17}$ & - \\
$\alpha_{a}$[END] & $2.6113 \times 10^{-5}$ & $2.6113 \times 10^{-5}$ & $7.736 \times 10^{-24}$ & - \\
$\alpha_{b}$[END] & $-2.2893 \times 10^{-4}$ & $-2.2893 \times 10^{-4}$ & $1.156 \times 10^{-22}$ & - \\
$\eta_{a}$[END]   & $7.2865 \times 10^{-2}$  & $6.3143 \times 10^{-2}$  & $9.452 \times 10^{3}$ & - \\
$\eta'_{a}$[END]  & $-7.6926 \times 10^{-7}$ & $-6.2086 \times 10^{-4}$ & $3.845 \times 10^{1}$ & - \\
$R_{56}$[END]     & $2.8081 \times 10^{-1}$  & $2.8081 \times 10^{-1}$  & 0 & - \\
max($\beta_a$) & $7.0000 \times 10^{2}$ & $2.9889 \times 10^{2}$ & 0 & M19Q2 \\
max($\beta_b$) & $7.0000 \times 10^{2}$ & $2.7170 \times 10^{2}$ & 0 & MCB4B \\

\end{longtable}

\sisetup{exponent-product = \cdot} 

\begin{longtable}{lS[table-format=2.2]S[table-format=-1.4] @{\hskip 1cm} lS[table-format=2.2]S[table-format=-1.4]}
\caption{Optimized integrated quadrupole strengths ($K_1$) for each of the six passes. The longitudinal position of the center of each magnet is given by the S-coordinate. All quadrupole strengths were limited to a range of $\pm \SI{0.6}{\per\meter\squared}$ during the optimization.}
\label{tab:WFT_R56_Twiss_quad_strengths} \\

\toprule
\textbf{Element} & {\textbf{S-Pos. (m)}} & {\textbf{$K_1$ (\si{\per\meter\squared})}} & \textbf{Element} & {\textbf{S-Pos. (m)}} & {\textbf{$K_1$ (\si{\per\meter\squared})}} \\
\midrule
\endfirsthead

\caption{(Continued)} \\
\toprule
\textbf{Element} & {\textbf{S-Pos. (m)}} & {\textbf{$K_1$ (\si{\per\meter\squared})}} & \textbf{Element} & {\textbf{S-Pos. (m)}} & {\textbf{$K_1$ (\si{\per\meter\squared})}} \\
\midrule
\endhead

\bottomrule
\endfoot

\multicolumn{6}{c}{\textbf{First Pass (P9) Quadrupoles}} \\
\midrule
M9Q1 & 20.66 & -0.4724 & M9Q11 & 57.28 & 0.5236 \\
M9Q2 & 24.16 & 0.6000 & M9Q12 & 58.28 & 0.5403 \\
M9Q3 & 27.66 & -0.2600 & M9Q13 & 64.78 & -0.1129 \\
M9Q4 & 31.16 & -0.6000 & M9Q14 & 65.78 & -0.3928 \\
M9Q5 & 32.16 & -0.4271 & M9Q15 & 69.78 & -0.1660 \\
M9Q6 & 40.36 & -0.1253 & M9Q16 & 70.78 & 0.5417 \\
M9Q7 & 41.36 & 0.5453 & M9Q17 & 79.88 & 0.2857 \\
M9Q8 & 46.66 & 0.1583 & M9Q18 & 80.78 & 0.2811 \\
M9Q9 & 51.66 & -0.6000 & M9Q19 & 81.68 & -0.1065 \\
M9Q10 & 52.66 & 0.0714 & M9Q20 & 82.58 & -0.5491 \\
\midrule
\multicolumn{6}{c}{\textbf{Second Pass (P11) Quadrupoles}} \\
\midrule
M11Q1 & 32.11 & 0.1350 & M11Q8 & 63.28 & -0.0939 \\
M11Q2 & 36.36 & -0.0002 & M11Q9 & 67.54 & -0.0687 \\
M11Q3 & 37.36 & 0.2417 & M11Q10 & 71.80 & -0.1106 \\
M11Q4 & 42.86 & -0.5269 & M11Q11 & 77.21 & -0.1655 \\
M11Q5 & 48.50 & 0.6000 & M11Q12 & 78.21 & -0.1339 \\
M11Q6 & 57.86 & 0.5963 & M11Q13 & 81.21 & -0.1222 \\
M11Q7 & 58.86 & -0.4928 & M11Q14 & 82.21 & 0.6000 \\
\midrule
\multicolumn{6}{c}{\textbf{Third Pass (P13) Quadrupoles}} \\
\midrule
M13Q1 & 34.16 & -0.3901 & M13Q7 & 60.66 & -0.0095 \\
M13Q2 & 35.16 & 0.4097 & M13Q8 & 61.66 & -0.1809 \\
M13Q3 & 43.66 & 0.1338 & M13Q9 & 67.28 & -0.3596 \\
M13Q4 & 50.55 & -0.0937 & M13Q10 & 72.46 & -0.5005 \\
M13Q5 & 55.30 & -0.6000 & M13Q11 & 73.46 & 0.0706 \\
M13Q6 & 59.66 & -0.2304 & M13Q12 & 74.46 & 0.6000 \\
\midrule
\multicolumn{6}{c}{\textbf{Fourth Pass (P15) Quadrupoles}} \\
\midrule
M15Q1 & 30.86 & 0.1283 & M15Q6 & 62.04 & -0.0503 \\
M15Q2 & 31.66 & -0.3726 & M15Q7 & 63.04 & 0.3662 \\
M15Q3 & 42.35 & 0.3261 & M15Q8 & 68.11 & -0.0253 \\
M15Q4 & 48.36 & -0.2382 & M15Q9 & 72.70 & -0.6000 \\
M15Q5 & 54.05 & -0.4196 & M15Q10 & 76.62 & 0.3542 \\
\midrule
\multicolumn{6}{c}{\textbf{Fifth Pass (P17) Quadrupoles}} \\
\midrule
M17Q1 & 36.56 & 0.2012 & M17Q8 & 64.50 & -0.0014 \\
M17Q2 & 37.36 & -0.0837 & M17Q9 & 68.63 & -0.1290 \\
M17Q3 & 38.16 & -0.0721 & M17Q10 & 69.63 & -0.1160 \\
M17Q4 & 49.23 & -0.0697 & M17Q11 & 74.48 & 0.1427 \\
M17Q5 & 49.87 & 0.4606 & M17Q12 & 75.48 & 0.2015 \\
M17Q6 & 54.09 & -0.4186 & M17Q13 & 77.48 & 0.0679 \\
M17Q7 & 62.50 & 0.3701 & M17Q14 & 78.48 & -0.0709 \\
\midrule
\multicolumn{6}{c}{\textbf{Sixth Pass (P19) Quadrupoles}} \\
\midrule
M19Q1 & 33.36 & -0.4822 & M19Q9 & 73.75 & -0.4762 \\
M19Q2 & 33.96 & 0.5789 & M19Q10 & 74.75 & 0.4508 \\
M19Q3 & 43.86 & 0.1200 & M19Q11 & 78.75 & 0.5477 \\
M19Q4 & 44.86 & -0.1308 & M19Q12 & 79.75 & -0.5497 \\
M19Q5 & 45.86 & -0.1168 & M19Q13 & 80.75 & 0.1366 \\
M19Q6 & 61.81 & 0.6000 & M19Q14 & 86.75 & 0.5520 \\
M19Q7 & 62.81 & 0.3871 & M19Q15 & 87.75 & -0.4565 \\
M19Q8 & 63.81 & -0.5773 & & & \\

\end{longtable}
\FloatBarrier

%% file: Appendices/AppendixB.tex
\section{Aperture Analysis and Beam Envelope}
\label{sec:AppB_Aperture}

A key consideration for the viability of the optics solutions presented in this document is whether the resulting beam envelope remains within a reasonable physical aperture, particularly given the large peak $\beta$-functions observed in some cases (exceeding \SI{1000}{\meter}). This appendix provides a preliminary analysis to address this concern.

The beam size ($\sigma$) is determined by the geometric emittance ($\epsilon_{\text{geom}}$) and the $\beta$-function ($\beta$) according to $\sigma = \sqrt{\beta \epsilon_{\text{geom}}}$. The geometric emittance is related to the normalized emittance ($\epsilon_n$) and the relativistic factor ($\gamma_r$) by $\epsilon_{\text{geom}} = \epsilon_n / \gamma_r$.

For this analysis, we adopt the projected normalized emittance for the FFA@CEBAF upgrade, $\epsilon_n = \SI{80}{\milli\meter\cdot\milli\radian}$ \cite{deitrickCEBAF22GeV23}. We then examine the maximum $\beta$-function ($\beta_{\text{max}}$) encountered for each of the six passes across all matching solutions presented. The required beam-stay-clear (BSC) is calculated based on the standard CEBAF operational requirement of a $10\sigma$ envelope.

Table \ref{tab:aperture_analysis} summarizes the results. The worst-case beam envelope occurs at the location of the maximum value of $\beta_{\text{max}} / \gamma_r$.

\begin{table}[htbp]
\centering
\caption{Calculation of the $10\sigma$ beam-stay-clear (BSC) envelope for each pass, using the projected FFA@CEBAF normalized emittance and the maximum $\beta$-function found in any solutions presented in this work.}
\label{tab:aperture_analysis}
\begin{tabular}{@{} c  S[table-format=5.0] S[table-format=4.1] S[table-format=1.2] S[table-format=2.1] @{}}
\toprule
\multicolumn{1}{c}{\textbf{Energy (GeV)}} & \multicolumn{1}{c}{\textbf{$\approx \gamma_r$}} & \multicolumn{1}{c}{\textbf{$\beta_{\text{max}}$ (m)}} & \multicolumn{1}{c}{\textbf{$1\sigma$ Size (mm)}} & \multicolumn{1}{c}{\textbf{$10\sigma$ BSC (mm)}} \\
\midrule
10.55 & 20646 & 1134.2 & 2.10 & 21.0 \\
12.95 & 25342 &  710.2 & 1.50 & 15.0 \\
14.95 & 29256 & 1309.7 & 1.89 & 18.9 \\
17.15 & 33562 & 1075.5 & 1.60 & 16.0 \\
19.35 & 37867 & 1459.0 & 1.76 & 17.6 \\
21.55 & 42172 &  869.2 & 1.28 & 12.8 \\
\bottomrule
\end{tabular}
\end{table}

The analysis indicates that the largest required beam envelope occurs in Pass 1, reaching a $10\sigma$ stay-clear radius of approximately \SI{21.0}{\milli\meter}. This maximum occurs in the solution matching the SFT LINAC input to the "Beginning" FFA match point without $R_{56}$ compensation (Table \ref{tab:SFT_Beginning_Twiss_full_summary} in Appendix~\ref{sec:AppA}).

A required aperture radius of \SI{21}{\milli\meter} is well within the typical dimensions of accelerator vacuum chambers (e.g., \SIrange{40}{50}{\milli\meter} diameter). Therefore, this preliminary analysis confirms that even the largest $\beta$-function excursions found in the presented optics solutions do not pose an aperture limitation for the beam. Further detailed tracking studies including misalignment and momentum offsets are planned for future work.

%% file: Appendices/Appendix_AI_Statement.tex
\section{Declaration of AI Assistance in this Work}
\label{sec:App_AI_Statement}

During the preparation of this work the authors used AI tools, mostly Google's Gemini, in order to assist with various elements of the preparation of this work. This assistance includes the formatting of tables and text, checking for overall consistency of the paper, and other small, editorial tasks. After using this tool/service, the authors reviewed and edited the content as needed and take full responsibility for the content of the published article.

%% file: Bibliography/SplitterBib.bib
@article{bartnikCBETAFirstMultipass2020,
  title = {{{CBETA}}: {{First Multipass Superconducting Linear Accelerator}} with {{Energy Recovery}}},
  shorttitle = {{{CBETA}}},
  author = {Bartnik, A. and Banerjee, N. and Burke, D. and Crittenden, J. and Deitrick, K. and Dobbins, J. and Gulliford, C. and Hoffstaetter, G. H. and Li, Y. and Lou, W. and Quigley, P. and Sagan, D. and Smolenski, K. and Berg, J. S. and Brooks, S. and Hulsart, R. and Mahler, G. and Meot, F. and Michnoff, R. and Peggs, S. and Roser, T. and Trbojevic, D. and Tsoupas, N. and Miyajima, T.},
  date = {2020-07-23},
  journaltitle = {Physical Review Letters},
  shortjournal = {Phys. Rev. Lett.},
  volume = {125},
  number = {4},
  pages = {044803},
  issn = {0031-9007, 1079-7114},
  doi = {10.1103/PhysRevLett.125.044803},
  url = {https://link.aps.org/doi/10.1103/PhysRevLett.125.044803},
  urldate = {2025-11-13},
  langid = {english}
}

@report{beneschConventionalDipoleFFA,
  title = {Conventional {{Dipole}} for {{FFA Splitters}}, {{JLAB-TN-23-016}}},
  author = {Benesch, Jay},
  url = {https://jlabdoc.jlab.org/docushare/dsweb/Get/Document-271507/23-016.pdf},
  urldate = {2025-11-13}
}

@report{beneschCorrectorConceptFFA,
  title = {Corrector {{Concept}} for {{FFA Arcs}}, {{JLAB-TN-22-034}}},
  author = {Benesch, Jay},
  url = {https://jlabdoc.jlab.org/docushare/dsweb/Get/Document-259373/22-034.pdf},
  urldate = {2025-11-17}
}

@report{beneschExtendedLambertsonFFA,
  title = {Extended {{Lambertson}} for {{FFA}}, {{JLAB-TN-22-041}}},
  author = {Benesch, Jay},
  url = {https://jlabdoc.jlab.org/docushare/dsweb/Get/Document-261003/22-041.pdf},
  urldate = {2025-11-13}
}

@report{beneschFirstAttemptConductivelycooled,
  title = {First Attempt at a Conductively-Cooled Superconducting Septum Magnet for the {{FFA}} Upgrade, {{JLAB-TN-22-033}}},
  author = {Benesch, Jay},
  url = {https://jlabdoc.jlab.org/docushare/dsweb/Get/Document-259326/22-033.pdf},
  urldate = {2025-11-13}
}

@report{beneschMagnetAllow11,
  title = {C Magnet to Allow 11 {{GeV FFA}} Tests in {{BSY}} Dump Line, {{JLAB-TN-23-017}}},
  author = {Benesch, Jay},
  url = {https://jlabdoc.jlab.org/docushare/dsweb/Get/Document-271580/23-017.pdf},
  urldate = {2025-11-13}
}

@report{beneschRectangularCommonDipoles,
  title = {Rectangular Common Dipoles for {{FFA}}, {{JLAB-TN-23-041}}},
  author = {Benesch, Jay},
  url = {https://jlabdoc.jlab.org/docushare/dsweb/Get/Document-272883/23-041.pdf},
  urldate = {2025-11-13}
}

@report{beneschSecondAttemptConductivelycooled,
  title = {Second Attempt at a Conductively-Cooled Superconducting Septum Magnet for the {{FFA}} Upgrade, {{JLAB-TN-22-037}}},
  author = {Benesch, Jay},
  url = {https://jlabdoc.jlab.org/docushare/dsweb/Get/Document-259879/22-037.pdf},
  urldate = {2025-11-13}
}

@online{beneschSimpleModificationDC2023a,
  title = {A {{Simple Modification}} of {{DC Current Septa}} to {{Reduce Current Density}} by {{Half}}},
  author = {Benesch, Jay},
  date = {2023-09-06},
  eprint = {2309.00619},
  eprinttype = {arXiv},
  eprintclass = {physics},
  doi = {10.48550/arXiv.2309.00619},
  url = {http://arxiv.org/abs/2309.00619},
  urldate = {2025-11-13},
  pubstate = {prepublished}
}

@report{beneschThirdAttemptConductivelycooleda,
  title = {Third Attempt at a Conductively-Cooled Superconducting Septum Magnet for the {{FFA}} Upgrade, {{JLAB-TN-22-052}}},
  author = {Benesch, Jay},
  url = {https://jlabdoc.jlab.org/docushare/dsweb/Get/Document-262245/22-052.pdf},
  urldate = {2025-11-13}
}

@report{beneschWatercooledCopperSeptum,
  title = {Water-Cooled Copper Septum for {{FFA}}, {{JLAB-TN-22-054}}},
  author = {Benesch, Jay},
  url = {https://jlabdoc.jlab.org/docushare/dsweb/Get/Document-262317/22-054.pdf},
  urldate = {2025-11-13}
}

@report{beneschZAModificationConcept,
  title = {{{ZA}} Modification Concept, {{JLAB-TN-22-051}}},
  author = {Benesch, Jay},
  url = {https://jlabdoc.jlab.org/docushare/dsweb/Get/Document-261885/22-051.pdf},
  urldate = {2025-11-13}
}

@report{bodensteinHorizontalSplitterDesign,
  title = {Horizontal {{Splitter Design}} for {{FFA}}@{{CEBAF}}: {{Focus}} on {{Geometry JLAB-TN-23-069}}},
  author = {Bodenstein, Ryan},
  url = {https://jlabdoc.jlab.org/docushare/dsweb/Get/Document-277999/23-069.pdf},
  langid = {english}
}

@article{brooksOpenmidplaneGradientPermanent23,
  title = {Open-Midplane Gradient Permanent Magnet with 1.53 {{T}} Peak Field},
  author = {Brooks, S.J.},
  editor = {{Assmann,Ralph} and {McIntosh,Peter} and {Fabris,Alessandro} and {Bisoffi,Giovanni} and {Andrian,Ivan} and {Vinicola,Giulia}},
  date = {0023-09-26},
  pages = {3870-3873 pages, 1.6 MB},
  publisher = {JACoW Publishing},
  issn = {2673-5490},
  doi = {10.18429/JACOW-IPAC2023-WEPM128},
  url = {https://jacow.org/ipac2023/doi/jacow-ipac2023-wepm128},
  urldate = {2025-11-13},
  isbn = {9783954502318},
  langid = {english}
}

@article{brooksPermanentMagnetsCEBAF2022,
  title = {Permanent {{Magnets}} for the {{CEBAF 24 GeV Upgrade}}},
  author = {Brooks, Stephen and Bogacz, Alex},
  editor = {Zimmermann, Frank and Tanaka, Hitoshi and Sudmuang, Porntip and Klysubun, Prapong and Sunwong, Prapaiwan and Chanwattana, Thakonwat and Petit-Jean-Genaz, Christine and Schaa, Volker R.W.},
  date = {2022},
  pages = {4 pages, 0.715 MB},
  publisher = {JACoW Publishing, Geneva, Switzerland},
  issn = {2673-5490},
  doi = {10.18429/JACOW-IPAC2022-THPOTK011},
  url = {https://jacow.org/ipac2022/doi/JACoW-IPAC2022-THPOTK011.html},
  urldate = {2025-11-13},
  isbn = {9783954502271},
  langid = {english}
}

@article{coxeBeamCorrectionMultipass2024,
  title = {Beam Correction for Multi-Pass Arcs in {{FFA}}@{{CEBAF}}: Status Update},
  shorttitle = {Beam Correction for Multi-Pass Arcs in {{FFA}}@{{CEBAF}}},
  author = {Coxe, Alexander and {Benesch,Jay} and {Price,Katheryne} and {Deitrick,Kirsten} and {Bodenstein,Ryan} and {Satogata,Todd}},
  editor = {{Pilat,Fulvia} and {Fischer,Wolfram} and {Saethre,Robert} and {Anisimov,Petr} and {Andrian,Ivan}},
  date = {2024-07-01},
  pages = {1057-1059 pages, 0.44 MB},
  publisher = {JACoW Publishing},
  issn = {2673-5490},
  doi = {10.18429/JACOW-IPAC2024-TUPC23},
  url = {https://jacow.org/ipac2024/doi/jacow-ipac2024-tupc23},
  urldate = {2025-11-13},
  isbn = {9783954502479},
  langid = {english}
}

@article{coxeErrorCorrectionAnalysis2024,
  title = {Error and {{Correction Analysis}} for the {{FFA}}@{{CEBAF Energy Upgrade}}},
  author = {Coxe, Alexander},
  date = {2024-10-01},
  journaltitle = {Physics Theses \& Dissertations},
  doi = {10.25777/q9mc-vk53},
  url = {https://digitalcommons.odu.edu/physics_etds/212}
}

@report{coxeFFACEBAFAlignmentCorrector,
  title = {{{FFA}}@{{CEBAF}}: {{Alignment}} of {{Corrector Magnets}} \& {{BPM Readings}} in the {{West FFA Arc}}, {{JLAB-TN-23-068}}},
  author = {Coxe, Alexander},
  url = {https://jlabdoc.jlab.org/docushare/dsweb/Get/Document-277958/23-068.pdf},
  langid = {english}
}

@article{coxeStatusErrorCorrection23,
  title = {Status of Error Correction Studies in Support of {{FFA}}@{{CEBAF}}},
  author = {Coxe, A.M. and Bodenstein, R.M. and Benesch, J. and Bogacz, S.A. and Deitrick, K.E. and Morozov, V and {Seryi,Andrei} and {Trbojevic,Dejan} and {Khan,Donish} and {Gamage,Bamunuvita} and {Brooks,Stephen} and {Price,Katheryne} and {Berg,J.}},
  editor = {{Assmann,Ralph} and {McIntosh,Peter} and {Fabris,Alessandro} and {Bisoffi,Giovanni} and {Andrian,Ivan} and {Vinicola,Giulia}},
  date = {0023-09-26},
  pages = {949-951 pages, 2.3 MB},
  publisher = {JACoW Publishing},
  issn = {2673-5490},
  doi = {10.18429/JACOW-IPAC2023-MOPL177},
  url = {https://jacow.org/ipac2023/doi/jacow-ipac2023-mopl177},
  urldate = {2025-11-13},
  isbn = {9783954502318},
  langid = {english}
}

@article{deitrickCEBAF22GeV23,
  title = {{{CEBAF}} 22 {{GeV FFA Energy Upgrade}}},
  author = {Deitrick, Kirsten and Bodenstein, R.M. and {Benesch,Jay} and {Morozov,Vasiliy} and {Bogacz,Alex} and {Seryi,Andrei} and {Trbojevic,Dejan} and {Satogata,Todd} and {Roblin,Yves} and {Krafft,Geoffrey} and {Brooks,Stephen} and {Gamage,Bamunuvita} and {Kazimi,Reza} and {Hoffstaetter,Georg} and {Khan,Donish} and {Coxe,Alexander} and {Price,Katheryne} and {Berg,J.}},
  editor = {{Assmann,Ralph} and {McIntosh,Peter} and {Fabris,Alessandro} and {Bisoffi,Giovanni} and {Andrian,Ivan} and {Vinicola,Giulia}},
  date = {0023-09-26},
  pages = {962-964 pages, 0.66 MB},
  publisher = {JACoW Publishing},
  issn = {2673-5490},
  doi = {10.18429/JACOW-IPAC2023-MOPL182},
  url = {https://jacow.org/ipac2023/doi/jacow-ipac2023-mopl182},
  urldate = {2025-11-14},
  isbn = {9783954502318},
  langid = {english}
}

@inproceedings{gamage:napac2025-tup029,
  title = {Design {{Update}} on the {{Transition Beamline}} for the {{CEBAF Energy Upgrade}}},
  booktitle = {Proc. {{NAPAC2025}}},
  author = {Gamage, B.R. and Benesch, J. and Bodenstein, R.M. and Bogacz, S.A. and Deitrick, K.E. and Kazimi, R. and Khan, D. and Krafft, G.A. and Nissen, E. and Roblin, Y.R. and Satogata, T. and Seryi, A. and Morozov, V. and Berg, J.S.},
  date = {2025-08},
  series = {North American Particle Accelerator Conference},
  number = {2025},
  pages = {418--421},
  publisher = {JACoW Publishing, Geneva, Switzerland},
  issn = {2673-7000},
  doi = {10.18429/JACoW-NAPAC2025-TUP029},
  url = {https://indico.jacow.org/event/97/contributions/10854},
  isbn = {978-3-95450-261-5},
  langid = {english},
  paper = {TUP029},
  venue = {Sacramento, CA, USA}
}

@article{gamageResonantMatchingSection2024,
  title = {Resonant Matching Section for {{CEBAF}} Energy Upgrade},
  author = {Gamage, B.R. and {Bogacz,Alex} and {Coxe,Alexander} and {Seryi,Andrei} and {Trbojevic,Dejan} and {Turner,Dennis} and {Khan,Donish} and {Nissen,Edith} and {Meot,Francois} and {Krafft,Geoffrey} and {Hoffstaetter,Georg} and {Berg,J.} and {Price,Katheryne} and {Deitrick,Kirsten} and {Kazimi,Reza} and {Bodenstein,Ryan} and {Brooks,Stephen} and {Satogata,Todd} and {Morozov,Vasiliy} and {Roblin,Yves}},
  editor = {{Pilat,Fulvia} and {Fischer,Wolfram} and {Saethre,Robert} and {Anisimov,Petr} and {Andrian,Ivan}},
  date = {2024-07-01},
  pages = {3075-3078 pages, 1.1 MB},
  publisher = {JACoW Publishing},
  issn = {2673-5490},
  doi = {10.18429/JACOW-IPAC2024-THPC37},
  url = {https://jacow.org/ipac2024/doi/jacow-ipac2024-thpc37},
  urldate = {2025-11-13},
  isbn = {9783954502479},
  langid = {english}
}

@ARTICLE{helmEvaluationSynchrotronRadiation1973,
  author={Helm, R. H. and Lee, M. J. and Morton, P. L. and Sands, M.},
  journal={IEEE Transactions on Nuclear Science}, 
  title={Evaluation of Synchrotron Radiation Integrals}, 
  year={1973},
  volume={20},
  number={3},
  pages={900-901},
  doi={10.1109/TNS.1973.4327284}
}

@article{jowettIntroductoryStatisticalMechanics1987,
    author = {Jowett, John M.},
    title = {Introductory statistical mechanics for electron storage rings},
    journal = {AIP Conference Proceedings},
    volume = {153},
    number = {1},
    pages = {864-970},
    year = {1987},
    month = {02},
    issn = {0094-243X},
    doi = {10.1063/1.36374},
    url = {https://doi.org/10.1063/1.36374},
    eprint = {https://pubs.aip.org/aip/acp/article-pdf/153/1/864/11694813/864_1_online.pdf},
}

@online{gramesPositronBeamsCeBAF2023,
  title = {Positron {{Beams At Ce}}\$\textasciicircum +\${{BAF}}},
  author = {Grames, J. and Benesch, J. and Bruker, M. and Cardman, L. and Covrig, S. and Ghoshal, P. and Gopinath, S. and Gubeli, J. and Habet, S. and Hernandez-Garcia, C. and Hofler, A. and Kazimi, R. and Lin, F. and Nagaitsev, S. and Poelker, M. and Rimmer, B. and Roblin, Y. and Lizarraga-Rubio, V. and Seryi, A. and Spata, M. and Sy, A. and Turner, D. and Ushakov, A. and Valerio-Lizarraga, C. A. and Voutier, E.},
  date = {2023-09-28},
  eprint = {2309.15581},
  eprinttype = {arXiv},
  eprintclass = {physics},
  doi = {10.48550/arXiv.2309.15581},
  url = {http://arxiv.org/abs/2309.15581},
  urldate = {2025-11-14},
  pubstate = {prepublished}
}

@article{hiattFabricationMeasurement12,
  title = {Fabrication and {{Measurement}} of 12 {{GeV Prototype Quadrupoles}} at {{Thomas Jefferson National Accelerator Facility}}},
  author = {Hiatt, T and Baggett, K S and Beck, J M and Dail, J G and Harwood, L and Meyers, J and Wiseman, M},
  journaltitle = {Proc. PAC'09},
  series = {Particle {{Accelerator Conference}}},
  pages = {223--225},
  url = {https://jacow.org/PAC2009/papers/MO6PFP037.pdf},
  langid = {english}
}

@inproceedings{kazimi:ipac2025-tupm114,
  title = {An Extraction Scheme for Future {{CEBAF FFA}} Based Energy Upgrade},
  booktitle = {Proc. {{IPAC}}'25},
  author = {Kazimi, R.},
  date = {2025-06},
  series = {{{IPAC}}'25 - 16th International Particle Accelerator Conference},
  number = {16},
  pages = {1407--1410},
  publisher = {JACoW Publishing, Geneva, Switzerland},
  issn = {2673-5490},
  doi = {10.18429/JACoW-IPAC2025-TUPM114},
  url = {https://indico.jacow.org/event/81/contributions/7637},
  eventdate = {2025-06-01/2025-06-06},
  isbn = {978-3-95450-248-6},
  langid = {english},
  paper = {TUPM114},
  venue = {Taipei, Taiwan}
}

@article{nissenDesignProgress222025,
  title = {Design Progress for the 22 {{GeV CEBAF}} Energy Upgrade},
  author = {Nissen, E. and Bogacz, A and Coxe, A and Seryi, A and Gamage, B and Trbojevic, D and Khan, D and Hoffstaetter, G and Neththikumara, I and Berg, J and Deitrick, K and Sereno, N and Kazimi, R and Ruber, R and Bodenstein, R and Ogur, S and Brooks, S and Satogata, T and Morozov, V and Roblin, Y},
  date = {2025-08},
  journaltitle = {Proc. North American Particle Accelerator Conference (NAPAC2025)},
  series = {North {{American Particle Accelerator Conference}}},
  pages = {664--667},
  issn = {2673-7000},
  doi = {10.18429/JACoW-NAPAC2025-WEZN01},
  url = {https://indico.jacow.org/event/97/contributions/9576},
  langid = {english}
}

@article{saganBmadRelativisticCharged2006,
  title = {Bmad: {{A}} Relativistic Charged Particle Simulation Library},
  shorttitle = {Bmad},
  author = {Sagan, D.},
  date = {2006-03},
  journaltitle = {Nuclear Instruments and Methods in Physics Research Section A: Accelerators, Spectrometers, Detectors and Associated Equipment},
  shortjournal = {Nuclear Instruments and Methods in Physics Research Section A: Accelerators, Spectrometers, Detectors and Associated Equipment},
  volume = {558},
  number = {1},
  pages = {356--359},
  issn = {01689002},
  doi = {10.1016/j.nima.2005.11.001},
  url = {https://linkinghub.elsevier.com/retrieve/pii/S0168900205020371},
  urldate = {2025-11-13},
  langid = {english}
}

@online{saganTaoSimulationProgram2025,
  title = {Tao {{Simulation Program}}},
  author = {Sagan, David},
  date = {2025-10-25},
  url = {https://www.classe.cornell.edu/bmad/tao.html},
  urldate = {2025-10-27},
  organization = {Tao Manual}
}

@online{voutierJeffersonLabPositron2024,
  title = {The {{Jefferson Lab}} Positron Physics Program},
  author = {Voutier, Eric},
  date = {2024-01-30},
  eprint = {2401.16223},
  eprinttype = {arXiv},
  eprintclass = {nucl-ex},
  doi = {10.48550/arXiv.2401.16223},
  url = {http://arxiv.org/abs/2401.16223},
  urldate = {2025-11-14},
  pubstate = {prepublished}
}
